\documentstyle[prd,aps,preprint,tighten,eqsecnum,epsf]{revtex}
\newcommand{\RR}{\hbox{$I$\kern-3.8pt $R$}}
\newcommand{\rr}{\hbox{$\scriptstyle I$\kern-2.4pt $\scriptstyle R$}}
\newcommand{\be}{\begin{equation}}
\newcommand{\ee}{\end{equation}}
\newcommand{\bea}{\begin{eqnarray}}
\newcommand{\eea}{\end{eqnarray}}
\newcommand{\ab}{{\alpha\beta}}
\newcommand{\sss}{\scriptscriptstyle}
\newcommand{\diffsigma}{{\rm Diff}\Sigma}
\newcommand{\diffs}{{\rm Diff}\S}
\newcommand{\ldiffsigma}{{\rm LDiff}\Sigma}

\newcommand{\ldifftxs}{{\rm LDiff}({\cal T}\times{\cal S})}

\newcommand{\T}{{\cal T}}
\renewcommand{\S}{{\cal S}}
\newcommand{\M}{{\cal M}}
\newcommand{\F}{{\cal F}}
\renewcommand{\P}{{\cal P}}
\renewcommand{\H}{{\cal H}}
\renewcommand{\zeta}{z}
\newcommand{\boldV}{{\bbox{V}}}
\newcommand{\boldW}{{\bbox{W}}}
\newcommand{\boldU}{{\bbox{U}}}
\newcommand{\boldH}{{\bbox{H}}}
\newcommand{\boldG}{{\bbox{G}}}

\newcommand{\boldF}{{\bbox{F}}}
\newcommand{\boldh}{{\bbox{h}}}
\newcommand{\bXi}{{\bbox{\Xi}}}
\newcommand{\bTh}{{\bbox{\Theta}}}
\newcommand{\bOm}{{\bbox{\Omega}}}
\newcommand{\bPsi}{{\bbox{\Psi}}}
\newcommand{\bPhi}{{\bbox{\Phi}}}
\newcommand{\bpsi}{{\bbox{\psi}}}
\newcommand{\bphitoo}{{\bbox{\phi}}}
\newcommand{\bxi}{{\bbox{\xi}}}
\newcommand{\bth}{{\bbox{\vartheta}}}

\newcommand{\bg}{{\bbox{g}}}
\newcommand{\bp}{{\bbox{p}}}
\newcommand{\bT}{{\bbox{T}}}
\newcommand{\bP}{{\bbox{P}}}

\newcommand{\dsurface}{
  {\hbox{ $d$\kern-0.3pt
  \lower1.25ex\hbox{${\stackrel{\textstyle\Sigma}{\sim}}$}
  \kern-2.5pt}}{}_\alpha }
\newcommand{\alternate}{
  {\hbox{ \lower1.25ex\hbox{${\stackrel{\textstyle\delta}{\sim}}$}
  \kern-2.8pt}}{}_{\alpha\beta\gamma\delta} }
\newcommand{\Jtilde}{
   {\stackrel{\hbox{\lower1.7ex\hbox{$\sim$}}}{J} }{}^\alpha   }
\newcommand{\Jtildeb}{
   {\stackrel{\hbox{\lower1.7ex\hbox{$\sim$}}}{J} }{}^\beta   }
\begin{document}
\preprint{August 1994} 

\title{Dust as a Standard of Space and Time\\
   in Canonical Quantum Gravity}
\author{J. David Brown}
\address{Department of Physics and
    Department of Mathematics,\\ North Carolina State University,
    Raleigh, NC 27695--8202}
\author{Karel V. Kucha\v r}
\address{Department of Physics, University of Utah,
   Salt Lake City, UT 84112}
\maketitle

\begin{abstract}
The coupling of the metric to an incoherent dust introduces into
spacetime a privileged dynamical reference frame and time
foliation. The comoving coordinates of the dust particles and the
proper time along the dust worldlines become canonical coordinates in
the phase space of the system.  The Hamiltonian constraint can be
resolved with respect to the momentum that is canonically conjugate to
the dust time. Imposition of the resolved constraint as an operator
restriction on the quantum states yields a functional Schr\"{o}dinger
equation. The ensuing Hamiltonian density has an extraordinary
feature: it depends only on the geometric variables, not on the dust
coordinates or time. This has three important consequences. First, the
functional Schr\"{o}dinger equation can be solved by separating the
dust time from the geometric variables.  Second, the Hamiltonian
densities strongly commute and therefore can be simultaneously defined
by spectral analysis. Third, the standard constraint system of vacuum
gravity is cast into a form in which it generates a true Lie
algebra. The particles of dust introduce into space a privileged
system of coordinates that allows the supermomentum constraint to be
solved explicitly. The Schr\"{o}dinger equation yields a conserved
inner product that can be written in terms of either the instantaneous
state functionals or the solutions of constraints. Gravitational
observables admit a similar dual representation. Examples of
observables are given, though neither the intrinsic metric nor the
extrinsic curvature are observables. Disregarding the standard
factor--ordering difficulties, the introduction of dust provides a
satisfactory phenomenological approach to the problem of time in
canonical quantum gravity.
\end{abstract}
\pacs{???}
\section{Introduction}

The Dirac constraint quantization of vacuum Einstein gravity yields the
Wheeler--DeWitt equation for the quantum state of the intrinsic three
geometry
of space \cite{Wheeler,DW}.  One can view this equation as a statement that
only two out
of three independent components of the intrinsic geometry are dynamical. The
third component is an intrinsic time that specifies the location of space as
a hypersurface in spacetime.  The Wheeler--DeWitt equation is then
interpreted as an evolution equation for the state in the intrinsic time.

The Wheeler--DeWitt equation is a second--order variational differential
equation.  The space of its solutions carries no obvious Hilbert space
structure \cite{K-Ox,k-nosymmetry}. This has prompted numerous attempts
aimed at replacing the
Wheeler--DeWitt equation by a first--order Schr\"{o}dinger equation. In
order to do that, one should identify the intrinsic time at the classical
level,
solve the Hamiltonian constraint for the momentum conjugate to time, and
impose the resolved constraint as an operator restriction on the quantum
states. Unfortunately, there is no natural candidate for the intrinsic
time, and the procedure is beset by a number of conceptual and technical
difficulties \cite{KonT}.

Intrinsic clocks are strange contraptions.  From the early days of general
relativity, most researchers felt that spacetime intervals are to be measured
not by internal clocks but by material devices, analogous to the measurement
of electromagnetic fields by test charges. At the most basic level,  matter
splits the spacetime manifold into space and time. The phenomenological
medium introduced for this purpose is the reference fluid
\cite{Einstein,Hilbert,Moller,Landau}. The particles of
the reference fluid identify the points of space, and clocks carried by these
particles identify the instants of time. In this way, the fluid fixes the
reference frame (the space) and the time foliation (the time). In that frame
and on that foliation, the entire intrinsic metric, not just two selected
components of the intrinsic geometry, becomes dynamical.

The reference fluid is traditionally considered to be a tenuous material
medium whose back reaction on the geometry can be neglected. There are just
enough fluid particles to discern the points of space from one another, but
not enough to disturb the geometry. Instead of deriving the motion of the
fluid from its action, one encodes it in coordinate conditions.
Unfortunately, such a standpoint makes it difficult to view the reference
fluid as physical matter.

There are two alternative routes through which the reference fluid can be
turned into a physical system. The first is to impose the coordinate
conditions
before variation by adjoining them to the Hilbert action with Lagrange
multipliers \cite{K+T,K+S+T}. Equivalently, a Lagrangian might be devised to
enforce  these conditions \cite{Stone+K,K=con}.  The additional terms in the
action can be
interpreted as a matter source coupled to the gravitational field. A
disadvantage of this approach is that simple coordinate conditions often
lead to contrived matter sources. The second route is to select a realistic
material medium for the fluid and to describe its properties by a physical
Lagrangian \cite{DW-medium,DW}. Again, by adding this Lagrangian to the
Hilbert Lagrangian,
the fluid becomes coupled to gravity. The challenge is to find a medium whose
space and time variables can be turned into canonical coordinates that lead,
upon the Dirac constraint quantization, to a simple Schr\"{o}dinger equation
for canonical quantum gravity.

Our goal is to show that the most simple type of matter conceivable,
incoherent dust, is just what one needs to obtain a satisfactory
framework for interpreting quantum gravity. The particles of dust follow
timelike geodesics, and in this sense dust is quite special: Dust is the
`purest' form of matter, matter that interacts with itself only via the
gravitational field.

The canonical scheme offered by incoherent dust has an extraordinary
feature: The many--fingered Hamiltonian of the ensuing functional
Schr\"{o}dinger equation does not depend on the dust variables. This fact
has a number of important consequences. First, the Hamiltonian densities
strongly commute (modulo factor--ordering difficulties), and thus can be
simultaneously defined by spectral analysis. Second, the functional
Schr\"{o}dinger equation can be solved by separating the dust time from the
gravitational variables.  Third, the
fact that the Hamiltonian depends only on the gravitational
variables leads to the discovery of a form for the constraint system of
vacuum gravity in which the constraints generate a true Lie algebra.

The dust particles also endow space
with a privileged, physical system of coordinates. Canonical variables
such as the spatial metric can be specified as tensors in these dust
coordinates, and in this way the combined system of gravity and dust can
be expressed in a form that does not rely on arbitrary systems of
spatial coordinates \cite{Brown2}. As a consequence, the supermomentum
constraint, which is the canonical generator of spatial diffeomorphisms,
becomes trivial and is exactly solvable in the quantum theory.

All of the above features combine into an intriguing scheme for
interpreting canonical quantum gravity. Although we focus on the case of
gravity coupled only to dust, the scheme can be applied in the presence
of other matter couplings as well---as long as dust is included in the
list of matter fields, all of our basic conclusions remain intact. Here
is how the scheme works:

We design a spacetime action whose variation yields the Euler equations of
motion for an incoherent dust (Section 2a).  All of the state variables have
a direct physical significance. In particular, four of these variables are
the proper time and the dust frame (Section 2b); these are the desired tools
for interpreting quantum gravity. The action is unchanged by a subgroup of
diffeomorphisms in the dust spacetime manifold (Appendix C). These global
symmetries yield an infinite set of Noether charges (Section 2e) that are
conserved by virtue of continuity equations (Section 2c). By varying the
dust action with respect to the spacetime metric, the standard
energy--momentum tensor is obtained (Section 2d).  The dust action can
be cast into several equivalent forms (Appendix B).

The ADM decomposition of the dust action, followed by a Legendre dual
transformation, yields the canonical description of dust (Section 3a).
The canonical coordinates coincide with the dust time and frame variables,
while the conjugate momenta are simply related to the mass density and
velocity of the dust.  The canonical form of the dust action can be
derived from the action for a continuously infinite number of
relativistic point particles (Appendix D), and is a special case of the
canonical action for arbitrary perfect fluids \cite{FluidS}. The Noether
charges are represented by linear functionals of the dust momenta (Section
3b).

Dust is coupled to gravity by adding its action to the Dirac--ADM action.
The dynamics is then generated by the standard super--Hamiltonian and
supermomentum constraints on the phase space of the coupled system
(Section 4a). The super--Hamiltonian
and supermomentum constraints can be brought to a form in which they are
resolved in the dust momenta. These new constraints $H_\uparrow(x)$ and
$H_{\uparrow k}(x)$ generate an Abelian
algebra (Section 4b) that allows us to represent diffeomorphisms
in the dust spacetime by canonical transformations (Section 4c).

The new super--Hamiltonian constraint $H_\uparrow(x)$ generates dynamics
along the flow lines of the dust, while the new supermomentum constraint
$H_{\uparrow k}(x)$ generates motion along the hypersurfaces of constant
dust proper time. The true Hamiltonian density in the super--Hamiltonian
constraint is the square root of a quadratic combination $G(x)$ of the
gravitational super--Hamiltonian and supermomentum. As such, it does
not depend on the dust variables. The expressions $G(x)$ have vanishing
Poisson brackets among themselves (Appendix E). When complemented by
the gravitational supermomenta, they provide an alternative set of
constraints for vacuum gravity which generate a true Lie algebra
(Section 4d).

The dust frame variables are fields $Z^k(x)$, $k=1$, $2$, $3$, whose values
at a given space point $x$ constitute a set of Lagrangian coordinate labels
for the dust flow line that passes through $x$. $Z$ can be viewed as a mapping
from the space manifold to the `dust space' manifold whose points $\zeta$ are
the individual flow lines (Section 2b). Correspondingly, $Z$ induces mappings
of tensor fields. In particular, the spatial metric, the dust proper time
variable, and their canonical conjugates can be mapped from space to the
dust space. This mapping of tensors is completed into a canonical
transformation by replacing the momentum conjugate to the dust frame
$Z^k(x)$ with the new supermomentum generator $H_{\uparrow k}(x)$
(Section 5a). Then the constraint $H_{\uparrow k}(x) = 0$ is simply the
statement that the momentum conjugate to $Z^k(x)$ should vanish.
Because the new super--Hamiltonian constraint $H_\uparrow(x)$ is
independent of the dust frame $Z$ and its conjugate, it can be expressed
in terms of the new canonical variables by mapping the space scalar
density $H_\uparrow(x)$ to the dust space. This turns the
super--Hamiltonian constraint into a dust space scalar density
$\boldH_\uparrow(\zeta)$. The true Hamiltonian density in this constraint
is the square root of a dust space density $\boldG(\zeta)$ that is
constructed solely from the gravitational variables. The Noether charges
also can be expressed as functionals of the new canonical variables
(Section 5b).

We quantize gravity coupled to dust by imposing the supermomentum constraint
$H_{\uparrow k}(x) = 0$ and the new super--Hamiltonian constraint
$\boldH_\uparrow(\zeta) = 0$ as operator
restrictions on the states (Section 6). The supermomentum operator constraint
implies that the state functional is independent of $Z$, while the new
Hamiltonian constraint yields a functional Schr\"{o}dinger equation in the
dust time (Section 6a). Since the Hamiltonian density in this equation does
not depend on the dust variables, the dust time can be separated from the
gravitational variables. This leads to a time--independent functional
Schr\"{o}dinger equation (Section 6b). If the operators ${\hat\boldG}(\zeta)$
can be factor ordered and regularized so that they commute, then they possess
joint eigenfunctionals (Section 6c). We show
that the spectrum of ${\hat\boldG}(\zeta)$ is continuous and degenerate
(Appendix F). The square--root Hamiltonian density can be defined by spectral
analysis on the Hilbert space ${\cal H}^{+}$ spanned by the eigenfunctionals
labeled by positive eigenvalues ${\boldG}(\zeta)>0$. The Noether charges
appear in the quantum theory as generators for unitary operators that
transform the quantum states by the corresponding symmetries (Section 6d).
The quantum theory of gravity and dust can be constructed using either
the original set or the new set of canonical variables (Section 6g).

The Schr\"{o}dinger equation yields a conserved inner product that can be
written either in terms of the instantaneous state functionals, or as a
product in the space of solutions (Section 6e). Gravitational observables
admit a similar dual representation (Section 6f). Observables must keep
the state in the Hilbert space ${\cal H}^{+}$. Examples of such
observables are given; unfortunately, neither the intrinsic metric nor the
extrinsic curvature are gravitational observables. A similar difficulty
occurs already in a simple model system, namely, the relativistic particle
in the proper--time formalism (Appendix G).

The classical Noether charges generate canonical transformations that
transform the system from one dust time to another (Section 3b). They carry
a solution of the constraints and of the Hamilton equations of motion again
into a solution. If the dust does not rotate, the solution can be
transformed to the time foliation that is orthogonal to the dust frame
(Section 7a). The restriction to non--rotating dust in the
frame--orthogonal time foliation can be enforced by an additional
first--class constraint.  This constraint can be used to eliminate the
inconvenient square root in the Hamiltonian density $H_\uparrow(x)$
(Section 7b). The resulting Hamiltonian constraint $H_{\uparrow 0}(x)$
underlies our previous study of Gaussian reference fluids \cite{K+T}. Due to
factor--ordering ambiguities, the quantum theories constructed from
$H_\uparrow(x)$ and $H_{\uparrow 0}(x)$ do not necessarily coincide
(Section 7c).

Our notation is summarized in Appendix A.

The introduction of dust alleviates several notorious problems of time
in vacuum gravity:\hfil\break
$\bullet$ When passing to the Schr\"{o}dinger equation, one can rely on
the privileged time and space brought into spacetime by the state of a
material system.\hfil\break
$\bullet$ The dust time is a spacetime scalar. Its value does not depend
on the hypersurface passing through a given event. No intrinsic time in
vacuum gravity has this property.\hfil\break
$\bullet$ The Hamiltonian densities have vanishing Poisson brackets.
If---and this is an overwhelming if---a factor ordering and regularization
can be found for which the corresponding operators ${\hat\boldG}(\zeta)$
commute, then the Hamiltonian density operators can be defined by spectral
analysis and the evolution of states will be independent of the foliation
connecting the final and initial embeddings.
One important problem remains to be solved:  What operators, if any,
correspond to the measurements of the intrinsic metric and extrinsic
curvature on an embedding specified with respect to the dust time and the
dust frame?

To the extent limited by our cautionary remarks, the
introduction of incoherent dust provides a satisfactory phenomenological
approach to the problems of time in canonical quantum gravity. The
corresponding problems at the level of vacuum gravity
remain, of course, unresolved.

Preliminary accounts of our work have been given in
Refs.~\cite{Misnerfest}.
\section{Incoherent Dust}
\subsection{Spacetime Action and the Euler Equations}

Dust is described by eight spacetime scalars,  $T$, $Z^{k}$, $M$,
and $W_{k}$. (Indices $k$, $\ell$ from the middle of the Latin alphabet
take the values $1$, $2$, $3$.) The physical interpretation of these
state variables follows from the analysis of the action and the ensuing
equations of motion. In order to describe the structure of the action in
physical terms, we shall anticipate the results of this analysis. The values
of the variables $Z^k$ are the comoving coordinates of the dust particles,
and the value of the variable $T$ is the proper time measured along the
particle flow lines.  The scalars $Z^{K} = (T,Z^{k})$ are assumed to be four
independent functions of the spacetime coordinates $y^{\alpha}$,
$\alpha = 0$, $1$, $2$, $3$:
\be
\det\left( {Z^K}_{,\alpha} \right) \neq 0 \ . \label{2.1}
\ee
The four--velocity $U_{\alpha}$ of the dust is defined by its decomposition
in the cobasis ${Z^{K}}_{,\alpha}$ as
\begin{equation} U_\alpha = -T_{,\alpha} + W_k {Z^k}_{,\alpha} \ .
        \label{2.2} \end{equation}
The three spatial components $W_{k}$ of the four--velocity in the dust frame
$Z^{k}$ are also state variables.  Equation (2.2) expresses the
four--velocity one--form ${\boldU} = U_{\alpha} dy^{\alpha}$ as a Pfaff form
\cite{Caratheodory}
\begin{equation} {\boldU} = -dT + W_k dZ^k
         \label{2.3} \end{equation}
of the seven scalar fields $T,Z^{k}$ and $W_{k}$.  The eighth state
variable, $M$, is the rest mass density of the dust.

The dust action is a functional of the dust variables and of the spacetime
metric $\gamma _{\ab}$,
\begin{equation} S^D[T,Z^k;M,W_k;\gamma_{\ab}]=\int_{\cal M}d^4y\,
L^D(y) \ , \label{2.4} \end{equation}
with the Lagrangian density
\begin{equation}  L^D = -\frac{1}{2} |\gamma|^{1/2} M \left( \gamma^{\ab}
          U_\alpha U_\beta + 1\right) \ .
          \label{2.5} \end{equation}
It can be cast into many alternative forms (Appendix B).  The
matter equations of motion are obtained by varying the dust action with
respect
to the state variables $M$, $W_{k}$, $T$, and $Z^{k}$:
\begin{eqnarray}  0 & = & \frac{\delta S^D}{\delta M} = -\frac{1}{2}
          |\gamma|^{1/2} \left( \gamma^{\ab} U_\alpha U_\beta + 1\right)
          \ ,\label{2.6}\\
 0 & = & \frac{\delta S^D}{\delta W_k} = -
 |\gamma|^{1/2} M {Z^k}_{,\alpha} U^\alpha \ ,\label{2.7}\\
    0 & = & \frac{\delta S^D}{\delta T} = - \left( |\gamma|^{1/2}
          M U^\alpha \right) _{,\alpha} \ ,\label{2.8}\\
    0 & = & \frac{\delta S^D}{\delta Z^k} = \left( |\gamma|^{1/2}
          M W_k U^\alpha \right) _{,\alpha} \ .\label{2.9}
    \end{eqnarray}
These equations corroborate our interpretation of the dust variables:
\subsection{Dust As a Standard of Space and Time}

Equation (2.6) ensures that the four--velocity $U^{\alpha}$ is a unit timelike
vector field.  According to Eq.~(2.7), the three scalar fields $Z^{k}$ are
constant along the flow lines $U^{\alpha}$ and therefore their values
$\zeta^{k}$
can be interpreted as Lagrangian coordinates for the dust. That is, the
values $\zeta^{k}= Z^{k}(y)$ of the mappings $Z^{k}:{\cal M}\rightarrow{\RR}$
serve as
labels specifying which flow line passes through a given spacetime event $y$.
Such a set of Lagrangian coordinates can be generated by choosing an
arbitrary spacelike hypersurface $\Sigma$ and specifying a coordinate system
$\zeta^{k}$ on that hypersurface.  Then each flow line is labeled by the
coordinates of
the point where it intersects the hypersurface. Inversely, knowing the
congruence ${\cal S}$ of flow lines, we can define the mappings $Z^{k}$ of
${\cal M}$ into $\RR$ by assigning to each event $y$ in ${\cal M}$ the
coordinates $\zeta^{k}$ of the point in $\Sigma$ where the
flow line that passes through $y$ intersects $\Sigma$.

The congruence ${\cal S}$ of dust flow lines introduces a privileged
reference frame into the spacetime manifold ${\cal M}$.  This congruence
can be
viewed as an abstract three-dimensional space, the `dust space', whose points
are the individual flow lines.  As a manifold, the dust space $\cal S$ is
isomorphic to any spacelike hypersurface. (We assume that the spacetime
(${\cal M}$, $\gamma$) is globally hyperbolic with ${\cal M}=
\RR\times\Sigma$.)
A system of coordinates on that hypersurface induces a system of coordinates
$\zeta^{k}$ on ${\cal S}$. It may be impossible to cover the dust space with
a single coordinate chart, so the Lagrangian coordinates $\zeta^{k}$ are
generally defined only in open subsets of the dust space. Correspondingly,
the scalar fields $Z^{k}$ are generally defined only in open tubes of the
space--time manifold. Our interpretation of the fields $Z^{k}$ presupposes
that their values constitute a good set of Lagrangian coordinates,
{\it i.e.\/}, that within the appropriate open tubes each flow line carries
a unique set of labels $\zeta^{k}$. As a consequence, the three gradients
${Z^k}_{,\alpha}$ must be linearly independent covectors. This is ensured
by the assumption (\ref{2.1}).

Our discussion reveals that the three scalar fields $Z^{k}(y)$ are not
necessarily globally defined on the whole spacetime manifold ${\cal M}$.
However, these fields appear in the action (2.4) only in the combination
$W_k {Z^k}_{,\alpha}$, so this defect can be overcome by a slight
reinterpretation of the formalism. Let $Z:{\cal M}\to {\cal S}$ denote a
globally defined mapping of the spacetime ${\cal M}$ into the dust space
${\cal S}$, and view $Z^k$ as the composition of the mapping $Z$ with
a local coordinate chart mapping on $\S$. With the indices $k$ and $\alpha$
interpreted as abstract tensor indices, ${Z^k}_{,\alpha}$ is then globally
defined as the mapping ${Z^k}_{,\alpha} : \M\to T\S\otimes T^*\M$.
Likewise, $W_k$ is defined as the mapping $W_k : \M\to T^*\S$ where, again,
$k$ is viewed as an abstract index. In this way, $W_k{Z^k}_{,\alpha}$ is
understood to be a globally defined element of $T^*\M$.

The dust enables us to introduce not only the privileged matter space, but
also a matter time.  By multiplying Eq.~(2.2) by $U^{\alpha}$ and using the
field equations (2.6)--(2.7), we learn that
\begin{equation} \pounds_{U} T = T_{,\alpha} U^\alpha = 1 \ ,
     \label{2.10} \end{equation}
{\it i.e.\/}, that $T$ is the proper time between a fiducial hypersurface
$T = 0$ and an arbitrary hypersurface $T = {\it const\/}$ along the flow lines
of the dust particles. The fiducial hypersurface can be chosen to be
spacelike.
The foliation of the spacetime manifold ${\cal M}$ by hypersurfaces of
constant
$T$ can be viewed as an abstract one--dimensional space, the `dust time'
${\cal
T}$, whose points (instants) are the individual hypersurfaces.  As a
manifold, the dust time ${\cal T}$ is isomorphic to $\RR$.  The value $\tau =
T(y)$ of the time mapping $T :{\cal M} \rightarrow {\cal T}$ can serve as the
time coordinate of an instant of time that passes through the event $y$.

In our analysis, we must assume that the three covectors ${Z^k}_{,\alpha}$
are linearly independent. That the fourth covector, $T_{,\alpha}$, is linearly
independent of ${Z^k}_{,\alpha}$ follows from the equations of motion
(2.7) and (2.10). As a result, $(T_{,\alpha}, {Z^k}_{,\alpha})$ form a basis
in $T^{*}{\cal M}$, Eq. (\ref{2.1}).  The four--velocity $U^{\alpha}$ of the
dust is a privileged timelike future--pointing vector. Equation (2.10) tells
us that the time function $T(y)$ grows from the past to the future. The
covectors ${Z^k}_{,\alpha}$ are orthogonal to the four--velocity
$U^{\alpha}$ of the
dust and hence spacelike. These relations are sketched in Fig.~1.

Because $(T_{,\alpha},{Z^k}_{,\alpha})$ are linearly independent, the values
$\tau$ and $\zeta^{k}$ of the four mappings $T$ and $Z^{k}$ can be used as
coordinates on ${\cal M}$.  By inverting the mappings
\begin{equation}  \tau = T(y) \ ,\qquad \zeta = Z(y)  \label{2.11}
\end{equation}
we obtain the mapping
\begin{equation}  \Upsilon:\T\times\S \to \M \quad{\rm by}\quad (\tau,\zeta)
     \mapsto
     y = \Upsilon(\tau,\zeta) \ . \label{2.12}
\end{equation}
The four vectors
\begin{equation} U^\alpha := {\Upsilon^\alpha}_{,\tau} \quad {\rm and}\quad
     {\Upsilon^\alpha}_{,k} \label{2.13} \end{equation}
form a basis in $T\M$ dual to the cobasis
\begin{equation} {Z^K}_{,\alpha} = (T_{,\alpha} , {Z^k}_{,\alpha})
\label{2.14}
      \end{equation}
in $T^{*}{\cal M}$.  The basis (2.13) and the cobasis (2.14) satisfy the
standard orthonormality and
completeness relations. As a consequence of the
equations of motion (2.7) and (2.10), these basis and cobasis vectors are
conserved along the flow lines of $U^\alpha$; that is, $\pounds_{U}
T_{,\alpha} = 0 = \pounds_{U} {Z^k}_{,\alpha}$ and
$\pounds_{U} U^\alpha = 0 = \pounds_{U} {Y^\alpha}_{,k}$. The basis
vector $U^{\alpha} : = {\Upsilon^{\alpha}}_{,\tau}$ coincides with the
four--velocity of the dust and the remaining three vectors,
${\Upsilon^\alpha}_{,k}$, are tangent to the $T = {\it const\/}$
hypersurfaces.
These relations are again sketched in Fig.~1.

To summarize, any solution of the dust equations (\ref{2.6}) and (2.7) which
satisfies the condition (\ref{2.1}) describes a motion of the dust that
allows the spacetime manifold ${\cal M}$ to be split into the space and
time manifolds, ${\S}$ and ${\T}$. An instant of ${\T}$ and a flow line
of ${\S}$ intersect at a
unique event $y\in{\M}$. The spacetime ${\M}$ is thus a Cartesian product
${\T}\times {\S}$ of the space and time manifolds. The mapping $T(y) \times
Z(y)$ takes us from ${\M}$ to ${\T} \times {\S}$, the inverse mapping
$\Upsilon(\tau,\zeta)$ takes us from ${\T} \times {\S}$ to ${\M}$. In this
sense, dust serves as a standard of space and time.
\subsection{Conservation Laws}

Let us now turn to the second set of Euler equations. Equation (2.8) is a
continuity equation for the rest mass current
\begin{equation}  J^\alpha := MU^\alpha \ .\label{2.15} \end{equation}
The total mass is thus conserved from one spacelike hypersurface to another.
Similarly, the Euler equation (2.9) is a continuity equation for
\begin{equation}  J^\alpha_k := W_k J^\alpha \ ,\label{2.16}
     \end{equation}
which we refer to as the momentum current.
These two continuity equations imply
\begin{equation} \pounds_{U} W_k = W_{k,\alpha} U^\alpha = 0
        \ , \label{2.17} \end{equation}
so that $W_{k}$ stays the same along each flow line.  As a result, the
velocity $W_{k}$ can depend on $y$ only through the Lagrangian coordinates
$\zeta = Z(y)$:
\begin{equation} W_k(y) = {\boldW}_k(Z(y)) \ . \label{2.18} \end{equation}
This shows that $W_{k}$ can be viewed as a covector, or one--form, in
the dust space $\S$.
\subsection{The Energy--Momentum Tensor}

Dust is coupled to the metric field by adding its action $S^{D}$ to the
Hilbert action $S^{G}$, which is given by
\begin{equation} S^G[\gamma_\ab ] = \int_\M d^4y\,L^G \quad,\quad
   L^G = |\gamma|^{1/2} R(y;\gamma_\ab] \ . \label{2.19}
  \end{equation}
The variation of the total action $S$ with respect to $\gamma
_{\alpha \beta}$ yields the Einstein law of gravitation
\begin{equation}R^\ab - \frac{1}{2} R \gamma^\ab = \frac{1}{2} T^\ab
      \label{2.20} \end{equation}
with the energy--momentum tensor
\begin{eqnarray} T^\ab := 2 |\gamma|^{1/2}
      \frac{\delta S^D}{\delta\gamma_\ab} & = & MU^\alpha U^\beta +
      \frac{1}{2}\gamma^\ab M(\gamma^{\mu\nu}U_\mu U_\nu + 1) \nonumber \\
      & \approx & M U^\alpha U^\beta \ .\label{2.21} \end{eqnarray}
Modulo the normalization equation (\ref{2.6}), this tensor has the form
required for the description of incoherent dust with rest mass density $M$
and four--velocity $U^{\alpha}$.

By the well--known argument, the invariance of the action (2.4)--(2.5)
under spacetime diffeomorphisms $\varphi$, namely,
\begin{equation}  S^D[\varphi^*T,\; \varphi^*Z^k;\;
 \varphi^*M,\; \varphi^*W_k; \; \varphi^*\gamma_\ab] = S^D[T, Z^k; M,
     W_k; \gamma_\ab] \ , \label{2.22} \end{equation}
leads to the identity
\begin{equation}  |\gamma|^{1/2}\nabla_\beta T^\beta_\alpha \equiv
   \frac{\delta S^D}{\delta T} T_{,\alpha} + \frac{\delta S^D}{\delta Z^k}
      {Z^k}_{,\alpha} + \frac{\delta S^D}{\delta M} M_{,\alpha} +
      \frac{\delta S^D}{\delta W_k} W_{k,\alpha} \ .
      \label{2.23} \end{equation}
It follows that if the dust satisfies the field equations
(2.6)--(2.9), the energy--momentum tensor is covariantly conserved:
\begin{equation}  \nabla_\beta T^\beta_\alpha = 0 \ .
     \label{2.24} \end{equation}
The conservation law (2.24) also follows independently of the Euler
equations (2.6)--(2.9) from the Einstein law of gravitation (2.20) through the
contracted Bianchi identities. The identity (2.23) shows that if we postulate
the conservation law (2.24), the Euler equations (2.6)--(2.9) are not
independent. It suffices to impose the first set of equations, (2.6) and
(2.7), obtained by varying the multipliers $M$ and $W_{k}$.  The second
set of Euler equations, (2.8) and (2.9), obtained by varying the dynamical
variables $T$ and
$Z^{k}$, then follows because $(T_{,\alpha} , Z^{k} \, _{,\alpha})$  is
a regular cobasis.

The first and second sets of Euler equations have a complementary
character.  The first set, obtained by varying the multipliers $M$ and
$W_{k}$,
tells us that the dynamical variables $T$ and $Z^{k}$ have the properties
(2.10) and (2.7) of the proper time and the Lagrangian coordinates of the
dust.
The second set, obtained by varying the dynamical variables $T$ and $Z^{k}$,
tells us that the multipliers $M$ and $W_{k}$ satisfy conservation laws (2.8)
and (2.9) (or (2.17)).  In the canonical formalism, the state variables $T$
and
$Z^{k}$ are turned into canonical coordinates, while the Lagrange multipliers
$M$ and $W_{k}$ are combined to form the conjugate momenta.

{}From the energy--momentum conservation (2.24) and the mass conservation
(2.8) it follows that the flow lines of the dust particles are geodesics,
\begin{equation}  U^\beta\nabla_\beta U^\alpha = 0 \ . \label{2.25}
        \end{equation}
This demonstrates that the dust particles do not act on one another
by mechanical forces.
\subsection{Symmetries and Noether Charges}

The action (2.4)--(2.5) depends on the variables $T$, $Z^{k}$, and $W_{k}$
only through the combination that occurs in the Pfaff form (2.3). It follows
that invertible transformations of these variables that leave the Pfaff form
unchanged constitute symmetries of the dust action.
In Appendix C we identify two particular types of symmetry transformations
that mix the dust spacetime variables among themselves.
The first is expressed as
\be
Z^{k} = \bXi^{k}(Z^{m'}) \ ,\qquad W_{k'} =
      \bXi^\ell_{,k'}(Z^{m'})\,W_\ell \ .\label{2.26}
\ee
These transformations are diffeomorphisms of
the dust space ${\cal S}$, where the velocity $W_k$ transforms as a
covector. Under the transformations (2.26), the dust worldline that passes
through the spacetime point $y$ has its Lagrangian coordinate labels
changed from $\zeta^k = Z^k(y)$ to $\zeta^{k'} = Z^{k'}(y)$.

The second type of symmetry transformation is expressed as
\begin{equation} T' = T + \bTh(Z^\ell) \ ,\qquad W_{k'} = W_k +
       \bTh_{,k}(Z^\ell) \ . \eqnum{2.27} \end{equation}
The dust time ${\cal T}$ is not entirely determined by the dust
flow lines since the initial hypersurface can be chosen
arbitrarily. Once an initial hypersurface is chosen, all other
hypersurfaces of the time foliation are determined, according to
Eq.~(2.10), by Lie propagation of the initial hypersurface along the
flow lines of $U^{\alpha}$. If the initial hypersurface is changed, the
whole foliation ${\cal T}$ is changed. Equation (2.27) connects the
time function $T(y)$ of the old foliation $T(y) = \tau = {\it
const\/}$ with the time function $T'(y)$ of the new foliation $T'(y) =
\tau' = {\it const\/}$.  It also determines how the projection $W_{k}$
of $U^{\alpha}$ onto the leaves of the time foliation is affected by a
different tilt of the leaves with respect to the flow lines.

The transformations (2.26)--(2.27) are {\it global\/} symmetries
of the action, because they do not depend explicitly on $y$ and therefore
cannot be made to vanish in a compact region of ${\cal M}$. (Gauge
symmetries,
on the other hand, vary with $y$.) These symmetry transformations carry one
solution of the field equations into another solution. Indeed, it is easy to
check that if the field equations (\ref{2.7}) and (\ref{2.9})---the only ones
that contain the relevant state  variables---are satisfied for the original
variables, they are also satisfied for the transformed variables.

According to Noether's theorem, to each one--parameter group of symmetry
transformations connected to the identity there corresponds a conserved
charge. Let $\psi_r(y)$ denote a collection of field variables and $\psi'_r
= \Psi_r(\psi,\sigma)$ denote a one--parameter group of symmetries with
the identity at $\sigma=0$. The corresponding conserved charge $Q$ is given
by the integral
\begin{equation} Q = \int_\Sigma \dsurface \; {\Jtilde}
      \eqnum{2.28} \end{equation}
of the Noether current
\begin{equation} {\Jtilde} := \frac{\partial L}{\partial
     \psi_{r,\alpha}}  \left.
    \frac{\partial\Psi_r(\psi,\sigma)}{\partial\sigma} \right|_{\sigma=0}
     \ .\eqnum{2.29} \end{equation}
The integral in Eq.~(2.28) is over an arbitrary spacelike hypersurface
$Y(\Sigma)$ specified by an embedding
\begin{equation}  Y:\Sigma\to\M \quad{\rm by}\quad x\in\Sigma
     \mapsto   y = Y(x)  \eqnum{2.30} \end{equation}
and carrying the oriented surface element
\begin{equation} \dsurface :=
     \alternate {Y^\beta}_{,b} {Y^\gamma}_{,c} {Y^\delta}_{,d}
     \, dx^b\wedge dx^c\wedge dx^d \ .
     \eqnum{2.31} \end{equation}
(Note that ${\Jtilde}$ is a spacetime vector density of
weight $1$, and $\dsurface$ is a covector
density of the opposite weight $-1$ due to the presence of the alternating
symbol $\alternate$.)

The symmetry transformations (2.26)--(2.27) contain arbitrary functions
$\bXi^{k}(Z^{\ell'})$ and $\bTh(Z^\ell)$. The Lie algebra of their
one--parameter subgroups is thus labeled by the arbitrary functions
\be
\bxi^k(Z^{\ell}) := \left.
\frac{\partial\bXi^{k}(Z^{\ell'},\sigma)}{\partial\sigma}\right|_{\sigma=0}
\ ,\qquad
\bth(Z^\ell) := \left. \frac{\partial\bTh(Z^\ell,\sigma)}{\partial\sigma}
\right|_{\sigma=0}
\ .  \eqnum{2.32}
\ee
For the symmetry transformations (2.26), the corresponding Noether
currents (2.29) are
\be
{\Jtilde}[\vec\bxi] = |\gamma|^{1/2} J^\alpha_k \bxi^k(Z) \ ,\eqnum{2.33}
\ee
where $J^\alpha_k$ is the momentum current (2.16). For the symmetry
transformations (2.27), the corresponding Noether currents are
\be
{\Jtilde}[\bth] = |\gamma|^{1/2} J^\alpha  \bth(Z) \ ,\eqnum{2.34}
\ee
where $J^\alpha$ is the mass current (2.15). By virtue of the field
equations (2.7)--(2.9), the Noether currents satisfy the continuity
equations
\be
{{\Jtilde}}_{,\alpha}[\vec\bxi] = 0 = {{\Jtilde}}_{,\alpha}[\bth]  \ .
\eqnum{2.35}
\ee
The Noether charges (2.28) can be written in the form
\bea
Q[\vec\bxi] & = & -\int_\Sigma d^3x \; |g|^{1/2} n_\alpha J^\alpha_k
\bxi^k(Z(x))  \ , \eqnum{2.36}  \\
Q[\bth] & = & -\int_\Sigma d^3x \; |g|^{1/2} n_\alpha J^\alpha
\bth(Z(x))   \ ,\eqnum{2.37}
\eea
where $|g|^{1/2} d^3x$ is the proper volume element of
$Y(\Sigma)$ and $n_{\alpha}$ is the future--pointing unit normal.

In order to interpret the charge (2.37), choose $\bth$ to be unity for all
$Z^{k}$ in some ball ${\cal B}$ in the dust space $\S$, and zero outside
${\cal B}$.  Then $Q[\bth]$ becomes a proper volume integral over the
subspace of the hypersurface $Y(\Sigma)$ that contains flow lines in
${\cal B}$, with the integrand $M (-n_{\alpha} U^{\alpha})$.  The factor
$-n_{\alpha}U^{\alpha}$ is just the relativistic `gamma factor'
characterizing a boost from the Lagrangian observers whose four--velocities
are $U^{\alpha}$ to the Eulerian observers on $Y(\Sigma)$ whose
four--velocities are $n^{\alpha}$. Thus, $Q[\bth]$ is the rest mass in
${\cal B}$ as seen by the Eulerian observers. Conservation of the charge
(2.37) expresses the conservation of relativistic mass within a flow tube
defined by the bundle of flow lines crossing the ball ${\cal B}$.  For
${\cal B} = \Sigma$, conservation of the total relativistic mass is
obtained.

Similarly, choose $\bxi^k =  \delta_{\ell}^{k}$ in the ball ${\cal B}$,
and zero outside ${\cal B}$. The charge (2.36) has the meaning of minus the
component of dust momentum along the coordinate line $\zeta^\ell$ in the
$T = {\it const\/}$ surface within ${\cal B}$.
\section{Canonical Description of Dust}
\subsection{The Canonical Action}

The ADM decomposition of the dust action follows the familiar pattern.
Foliate
spacetime into spacelike hypersurfaces by
\begin{equation}  Y:\RR\times\Sigma \to  \M \quad{\rm by}\quad (t,x) \mapsto
     y = Y(t,x) \ ,\label{3.1}\end{equation}
a one--parameter family of embeddings of space $\Sigma$ into spacetime
${\M}$.  In local coordinates $x^{a}$ on $\Sigma$ and $y^{\alpha}$ on
${\M}$, the foliation (3.1) is represented by
\begin{equation}  (t,x^a) \mapsto y^\alpha = Y^\alpha(t,x^a) \ .
     \label{3.2} \end{equation}
A transition from one leaf of the foliation to another is described
by the deformation vector $N^{\alpha} := {\dot Y}^{\alpha}$. Its
decomposition
\begin{equation}  N^\alpha := {\dot Y}^\alpha = N^\perp n^\alpha +
   N^a {Y^\alpha}_{,a}   \label{3.3} \end{equation}
into the unit normal $n^{\alpha}$ and tangential ${Y^\alpha}_{,a}$
directions yields the lapse function $N^{\perp}$ and the shift vector
$N^{a}$. The spacetime metric $\gamma_{\ab}$ is determined by
\begin{equation}\gamma^\ab = -n^\alpha n^\beta + g^{ab} {Y^\alpha}_{,a}
   {Y^\beta}_{,b} \ ,\qquad |\gamma|^{1/2} = N^\perp |g|^{1/2} \ ,
   \label{3.4} \end{equation}
where $g^{ab}$ is the inverse of the spatial metric $g_{ab}$ induced on
$\Sigma$.

Scalar fields on ${\M}$, such as $Z^k$, can be pulled back to
$\RR \times\Sigma$ by the mapping (3.1). Then, according to Eq.~(3.3),
\begin{equation}  {Z^k}_{,\alpha} n^\alpha = (N^\perp)^{-1} (\dot Z^k
     - {Z^k}_{,a}N^a)
     \ . \label{3.5} \end{equation}
Similar equations hold for the scalar fields $T$, $M$, and $W_k$. From
a global point of view, the mapping (3.1) is used to pull back the mapping
$Z:\M\to\S$ to $\RR\times\Sigma$, which yields a one--parameter family of
mappings $Z_t:\Sigma\to\S$ from space $\Sigma$ into the dust space
$\S$. (We will often suppress the $t$--dependence and simply write
$Z:\Sigma\to\S$.) With the indices $k$ and $a$ viewed as abstract indices,
${\dot Z}^k$ denotes a one--parameter family of mappings from $\Sigma$ to
$T\S$. Likewise, ${Z^k}_{,a}$ denotes a one--parameter family of mappings
from $\Sigma$ to $T\S \otimes T^*\Sigma$ and $W_k$ denotes a one--parameter
family of mappings from $\Sigma$ to $T^*\S$.

The results above enable the
dust action (2.4)--(2.5) to be written as an integral over $\RR \times
\Sigma$,
\begin{equation} S^D[T,Z^k; \; M,W_k; \; g_{ab},N^\perp,N^a] = \int_{\rr} dt
\int_\Sigma
     d^3x \; L^D \ ,  \label{3.6} \end{equation}
where
\begin{equation}  L^D = \frac{1}{2} |g|^{1/2} M \left( \frac{1}{N^\perp}
    ( -\dot T + W_k {\dot Z}^k - U_a N^a )^2 - N^\perp ( g^{ab} U_a U_b +1)
    \right) \ , \label{3.7} \end{equation}
is the dust Lagrangian density and
$U_a := U_\alpha {Y^\alpha}_{,a} = - T_{,a} + W_k {Z^k}_{,a}$.

Let us introduce the momentum $P$ conjugate to $T$ by
\begin{equation} P := \frac{\partial L^D}{\partial {\dot T}} = |g|^{1/2}
     M (N^\perp)^{-1} (\dot T - W_k {\dot Z}^k + U_a N^a ) \
     .\label{3.8} \end{equation}
 The physical meaning of $P$ is revealed when we identify the
 relativistic `gamma factor' $- U_{\alpha} n^{\alpha}$ characterizing
 a boost from the Lagrangian observers whose four--velocities are
 $U^{\alpha}$, to the Eulerian observers on $Y(\Sigma)$ whose
 four-velocities are $n^{\alpha}$.  From Eq.~(2.2) for $U_{\alpha}$
 and Eqs.~(3.5) for $Z^{k}{}_{,\alpha} n^{\alpha}$ and
 $T_{,\alpha}n^{\alpha}$ we obtain
\begin{equation}
-U_{\alpha} n^{\alpha} = (N^{\perp})^{-1} \, \big( \dot{T} -
W_{k} \dot{Z}^{k} + U_{a} N^{a} \big) \, . \label{eq:gamma}
\end{equation}
This shows that the canonical momentum $P$ is equal to the projection
\begin{equation}  P = |g|^{1/2} (-J^\alpha n_\alpha)  \label{3.18}
\end{equation}
of the rest mass four--current (2.15) into the four--velocity of the
Eulerian observers.  Therefore $P$ is the density of rest mass on
$Y(\Sigma)$, per unit coordinate cell $d^{3}x$ (rather than per unit
proper volume $dV=|g|^{1/2}d^{3}x)$, as measured by the Eulerian
observers. Equation (3.10) also shows that the momentum $P$ is
necessarily positive,
\begin{equation}
P > 0 \ , \label{posP}
\end{equation}
since $J^{\alpha}$ and $n^{\alpha}$ are both future--pointing
timelike vectors.

Perform now the Legendre dual transformation $(T,\dot{T}) \to
(T,P)$ while leaving the variables $Z^{k}$, $M$, and
$W_{k}$ untouched.  At the same time, introduce the abbreviation
\begin{equation} P_k := -P W_k  \label{3.9} \end{equation}
for the Lagrange multiplier $W_{k}$ scaled by $-P$. The resulting
Hamiltonian $P\dot T - L^D$ is
\begin{equation}  P\dot T - L^D = -P_k {\dot Z}^k + N^\perp H^D_\perp
    + N^a H_a^D \ , \label{3.10} \end{equation}
with
\begin{equation} H^D_a := -PU_a = PT_{,a} + P_k {Z^k}_{,a}
      \label{3.11} \end{equation}
and
\begin{equation} H^D_\perp := \frac{1}{2}\frac{P^2}{M|g|^{1/2}} +
     \frac{1}{2} \frac{M|g|^{1/2}}{P^2} \Big(P^2 + g^{ab} H^D_a H^D_b
     \Big) \ . \label{3.12} \end{equation}
Equation (3.13) reveals that the Legendre transformed action
\begin{eqnarray}  \lefteqn{ S^D[T,P; \; Z^k,P_k; \; M; \;
 g_{ab},N^\perp,N^a] } \qquad\qquad\qquad\nonumber \\ & = & \int_{\rr}
        dt \int_\Sigma d^3x \left( P\dot T + P_k{\dot Z}^k - N^\perp
        H^D_\perp - N^a H^D_a \right) \label{3.13} \end{eqnarray}
is in canonical form not only in the variables $T$ and $P$, but also
in the variables $Z^{k}$ and $P_{k}$.  The variable (3.12) is thus the
momentum canonically conjugate to $Z^{k}$. As a pendant to
Eq. (3.10), the momentum $P_{k}$ of Eq. (3.12) can be
expressed as
\begin{equation} P_k = - |g|^{1/2} (- J^\alpha_k n_\alpha) \ ,
       \label{3.19} \end{equation}
and interpreted as (minus) the momentum density along
${\Upsilon^\alpha}_{,k}$ per unit coordinate cell $d^{3}x$, measured by
the Eulerian observers on $Y(\Sigma)$.

By varying the action (3.16) with respect to $M$ we get
\begin{equation}  \frac{\delta S^D}{\delta M} = - N^\perp
    \frac{\partial H^D_\perp}{\partial M} = 0 \quad\Longrightarrow\quad
    M = \frac{1}{|g|^{1/2}} \frac{P^2}{\sqrt{P^2 + g^{ab}H_a^D H_b^D}}
    \ .  \label{3.14} \end{equation}
The positive square root is dictated by the positivity of $M$. Substituting
the expression (3.18) back into $H^{D}_\perp$ gives
\begin{equation}  H^D_\perp = \sqrt{P^2 + g^{ab}H_a^D H_b^D} \ ,\label{3.15}
    \end{equation}
and eliminates the mass multiplier $M$ from the action.
Equations (3.12) and (3.18) connect the multiplier variables $M$, $W_{k}$ of
the spacetime action with the canonical momenta $P$, $P_{k}$.

The spatial densities $H_{a}^{D}$ and $H^{D}_\perp$ are, respectively, minus
the momentum density and energy density (in the coordinate basis $dx^{a}$) of
dust as seen by the Eulerian observers $n^\alpha$ on $Y(\Sigma)$.  The
canonical form (3.14) of $H_{a}^{D}$ is dictated by its role as the canonical
generator of spatial diffeomorphisms for the spatial scalars $T$, $Z^{k}$
and their conjugates $P$, $P_{k}$. Very few field theoretical systems have a
square--root energy density like $H^{D}_\perp$. (Two such systems known to us
are the Born--Infeld electrodynamics \cite{Born+Infeld} and the extrinsic
time reference fluid \cite{K=con}.)  The square--root structure of the dust
Hamiltonian (3.19)
arises from the square--root structure of the Hamiltonian of a single
relativistic particle. This connection is demonstrated in Appendix D, which
makes use of results from Sec.~5.

Because the dust action contains no derivatives of the metric, the ADM
decomposition of the total action leads to the super--Hamiltonian and
supermomentum constraints
\begin{eqnarray}  H_\perp & := & H^G_\perp + H^D_\perp = 0 \ ,
     \label{3.20} \\
    H_a & := & H_a^G + H^D_a = 0 \ , \label{3.21}
    \end{eqnarray}
whose gravitational contributions $H^{G}_\perp$ and $H_{a}^{G}$ have the
same form as in vacuum general relativity, namely,
\begin{eqnarray}  H^G_\perp(x;g_{ab},p^{ab}] & = & G_{abcd}(x;g)\;  p^{ab}(x)
           p^{cd}(x) - |g|^{1/2} R(x;g] \ ,  \eqnum{3.22a} \\
      G_{abcd} & = & \frac{1}{2} |g|^{-1/2} (g_{ac} g_{bd} + g_{ad}g_{bc}
            - g_{ab} g_{cd} ) \ , \eqnum{3.22b}\\
      H^G_a(x;g_{ab},p^{ab}] & = & -2D_b p^b_a(x) \ .
      \eqnum{3.23} \end{eqnarray}
The total  constraints (3.20) and (3.21) satisfy the usual Dirac
`algebra' (Sec.~4a).
\subsection{Symmetries and Charges}
By using Eqs. (3.10) and (3.17), the Noether charges (2.36) and (2.37) can be
expressed as functionals of the canonical variables,
\begin{eqnarray}  Q[\vec\bxi] & = & -\int_\Sigma d^3x\; \bxi^k(Z(x)) P_k(x)
     \ ,\eqnum{3.24} \\
     Q[\bth] & = & \int_\Sigma d^3x\; \bth(Z(x)) P(x) \ .
     \eqnum{3.25} \end{eqnarray}
The global symmetries of Section 2e appear in the canonical formalism as
canonical transformations of the dynamical variables generated through the
Poisson brackets by the Noether charges (3.24) and (3.25). For example, by
applying the canonical transformation generated by the charge (3.25)
to the dynamical variables $T(x)$ and
\begin{equation}  W_k(x) := - P_k(x)/P(x) \ , \eqnum{3.26} \end{equation}
the infinitesimal version of Eq.~(2.27) is recovered:
\begin{equation}  \{ T(x), Q[\bth] \} = \bth(Z(x)) \ ,\qquad
   \{ W_k(x), Q[\bth] \} = \bth_{,k}(Z(x))   \ .
   \eqnum{3.27} \end{equation}
Likewise, the canonical transformation generated by the charge (3.24)
yields the infinitesimal version of Eq.~(2.26).

The phase space functionals (3.24), (3.25) have vanishing Poisson brackets
with the constraints (3.20) and (3.21), so they indeed generate symmetries
of the theory. By the same token, they are constants of motion. The symmetry
generators (3.24) and (3.25) close under the Poisson brackets according
to the equations
\bea
\{ Q[{\vec\bxi}_1],Q[{\vec\bxi}_2] \} & = &
Q[\pounds_{\vec\bxi_1}{\vec\bxi}_2]
\ ,\eqnum{3.28a} \\
\{ Q[\vec\bxi],Q[\bth] \} & = & Q[\pounds_{\vec\bxi}\,\bth] \ ,
\eqnum{3.28b}\\
\{ Q[\bth_1],Q[\bth_2] \} & = & 0 \ . \eqnum{3.28c}
\eea
Thus, the $Q$'s form a realization of the Lie algebra of the subgroup
(2.26)--(2.27) of diffeomorphisms of the dust spacetime $\T\times\S$.
\section{The Constraint Algebra}
\subsection{The Dirac Algebra of Old Constraints}

The constraints (3.20)--(3.21) of dust coupled to gravity satisfy the usual
Dirac `algebra'
\begin{eqnarray} \{ H_\perp(x), H_\perp(x')\} & = & g^{ab}(x) H_b(x)
     \delta_{,a}(x,x')  - (x\leftrightarrow x') \ ,\label{4.1}\\
     \{ H_\perp(x), H_a(x')\} & = & - H_\perp(x') \delta_{,a'}(x,x')
      \ ,\label{4.2}\\
     \{ H_a(x), H_b(x')\} & = & H_b(x) \delta_{,a}(x,x') -
     (ax\leftrightarrow bx') \ .\label{4.3}   \end{eqnarray}
Equation (4.3) tells us that ${H}_{a}(x)$ represent the generators of the
spatial diffeomorphism group $\diffsigma$. Equation (4.2) expresses the fact
that the super--Hamiltonian ${H}_\perp(x)$ is a scalar density of weight $1$
under $\diffsigma$.  The right--hand--side of Eq.~(4.1) contains the dynamical
variable $g_{ab}(x)$. This implies that the total system (4.1)--(4.3) is not
a true Lie algebra \cite{I+K}.

The gravitational super--Hamiltonian $H^{G}_\perp(x)$ and supermomentum
$H_{a}^{G}(x)$ close according to the same relations (4.1)--(4.3) as do the
complete constraints $H_\perp(x)$ and $H_{a}(x)$. From here, one can deduce
the Poisson brackets among the matter parts $H^{D}_\perp(x)$ and
$H_{a}^{D}(x)$ of the constraints \cite{H+K+T}. The momentum densities
$H_{a}^{D}(x)$
still represent $\diffsigma$, Eq.~(4.3). For systems with non--derivative
gravitational coupling (which is the case for dust), the energy density
$H^{D}_\perp(x)$ depends only on the metric $g_{ab}(x)$, but not on its
derivatives. It follows that $H^{D}_\perp(x)$ and $H^{D}_\perp(x')$
close in the same manner, Eq.~(4.1), as the total constraints $H_\perp(x)$
and $H_\perp(x')$, or as the gravitational expressions $H^{G}_\perp(x)$ and
$H^{G}_\perp(x')$. The Poisson bracket $\{H^{D}_\perp(x),H_{a}^{D}(x')\}$
does not follow the pattern (4.2) because $H_{a}^{D}(x')$ does not act on the
metric in $H^{D}_\perp(x)$.  The concrete form of this bracket will not
concern us here.

The energy density $H^{D}_\perp(x)$ of the dust is the square root of the
expression
\begin{equation} D(x) := P^2(x) + g^{ab}(x) H^D_a(x) H^D_b(x) \ .
     \label{4.4} \end{equation}
{}From the closing relations (4.1) for $H^{D}(x)$ it follows that
\begin{equation}  \{ D(x),D(x') \} = 4 D(x) H^{Da}(x) \delta_{,a}(x,x')
    - (x\leftrightarrow x') \ .  \label{4.5} \end{equation}
\subsection{The Abelian Algebra of New Constraints}

The fact that the Dirac `algebra' is not a true Lie algebra creates
difficulties in canonical quantum gravity. Isham and Kucha\v{r} \cite{I+K}
found a way in which the Dirac `algebra' of gravity modified by the
imposition
of Gaussian coordinate conditions can be turned into a true Lie algebra.
Kucha\v{r} and Torre \cite{K+T} showed that this procedure is physically
equivalent to the coupling of gravity to a Gaussian reference fluid that has
the structure of a non--rotating, heat--conducting dust. We are presently
dealing with a simpler and more natural matter source, namely, a general
incoherent dust; however, the algorithm to be followed is essentially the
same.  The first step is to replace the old constraints (3.20)--(3.21) by an
equivalent set of constraints based on the resolution of the old constraints
with respect to the momenta $P$ and $P_{k}$ conjugate to the time $T$ and
frame $Z^{k}$ variables. This task is simplified by the fact that the energy
density $H^{D}_\perp(x)$ depends on $P_{k}(x)$ only through the momentum
density $H_{a}^{D}(x)$.  Hence, by using the supermomentum constraint (3.21),
the super--Hamiltonian constraint (3.20) can be cast into the form
\begin{equation} H_\uparrow(x) := P(x) + h(x; \, g_{ab},p^{ab}] = 0 \ ,
      \label{4.6} \end{equation}
with
\begin{eqnarray}  h(x) & = & -\sqrt{G(x)} \ , \eqnum{4.7a} \\
     G(x;\,g_{ab},p^{ab}] & := & \left( H^G_\perp(x)\right)^2 -
     g^{ab}(x) H^G_a(x) H^G_b(x) \ . \eqnum{4.7b} \end{eqnarray}
 The choice of the negative square root in Eq.~(4.7a) is mandated by
     the positivity of $P(x)$, Eq. (\ref{posP}).  Inversely, the
     positivity of $P(x)$ is a consequence of the new constraint
     (4.6).  By introducing the inverse $Z_{k}^{a}(x)$ to the matrix
     ${Z^k}_{,a}(x)$,
\begin{equation} {Z^\ell}_{,a}(x) Z^a_k(x) = \delta^\ell_k \ ,
      \eqnum{4.8} \end{equation}
the supermomentum constraint (3.21) can be rewritten in the form
\begin{eqnarray} H_{\uparrow k}(x) & := & H_a(x) Z^a_k(x) =
     P_k(x) + h_k(x;\,T,Z^k,g_{ab},p^{ab}] = 0  \ ,\quad\eqnum{4.9}\\
     h_k(x; \, T,Z^k,g_{ab},p^{ab}] & = & Z^a_k(x) H^G_a(x) +
    \sqrt{G(x)} T_{,a}(x) Z^a_k(x) \ , \eqnum{4.10} \end{eqnarray}
in which $P(x)$ has been eliminated by the use of Eq.~(4.6).

The old constraints (3.20)--(3.21) and the new constraints (4.6), (4.9) are
completely equivalent (in the region $G(x)>0$, $H^{G}_\perp(x)<0$ of the
gravitational phase space).  However, the momenta $P_{K}=(P,P_{k})$ in
Eqs.~(4.6) and (4.9) are separated from the rest of the canonical variables.
This has an important consequence for the constraint algebra, namely, that
the Poisson brackets of the new constraint functions $H_{\uparrow K}(x) =
(H_{\uparrow}(x), H_{\uparrow k}(x))$ must {\it strongly\/} vanish:
\begin{equation} \{ H_{\uparrow J}(x), H_{\uparrow K}(x') \} = 0  \ .
     \eqnum{4.11} \end{equation}
This remarkable fact follows by a simple argument [10] that circumvents the
tedious algebra:  Because the new constraints are equivalent to the
old constraints and the Poisson brackets of the old constraints weakly vanish,
the Poisson brackets of the new constraints also must weakly vanish.  However,
because the momenta $P_{K}$ appear in the constraints without any
coefficients,
these brackets cannot depend on $P_{K}$.  Thus, the new constraints cannot
help in any way to turn those brackets into zeros; this implies that the
brackets must vanish strongly.

Thus, the new constraints generate an Abelian algebra (4.11).  Let us smear
these constraints by a set $N^{\uparrow K}(x) = (N^{\uparrow}(x),
N^{\uparrow k}(x))$ of externally prescribed scalar functions:
\begin{eqnarray} H[N^\uparrow] & := & \int_\Sigma d^3x\, N^\uparrow(x)
     H_\uparrow(x) \ , \eqnum{4.12}\\
     H[{\vec N}^\uparrow] & := & \int_\Sigma d^3x\, N^{\uparrow k}(x)
     H_{\uparrow k}(x) \ .\eqnum{4.13}   \end{eqnarray}
The smeared constraints generate through the Poisson brackets the
dynamical changes of the field variables $T$, $Z^{k}$, $P$, $P_{k}$,
$g_{ab}$,
and $p^{ab}$. In particular, $H[N^{\uparrow}]$ generates the change
\bea
{\dot T}(x) & := & \{ T(x), H[N^\uparrow] \} = N^\uparrow(x)
\ , \eqnum{4.14a} \\
{\dot Z}^k(x) & := & \{ Z^k(x), H[N^\uparrow] \} = 0 \ , \eqnum{4.14b}
\eea
and $H[{\vec N}^{\uparrow}]$ generates the change
\bea
{\dot T}(x) & := & \{ T(x), H[{\vec N}^{\uparrow}] \} = 0 \ ,
\eqnum{4.15a} \\
{\dot Z}^k(x) & := & \{ Z^k(x),  H[{\vec N}^{\uparrow}] \}
= N^{\uparrow k}(x) \ , \eqnum{4.15b}
\eea
of the dust coordinates $T(x)$ and $Z^{k}(x)$. Equations (4.14) tell
us that $H[N^{\uparrow}]$ displaces the hypersurface by the proper time
$N^{\uparrow}(x)$ \ along the flow lines of the dust (Fig.~2).  The Poisson
bracket of any remaining variable
(like $g_{ab}(x)$) with $H[N^{\uparrow}]$
then gives the change of that variable under such a displacement. Equation
(4.15) tells us that $H[{\vec N}^{\uparrow}]$ does not change $T(x)$ but
generates a displacement to the flow line whose Lagrange coordinates differ
by the amount $N^{\uparrow k}(x)$ from the Lagrange coordinates
$\zeta^{k}=Z^{k}(x)$ of the flow line passing through $x$. This operation
does not leave the hypersurface $Y(\Sigma)$ fixed; rather, it displaces
each point of the hypersurface along the hypersurfaces of the dust
foliation ${\cal T}$ (Fig.~3). Again, the Poisson bracket of any remaining
variable with $H[{\vec N}^{\uparrow}]$ generates the change of this variable
under such a displacement of the hypersurface.

The new form of the constraints leads to the familiar form of the Heisenberg
equations of motion.  Let $F[Z^{K},P_{K},g,p]$ be any dynamical variable on
the phase space and choose a foliation $Z_{t}^{K}(x) = Z^{K}(t,x)$ of
$\T\times\S$. The gravitational variables $g_{ab}(x)$ and $p^{ab}(x)$ evolve
according to the Hamilton equations of motion from given initial data. These
equations along with the Hamilton equations (4.14) and (4.15) determine
$g_{ab}(x;Z_{t}^{K}]$ and $p^{ab}(x;Z_{t}^{K}]$; the constraints (4.6),
(4.9) then give $P_{K}(x;Z_{t}^{K}]$. In this way the dynamical
variable $F$ becomes a functional of $Z_{t}^{K}$,
\begin{equation} F[Z^K_t] := F\left[ Z^K_t, P_K[Z^L_t], g[Z^L_t],
     p[Z^L_t] \right] \ . \eqnum{4.16} \end{equation}
The rate of change of this functional with $t$ is given by
\begin{equation} {\dot F}[Z^K_t]  =  \int_\Sigma d^3x
    \frac{dF[Z^L_t]}{\delta Z^K_t(x)} {\dot Z}^K_t(x)
     =  \int_\Sigma d^3x \frac{dF[Z^L_t]}{\delta Z^K_t(x)}
    N^{\uparrow K}_t(x) \ . \eqnum{4.17} \end{equation}
The symbol $d/\delta Z_{t}^{K}(x)$ denotes the total variational
derivative of $F[Z_{t}^{K}]$ with respect to $Z_{t}^{K}(x)$, which takes
into account both the explicit dependence of $F[Z^{K},P_{K},g,p]$ on $Z^{K}$
and the implicit dependence through the dynamical variables $g$, $p$, and
$P_{K}$. On the other hand, $\dot{F}$ is also given by the Poisson bracket
of $F[Z^{K},P_{K},g,p]$ with the Hamiltonian $H[N^{\uparrow K}]$.
By comparing the two expressions for $\dot{F}$ for an arbitrary choice of
foliation, we obtain the many--fingered time Heisenberg equations of motion
\begin{equation} \frac{dF}{\delta Z^K(x)}  =  \{ F, H_{\uparrow K}(x) \}
       =  \frac{\delta F}{\delta Z^K(x)} + \{ F, h_K(x) \} \ .
      \eqnum{4.18} \end{equation}
Here, $h_K(x) = (h(x), h_k(x))$ are the true Hamiltonian and momentum
densities from Eqs.~(4.7) and (4.10).
\subsection{Representing ${\rm LDiff}(\T\times\S)$}

In Eqs.~(4.12)--(4.13), the new constraints were smeared by externally
prescribed multipliers $N^{\uparrow K}(x)$. The constraints also can be
smeared by vector fields ${\boldV_1}$, ${\boldV_2}$ on $\T\times\S$ in
the following way. Consider $\boldV_1$ and $\boldV_2$ as elements of
$\ldifftxs$ under the Lie bracket operation $-[\boldV_1,\boldV_2]$
which, up to a sign, is given by the commutator. When restricted to an
embedding $Z^{K}(x)$, these vector fields become dynamical variables on
the phase space of the dust. Thus, $\boldV_1$ and $\boldV_2$ can be used
to smear the new constraints on $\Sigma$; for example,
\begin{equation} \boldV_1 \mapsto H_\uparrow[\boldV_1] := \int_\Sigma
     d^3x \, \boldV_1^K \bigl( Z^L(x) \bigr) H_{\uparrow K}(x) \ .
    \eqnum{4.19} \end{equation}
In this way each vector field on $\T\times\S$ is mapped into a dynamical
variable over the phase space of dust coupled to gravity. Equations (4.6),
(4.9), and (4.11) then guarantee that
\begin{equation} \{ H_\uparrow[\boldV_1], H_\uparrow[\boldV_2] \} =
    H_\uparrow\left[ -[\boldV_1, \boldV_2] \right] \ ,
    \eqnum{4.20} \end{equation}
{\it i.e.\/}, that the mapping (4.19) is a homomorphism from $\ldifftxs$
into the Poisson bracket algebra over the phase space.  In this manner, the
introduction of dust enables us to represent the Lie algebra of the dust
spacetime diffeomorphism group by canonical transformations.
\subsection{New Constraints Lead to a True Algebra for Vacuum Gravity}

The new Hamiltonian constraint (4.6) has a feature that is quite astounding:
The Hamiltonian $h(x)$ of Eq.~(4.7) does not depend on the dust variables,
but only on the geometric variables $g_{ab}$, $p^{ab}$.  We have not
encountered such behavior in any other physical system.  Even for the
simplest parametrized systems, like a massless scalar field on a
Minkowskian background \cite{KucharScalar}, the Hamiltonian depends
on the many--fingered Minkowski time. In contrast, the Hamiltonian
(4.7) turns out to be independent of the time $T(x)$ and frame
$Z^{k}(x)$ variables.

This remarkable feature has an equally remarkable consequence: The Poisson
brackets $\{ H_\uparrow(x), H_\uparrow(x') \} = 0$ do not have any cross
terms between $P(x)$ and $h(x)$, and hence $h(x)$ must
have vanishing Poisson brackets among themselves:
\begin{equation} \{ h(x;\,g,p], h(x';\,g,p] \} = 0 \ .
       \eqnum{4.21} \end{equation}
Since the Hamiltonian $h(x)$ is a square root (4.7), it follows
that the expressions $G(x)$ under the square root must also have vanishing
Poisson brackets:
\begin{equation} \{ G(x;\,g,p], G(x';\,g,p] \} = 0 \ .
       \eqnum{4.22} \end{equation}
This property is vital for the definition of the Hamiltonian operator
$h(x;\;\hat{g},\hat{p}]$ by spectral analysis (Section 6c).

We proved Eq.~(4.22) by coupling geometry to dust. However, Eq.~(4.22) is
an identity involving only the geometric variables $g_{ab}$ and $p^{ab}$.
Thus, it must hold irrespective of whether geometry is coupled to dust, to
any other matter system, or whether it is left alone in vacuum.  Equation
(4.22) must simply hold by virtue of the Dirac `algebra'
(4.1)--(4.3) among the gravitational expressions $H^{G}_\perp(x)$ and
$H_{a}^{G}(x)$.  We verify this assertion directly in Appendix E.

We have found a simple quadratic combination $G(x)$ of the gravitational
super--Hamiltonian $H^{G}_\perp(x)$ and supermomentum $H_{a}^{G}(x)$ that
strongly commutes with itself, Eq.~(4.22).  Because $G(x)$ is a scalar
density of weight $2$ constructed solely from the geometric variables,
its Poisson bracket with the gravitational supermomentum is given by
\begin{equation} \{ G(x), H^G_a(x') \} = G_{,a}(x) \delta(x,x') +
    2 G(x) \delta_{,a}(x,x') \ . \eqnum{4.23} \end{equation}
Of course, the gravitational supermomenta represent $\ldiffsigma$,
\begin{equation} \{ H^G_a(x), H^G_b(x') \} = H^G_b(x) \delta_{,a}(x,x')
    - (ax \leftrightarrow bx') \ .  \eqnum{4.24} \end{equation}
Thus, the geometric expressions $G(x)$ and $H^{G}_a(x)$ form a true Lie
algebra that is a semidirect product of the Abelian algebra generated by
$G(x)$ and $\ldiffsigma$ generated by $H^{G}_a(x)$.

In vacuum gravity, the imposition of the usual super--Hamiltonian and
super\-momentum constraints $H^{G}_\perp(x)$ and $H_{a}^{G}(x)$ is clearly
{\it globally equivalent\/} to the imposition of the new set of
constraints, $G(x)$ and $H_{a}^{G}(x)$:
\be
H^G_\perp(x) = 0 = H^G_a(x) \quad\Longleftrightarrow\quad G(x) = 0
= H^G_a(x) \ . \eqnum{4.25}
\ee
Therefore, we have succeeded in casting vacuum gravity into a form in
which the constraints close according to a true Lie algebra.  This is
something that previously has been achieved only for systems in
two--dimensional spacetimes, such as strings \cite{HRTetc} and
2--dimensional gravity \cite{TandT}.

There is only one fly in the ointment.  In vacuum gravity, the Hamiltonian
vector field generated by $G(x)$ vanishes on the constraint surface
(4.25).  Therefore, $G(x)$ does not generate a motion on the constraint
surface, and it cannot be used for foliating the constraint surface into
orbits.  Such a difficulty does not exist for gravity coupled to dust: As long
as there is any dust at $x$,
\begin{equation} G(x) = P^2(x) > 0 \ .
     \eqnum{4.26} \end{equation}
The full significance of the new form of the constraints for vacuum
gravity remains to be investigated.
\section{Canonical Dust--Space Variables}
\subsection{The Dust--Space Variables and the Constraints}

Our description of dust coupled to gravity was based on the conjugate
pairs $Z^k(x)$ and $P_k(x)$, $T(x)$ and $P(x)$, and $g_{ab}(x)$ and
$p^{ab}(x)$ of canonical variables. The variables $T(x)$, $P(x)$,
$g_{ab}(x)$, and $p^{ab}(x)$ are ($t$--dependent) tensor fields
on $\Sigma$. The variable $Z^k(x)$ is the local coordinate chart
expression of a ($t$--dependent) mapping $Z:\Sigma\to\S$ from the
space manifold $\Sigma$ to the dust space $\S$. Its conjugate $P_k(x)$
is the coordinate expression of a ($t$--dependent) mapping from
$\Sigma$ to $T^*\S$, with density weight $1$ on $\Sigma$. The
Hamiltonian formulation of the combined system is characterized by the
constraints $H_\uparrow(x) = 0$ and $H_{\uparrow k}(x) = 0$ which
generate the abelian algebra (4.11).

Canonical quantization is vastly simplified by the introduction of an
alternative set of canonical variables which reflect the fact that the
dust particles naturally define a preferred set of spatial coordinates
on $\Sigma$. In this section, we construct such variables, and explain
how they are geared to the abelian algebra ~(4.11).

Start from the tensorial variables $T(x)$, $P(x)$, $g_{ab}(x)$, and
$p^{ab}(x)$ on $\Sigma$, and let the mapping $Z:\Sigma\to\S$ induce a
corresponding set of tensors $\bT(\zeta)$, $\bP(\zeta)$,
$\bg_{k\ell}(\zeta)$, and $\bp^{k\ell}(\zeta)$ on the dust space
$\S$. In terms of local coordinates, these $\S$--variables are
defined by
\bea
   \bT(\zeta) & := & T(X(\zeta)) \ ,\eqnum{5.1a}\\
   \bP(\zeta) & := & \left|\frac{\partial X(\zeta)}
      {\partial\zeta}\right| P(X(\zeta)) \ ,\eqnum{5.1b}\\
   \bg_{k\ell}(\zeta) & := & {X^a}_{,k}(\zeta) {X^b}_{,\ell}(\zeta)
       \, g_{ab}(X(\zeta)) \ ,\eqnum{5.1c} \\
    \bp^{k\ell}(\zeta) & := & \left|\frac{\partial X(\zeta)}
     {\partial\zeta}\right| {Z^k}_{,a}(X(\zeta)) {Z^\ell}_{,b}(X(\zeta))
     \, p^{ab}(X(\zeta)) \ .\eqnum{5.1d}
\eea
Here, the $t$--dependent mapping $X$ is simply the inverse of $Z$,
\begin{equation} X:\S\to\Sigma \ ,\qquad X := Z^{-1}
        \ ,\eqnum{5.2}
\end{equation}
and $|\partial X(\zeta) / \partial\zeta|$ is the Jacobian for the
change of variables $x^a = X^a(\zeta)$.

The $\S$--variable $\bT(\zeta)$ measures proper time along
the dust worldline whose Lagrangian coordinate label is $\zeta^k$,
and $\bP(\zeta)$ is the dust rest mass on $\Sigma$ per unit
coordinate cell $d^3\zeta$. The variables $\bg_{k\ell}(\zeta)$
are the components of the metric tensor
\begin{equation} ds^2 = \bg_{k\ell}(\zeta)\, d\zeta^k d\zeta^\ell
   \eqnum{5.3}\end{equation}
that measures proper distance $ds$ in $\Sigma$ between
neighboring dust particles with Lagrangian coordinate labels
$\zeta^k$ and $\zeta^k + d\zeta^k$.

We will now show that the $\S$--variables $\bT(\zeta)$, $\bP(\zeta)$,
$\bg_{k\ell}(\zeta)$, and $\bp^{k\ell}(\zeta)$, along with the dust
frame $Z^k(x)$ and the new supermomentum $H_{\uparrow k}(x)$, are
canonically conjugate. That is, we will argue that the only
nonvanishing Poisson brackets among these variables are
\bea
   \{ Z^k(x), H_{\uparrow\ell}(x') \} & = &
     \delta^k_\ell\delta(x,x')  \ , \eqnum{5.4a} \\
   \{ \bT(\zeta), \bP(\zeta') \} & = & \delta(\zeta, \zeta')
     \ ,\eqnum{5.4b} \\
   \{ \bg_{k\ell}(\zeta), \bp^{mn}(\zeta') \} & = &
      \frac{1}{2}\,(\delta^{m}_{k}\delta^{n}_{\ell}
      +\delta^{m}_{\ell}\delta^{n}_{k})\,
      \delta(\zeta,\zeta') \ .\eqnum{5.4c}
\eea
Recall (Section 4b) that the constraint functions
$H_{\uparrow k}(x)$ have strongly vanishing Poisson brackets among
themselves. They are thus natural candidates for a new set of
canonical momenta. Equation ~(5.4a) maintains that their conjugate
coordinates are the frame variables $Z^{k}(x)$. This is obvious
from the construction ~(4.9) of $H_{\uparrow k}(x)$.

 The Poisson brackets (5.4b) and (5.4c) exemplify a general rule: the
mapping $\Sigma\to\S$ turns canonically conjugate fields on $\Sigma$
into canonically conjugate fields on $\S$. This is most easily
established for a scalar field and its conjugate, such as the proper
time (5.1a) and the rest mass density (5.1b).  In this case, we first
write
\bea
   \bT(\zeta) & = & \int_\Sigma d^3x\,T(x)\,\delta(x-X(\zeta))
   \ ,\eqnum{5.5a}\\
   \bP(\zeta) & = & \int_\Sigma d^3x\, P(x)\, \delta(Z(x) - \zeta)
   \ ,\eqnum{5.5b}
\eea
and then directly verify that the Poisson brackets
\be
   \{ \bT(\zeta), \bP(\zeta') \} = \int _{\Sigma} d^3x \left(
   \frac{\delta\bT(\zeta)}{\delta T(x)}
   \frac{\delta\bP(\zeta')}{\delta P(x)}
   - \frac{\delta\bT(\zeta)}{\delta P(x)}
   \frac{\delta\bP(\zeta')}{\delta T(x)} \right) +\cdots
   \eqnum{5.6}
\ee
yield the desired result (5.4b). (The unwritten terms in Eq.~(5.6)
involve functional derivatives with respect to $Z^k(x)$, $P_k(x)$,
$g_{ab}(x)$, and $p^{ab}(x)$, and they all vanish.) It is also clear
that the Poisson brackets of $T({\zeta})$ among themselves and of
$P({\zeta})$ among themselves strongly vanish.
It is straightforward to extend these arguments to arbitrary
canonically conjugate tensorial variables and, in particular, to show
that $\bg_{k\ell}(\zeta)$ and $\bp^{k\ell}(\zeta)$ are canonically
conjugate.

The $\S$--variables $\bT(\zeta)$, $\bP(\zeta)$, $\bg_{k\ell}(\zeta)$,
and $\bp^{k\ell}(\zeta)$ obviously have vanishing Poisson brackets
with the frame variables $Z^{k}(x)$. The only remaining task is to show
that their brackets with the new momentum $H_{\uparrow k}(x)$ also
vanish. To do that, let us recall the relationship $H_{\uparrow k}(x)
= Z^a_k(x) H_a(x)$ between the new supermomentum $H_{\uparrow k}(x)$
and the original supermomentum $H_a(x)$. Since the matrix $Z^{a}_{k}$
depends only on the frame variables $Z^{k}(x)$ and the $\S$--variables
(5.1) are independent of $P_k(x)$, they have vanishing brackets with
$Z^a_k(x)$. Thus, the Poisson brackets of the $\S$--variables with
$H_{\uparrow k}(x)$ are proportional to their brackets with
$H_a(x)$.
However, the Poisson brackets of the $\S$--variables with $H_a(x)$
must vanish. Recall that the Poisson brackets of any dynamical
variable with the supermomentum smeared by an externally prescribed
vector field $N^a(x)$ on $\Sigma$,
\be
   H[{\vec N}] := \int_\Sigma d^3x \, N^a(x) H_a(x)
   \ ,\eqnum{5.7}
\ee
give the change in that variable under LDiff$\Sigma$. Since the
$\S$--variables are constructed so that they are invariant under
Diff$\Sigma$, their Poisson brackets with $ H[{\vec N}]$ and hence
with $H_{a}(x)$ vanish. It follows that their brackets with
$H_{\uparrow k}(x)$ also vanish. This completes our proof that the
${\S}$--variables (5.1) along with the pair $Z^k(x)$, $H_{\uparrow
k}(x)$ form a canonical chart.

In terms of the new canonical variables $\bT(\zeta)$, $\bP(\zeta)$,
$\bg_{k\ell}(\zeta)$, $\bp^{k\ell}(\zeta)$, and $Z^k(x)$, $H_{\uparrow
k}(x)$, the supermomentum constraint is simply the condition that the
canonical momentum $H_{\uparrow k}(x)$ vanishes.  The
super--Hamiltonian constraint $H_\uparrow(x) = 0$ can then be mapped
to the dust space $\S$, which yields an equivalent constraint
\be
    \boldH_\uparrow(\zeta) :=  \left|\frac{\partial X(\zeta)}
       {\partial\zeta}\right|  H_\uparrow(X(\zeta))
       =  \bP(\zeta) + \boldh(\zeta;\,\bg_{k\ell},\bp^{k\ell}] = 0
       \ .\eqnum{5.8}
\ee
Here, the Hamiltonian density
$\boldh(\zeta) = -\sqrt{\boldG(\zeta)}$ is obtained
from $h(x)$ of Eq.~(4.7) by replacing the $\Sigma$--tensors
$g_{ab}(x)$ and $p^{ab}(x)$ with the corresponding
$\S$--tensors $\bg_{k\ell}(\zeta)$ and
$\bp^{k\ell}(\zeta)$. The constraint $ \boldH_\uparrow(\zeta)$
is expressed entirely in terms of the $\S$--variables. Its Poisson
brackets with $H_{\uparrow k}(x)$ therefore vanish. We conclude that
the constraints $ \boldH_\uparrow(\zeta)$ and $H_{\uparrow k}(x)$
still form an abelian algebra.
\subsection{Noether Charges}

The Noether charge $Q[\bth]$ from Eq. (3.25) can be expressed in terms
of the new canonical variables by a simple change $x^a = X^a(\zeta)$ of
integration variables. This gives
\bea
Q[\bth] & = &  \int_\Sigma d^3x \; \bth(Z(x)) P(x) \eqnum{5.9a}\\
& = & \int_\S d^3\zeta \; \bth(\zeta) \bP(\zeta) \ .\eqnum{5.9b}
\eea
As discussed in Secs.~3b and 2e, $Q[\bth]$ generates a canonical
transformation that yields the change of any dynamical variable induced by
the change of the initial hypersurface of constant dust time $
T(y) = \tau = {\it const\/}$. Among the new canonical variables, only
the dust proper time $T(\zeta)$ is affected by $Q[\bth]$:
\be
\{ \bT(\zeta) , Q[\bth] \} = \bth(\zeta) \ .\eqnum{5.10}
\ee
The canonical transformation generated by $Q[\bth]$ produces a
translation of the proper time $\tau$ along the dust worldline
$\zeta^k$ by the amount $\bth(\zeta)$.

The expressions (4.6)--(4.10) can be used to write the Noether
charge $Q[\vec\bxi]$ from Eq. (3.24)
in terms of the new canonical variables. The result is
\bea
Q[\vec\bxi] & = & -\int_\Sigma d^3x \; \bxi^k \Big
(Z(x)\Big) P_k(x)
   \eqnum{5.11a}\\
& = & -\int_\Sigma d^3x \; \bxi^k\Big(Z(x)\Big) H_{\uparrow k}(x)
+ \int_\S d^3\zeta \; \bxi^k(\zeta) \biggl(
\boldH_k^G(\zeta) + \bP(\zeta) \bT_{,k}(\zeta)
\biggr) \ ,\eqnum{5.11b}
\eea
where $\boldH_k^G(\zeta)$ is obtained by mapping the
gravitational supermomentum $H_a^G(x)$ from space $\Sigma$
to the dust space $\S$:
\be
\boldH_k^G(\zeta) = \left|\frac{\partial X(\zeta)}
    {\partial\zeta}\right| {X^a}_{,k}(\zeta) H_a^G(X(\zeta))
    \ .\eqnum{5.12}
\ee
Thus, $\boldH_k^G(\zeta)$ depends only on $\bg_{k\ell}(\zeta)$
and $\bp^{k\ell}(\zeta)$.

The Noether charge $Q[\vec\bxi]$ can be recognized as
the canonical generator of dust space diffeomorphisms, $\diffs$.
That is, for any canonical variable $F$, the Poisson brackets
$\{ F , Q[\vec\bxi]\}$ give the change in $F$ due to an infinitesimal
diffeomorphism of $\S$ generated by the vector field $\vec\bxi(\zeta)$.
If $F$ is one of the $\S$--tensors $\bT$, $\bP$, $\bg$, or
$\bp$, then the Poisson brackets $\{ F , Q[\vec\bxi]\}$ equal the Lie
derivative $\pounds_{\vec\bxi} F$. For the dust frame variable
$Z^k(x)$ and its conjugate $H_{\uparrow k}(x)$, the
transformation generated by $Q[\vec\bxi]$ is
\bea
\{ Z^k(x), Q[\vec\bxi] \} & = & - \bxi^k(Z(x)) \ ,\eqnum{5.13a}\\
\{ H_{\uparrow k}(x), Q[\vec\bxi]\} & = & {\bxi^\ell}_{,k}(Z(x))
H_{\uparrow\ell}(x) \ .\eqnum{5.13b}
\eea
The charges $Q[\bth]$ and $Q[\vec\bxi]$ close under the Poisson
brackets according to the relations (3.28).

As noted in Sec.~3b, the Noether charges $Q[\bth]$ and $Q[\vec\bxi]$ have
vanishing Poisson brackets with the constraints $H_\perp(x)$ and $H_a(x)$.
Since the new set of constraints $H_{\uparrow k}(x) = 0$ and
$\boldH_\uparrow(\zeta) = 0$ are equivalent to the old set, the Poisson
brackets of the charges with the new constraints should vanish at least
weakly, that is, when the constraints hold. Indeed, the Poisson
brackets of $Q[\bth]$ with $H_{\uparrow k}(x)$ and
$\boldH_\uparrow(\zeta)$ vanish strongly. The Poisson brackets of
$Q[\vec\bxi]$ with $H_{\uparrow k}(x)$ vanish when the constraints
hold, as seen from Eq.~(5.13b). The remaining Poisson brackets are
\be
\{ \boldH_\uparrow(\zeta), Q[\vec\bxi] \} = \left(
\boldH_\uparrow(\zeta) \bxi^k(\zeta) \right)_{,k} \ , \eqnum{5.14}
\ee
and they also weakly vanish. Expression
(5.14) shows that $\boldH_\uparrow(\zeta)$ transforms as a scalar
density under Diff$\S$.
\section{Quantization}
\subsection{Operator Constraints and the Functional Schr\"{o}dinger Equation}

In the Dirac method of quantization, constraints are turned into operators
and
imposed as restrictions on the state functionals of the system. With the
standard form (3.20) of the super--Hamiltonian constraints, this procedure
leads
to the Wheeler--DeWitt equation for gravity coupled to a source.  This is a
second--order variational differential equation in the metric variables whose
solution space carries no obvious Hilbert space structure. If the source
happens to be dust, there is an additional difficulty:  the square--root
Hamiltonian density (3.19) is difficult to define by spectral analysis
because
the expressions under the square root do not commute, Eq.~(4.5).

In order for dust to serve as a clock, the super--Hamiltonian constraint
should be resolved with respect to the momentum conjugate to the dust time
variable prior to quantization. The constraint $H_\uparrow(x)$, or
equivalently
$\boldH_\uparrow(\zeta)$, has this desired form. We choose to work primarily
with the dust--space variables, so the quantum states of the system are
functionals $\bPsi[Z,\bT,\bg]$ of the canonical coordinates $Z^k(x)$,
$\bT(\zeta)$, and $\bg_{k\ell}(\zeta)$. From the expression (5.8) of the
classical constraint $\boldH_\uparrow(\zeta) = 0$, the operator condition
\be
\boldH_\uparrow(\zeta;\,\hat\bP,\hat\bg,\hat\bp]\, \bPsi[Z,\bT,\bg] = 0
\eqnum{6.1}
\ee
yields the functional Schr\"{o}dinger equation
\be
i\frac{\delta\bPsi[Z,\bT,\bg]}{\delta\bT(\zeta)} =
\boldh(\zeta;\,\hat\bg,\hat\bp]\, \bPsi[Z,\bT,\bg] \ .\eqnum{6.2}
\ee
In addition to Eq.~(6.2), the state functionals must obey the operator
version of the classical constraint $H_{\uparrow k}(x) = 0$,
namely,
\be
{\hat H}_{\uparrow k}(x)\, \bPsi[Z,\bT,\bg] = 0 \ .\eqnum{6.3}
\ee
Because $ H_{\uparrow k}(x)$ is canonically conjugate to
$Z^{k}(x)$, ${\hat H}_{\uparrow k}(x)$ is the momentum operator
\be
{\hat H}_{\uparrow k}(x) = -i \frac{\delta}{\delta Z^k(x)} \ .
\eqnum{6.4}
\ee
Equation (6.3) then implies that the quantum states must be independent
of the dust frame $Z^k(x)$. Therefore the state functionals
$\bPsi[\bT,\bg]$ satisfy the functional Schr\"{o}dinger equation
\be
i\frac{\delta\bPsi[\bT,\bg]}{\delta\bT(\zeta)} =
\boldh(\zeta;\,\hat\bg,\hat\bp]\, \bPsi[\bT,\bg] \ ,\eqnum{6.5}
\ee
with no further restrictions.

The functional differential equations (6.1), (6.3) are consistent
only if the commutators of the operators ${\hat\boldH}_{\uparrow}(\zeta)$
and ${\hat H}_{\uparrow k}(x)$ replicate the classical Poisson bracket
algebra
\bea
\{ \boldH_\uparrow(\zeta) , \boldH_\uparrow(\zeta') \} & = & 0
\ ,\eqnum{6.6a}\\
\{ \boldH_\uparrow(\zeta) , H_{\uparrow k}(x) \} & = & 0 \ ,\eqnum{6.6b}\\
\{ H_{\uparrow k}(x) , H_{\uparrow\ell}(x') \} & = & 0 \ .\eqnum{6.6c}
\eea
With the representation (6.4) for the momentum operator the subalgebra
(6.6c) is indeed carried over into the quantum theory.
The brackets (6.6b) are also satisfied quantum mechanically since each
of the operators $\hat\bP$, $\hat\bg$, $\hat\bp$ that appear in
${\hat\boldH}_\uparrow(\zeta)$ commutes with ${\hat H}_{\uparrow k}(x)$.
We shall proceed under the assumption that there exists a factor ordering
and regularization of $\boldh(\zeta;\;\hat\bg,\hat\bp]$ such that the
quantum operators ${\hat\boldH}_\uparrow(\zeta)$ commute among themselves,
and thereby replicate the vanishing of the classical Poisson brackets
(6.6a).

\subsection{Time--Independent Schr\"{o}dinger Equation}

${\hat\boldH}_\uparrow(\zeta)$ does not depend on $\bT(\zeta)$ so that
\be
[ {\hat\boldH}_\uparrow(\zeta) , {\hat\bP}(\zeta') ] = 0 \ .\eqnum{6.7}
\ee
Furthermore, the operators ${\hat\bP}(\zeta)$ commute among themselves,
and the super--Hamiltonian operators ${\hat\boldH}_\uparrow(\zeta)$
commute by virtue of Eq.~(6.6a). It follows that the Schr\"{o}dinger
equation (6.5) and the eigenvalue equation
\be
{\hat\bP}(\zeta)\; \bPsi = \bP(\zeta)\; \bPsi \eqnum{6.8}
\ee
have common solutions $\bPsi[\bP;\, \bT,\bg]$. These are functionals of
$\bT(\zeta)$ and $\bg_{k\ell}(\zeta)$ labeled by the eigenvalues $\bP(\zeta)$
of
the momentum operator ${\hat\bP}(\zeta)$. The general solution of
Eq.~(6.8) has the form
\be
\bPsi[\bP;\, \bT,\bg] = \bpsi[\bg] \exp\biggl( i \int_\S d^3\zeta \, \bP(\zeta)
\bT(\zeta) \biggr) \ .\eqnum{6.9}
\ee
By substituting expression (6.9) into Eq.~(6.5) we learn that
$\bpsi[\bg]$ must satisfy the time--independent functional Schr\"{o}dinger
equation
\be
\boldh(\zeta; \,\hat\bg,\hat\bp]\; \bpsi[\bg] = - \bP(\zeta)\, \bpsi[\bg]
\ .\eqnum{6.10}
\ee
Equations (6.9) and (6.10) accomplish the separation of the
many--fingered time $\bT(\zeta)$ from the metric $\bg_{k\ell}(\zeta)$.

\subsection{Spectral Analysis}

The classical Hamiltonian density $\boldh(\zeta)$ is the dust space
counterpart of the square root Hamiltonian density $h(x) = -\sqrt{G(x)}$,
as defined in Eq.~(4.7). The square root structure of $h(x)$ is
inherited by $\boldh(\zeta)$, so that $\boldh(\zeta) =
-\sqrt{\boldG(\zeta)} $ where
\be
\boldG(\zeta;\;\bg,\bp] := \Bigl( \boldH_\perp^G(\zeta) \Bigr)^2 -
\bg^{k\ell}(\zeta) \boldH_k^G(\zeta) \boldH_\ell^G(\zeta) \ .\eqnum{6.11}
\ee
As a consequence, the operators ${\hat\boldh}(\zeta)$ that appear in the
functional Schr\"{o}dinger equation must be defined by spectral analysis.
The first step is to turn the classical expressions (6.11) into operators
$\boldG(\zeta;\,\hat\bg,\hat\bp]$. The assumption that the operators
${\hat\boldH}_\uparrow(\zeta)$ commute translates into the assumption
that there exists a factor ordering of $\hat\bg$ and $\hat\bp$, and a
regularization of ${\hat\boldG}(\zeta)$, such that
\be
\bigl[\, \boldG(\zeta;\,\hat\bg,\hat\bp] , \boldG(\zeta';\,\hat\bg,\hat\bp]
\,\bigr] = 0 \ .\eqnum{6.12}
\ee
The operators ${\hat\boldG}(\zeta)$ then have common eigenfunctionals
$\bphitoo[\boldG;\,\bg]$,
\be
{\hat\boldG}(\zeta)\, \bphitoo[\boldG;\,\bg] = \boldG(\zeta)\,
\bphitoo[\boldG;\,\bg]
\ ,\eqnum{6.13}
\ee
with eigenvalues $\boldG(\zeta)$.

The classical variables (6.11) are not positive definite, so the eigenvalues
$\boldG(\zeta)$ are not necessarily positive everywhere on $\S$.
Let us define a Hilbert space ${\cal H}^{+}$ that is spanned by the
eigenfunctionals of ${\hat\boldG}(\zeta)$ with positive eigenvalues
$\boldG(\zeta) > 0$. On ${\cal H}^{+}$, we can define the operator
\be
{\hat\boldh}(\zeta) = -\sqrt{{\hat\boldG}(\zeta)} \eqnum{6.14}
\ee
by the requirement that it have eigenfunctionals $\bphitoo[\boldG;\,\bg]$ with
eigenvalues
\be
\boldh(\zeta) = -\sqrt{\boldG(\zeta)} \ .\eqnum{6.15}
\ee
Because ${\hat\boldh}(\zeta)$ is the Hamiltonian of the system, a state
functional in ${\cal H}^{+}$ will remain in ${\cal H}^{+}$ throughout its
dynamical evolution.

If an eigenfunctional $\bphitoo[\boldG;\,\bg]$ with eigenvalue $\boldG(\zeta)$
is known, then infinitely many eigenfunctionals corresponding to other
eigenvalues can be generated by the action of dust space diffeomorphisms
$\bXi\in\diffs$. This result is expressed by the following theorem, which
is proved in Appendix G:
Let $\bphitoo[\boldG;\,\bg]$ be an eigenfunctional of ${\hat\boldG}(\zeta)$
with eigenvalue $\boldG(\zeta)$. Then $\bphitoo[\boldG;\,\bXi^*\bg]$ is an
eigenfunctional of ${\hat\boldG}(\zeta)$ with eigenvalue
$(\bXi^{-1})^*\boldG(\zeta)$.

Disregarding global difficulties, a given dust space scalar density
${\boldG}(\zeta) > 0$ of weight 2 can be transformed by the action of an
appropriate diffeomorphism $\bXi^{-1}$ into any other arbitrary function
$\boldG_*(\zeta) > 0$. In particular, it can be transformed into
$\boldG_*(\zeta) = 1$. This means that ${\hat\boldG}(\zeta)$ has a
continuous spectrum, {\it i.e.\/}, that there is an eigenfunctional
$\bphitoo[\boldG;\,\bg]$ for any ${\boldG}(\zeta) > 0$.

These arguments also indicate that the spectrum of the collection
$\zeta\in\S$ of commuting operators ${\hat\boldG}(\zeta)$ is infinitely
degenerate. Indeed, let $\bXi_0\in{\rm Diff}_0\S$ be an arbitrary
volume--preserving diffeomorphism, $\det(\partial\bXi_0(\zeta)/
\partial\zeta) = 1$. For $\boldG(\zeta)=1$, we have also
$(\bXi_0^{-1})^*\boldG(\zeta) = 1$. Therefore, according to the theorem
stated above, if $\bphitoo[1;\bg]$ is an eigenfunctional of
${\hat\boldG}(\zeta)$ with eigenvalue $1$, then $\bphitoo[1;\,\bXi^*_0\bg]$
is another eigenfunctional of ${\hat\boldG}(\zeta)$ with eigenvalue $1$.
This shows that there are at least as many eigenfunctionals of
${\hat\boldG}(\zeta)$ with eigenvalue $1$ as there are volume--preserving
diffeomorphisms. When we act on the argument of this set of
eigenfunctionals by an arbitrary diffeomorphism $\bXi$, we turn it into
a set of eigenfunctionals
\be
\bphitoo[1;\,\bXi^*\bXi^*_0\bg] \ ,\qquad \bXi_0\in{\rm Diff}_0\S
\eqnum{6.16}
\ee
where each eigenfunctional has the eigenvalue
$\boldG(\zeta) := (\det(\partial\bXi(\zeta)/\partial\zeta))^{-2}$.
In this way, we conclude that an arbitrary level $\boldG(\zeta) > 0$
is infinitely degenerate.

It remains an open question whether there is more degeneracy of the
$\boldG(\zeta) > 0$ levels than that corresponding to the action of
volume--preserving diffeomorphisms, and, if so, how the additional
degeneracy should be labeled.  We do not need to answer this
question. Formally, the degeneracy of the levels $\boldG(\zeta) > 0$
can be taken into account by the introduction of a degeneracy index $d$.
The Hilbert space ${\cal H}^{+}$ is spanned by the eigenfunctionals
$\bphitoo_d[\boldG;\bg]$ which satisfy
\be
{\hat\boldG}(\zeta)\, \bphitoo_d[\boldG;\,\bg] = \boldG(\zeta)\,
\bphitoo_d[\boldG;\,\bg] \ .\eqnum{6.17}
\ee
An arbitrary state $\bPsi[\bg]$ in ${\cal H}^{+}$ can be written as a
linear superposition
\be
\bPsi[\bg] = \sum_d\int_{G(\zeta)>0} {\rm D}\boldG \  C_d[\boldG]\,
\bphitoo_d[\boldG;\,\bg] \eqnum{6.18}
\ee
of the eigenfunctionals $\bphitoo_d[\boldG;\,\bg]$ with coefficients
$C_d[\boldG]$. These eigenfunctionals
form a basis for the energy--density representation (6.14)--(6.15).

{}From the considerations above, the general solution of the functional
Schr\"{o}dinger equation (6.5) can be written in the form
\be
\bPsi[\bT,\bg] = {\boldU}[\bT - \bT_0;\;\hat\bg,\hat\bp] \, \bPsi[\bT_0,\bg]
\ ,\eqnum{6.19}
\ee
where
\be
{\boldU}[\bT - \bT_0;\;\hat\bg,\hat\bp] := \exp\biggl( -i \int_\S d^3\zeta
\Bigl( \bT(\zeta) - \bT_0(\zeta) \Bigr) {\hat\boldh}(\zeta) \biggr)
\ .\eqnum{6.20}
\ee
Here, $\bPsi[\bT_0,\bg]$ is an arbitrary initial state in
${\cal H}^{+}$, and the operator ${\hat\boldh}(\zeta)$
is defined by spectral analysis. The state $\bPsi[\bT,\bg]$ at the
dust time $\bT(\zeta)$ follows from the initial state $\bPsi[\bT_0,\bg]$
at the dust time $\bT_0(\zeta)$ by the action of the many--fingered--time
evolution operator ${\boldU}[\bT - \bT_0;\;\hat\bg,\hat\bp]$.

\subsection{The Action of Symmetries on States}

The wave functional $\bPsi[Z,\bT,\bg]$ characterizes the quantum state of
the gravitational field on the hypersurface specified by the dust time
$\bT(\zeta)$. Such a description is not unique. First of all, the dust time
is not unique because the  time foliation $\T$ depends on the choice of the
fiducial hypersurface $\bT(\zeta)=0$. By changing this hypersurface, the
time variable $\bT(\zeta)$ is changed according to
\be
\bT'(\zeta) = \bT(\zeta) + \bTh(\zeta) \ .\eqnum{6.21}
\ee
(This is Eq.~(2.27) expressed in the dust coordinates.) The
variables $\bT(\zeta)$ and $\bT'(\zeta)$ connected by Eq.~(6.21) specify
a given hypersurface relative to two choices of dust time, $\T$ and $\T'$.
Thus, the fiducial hypersurface $\bT(\zeta)=0$ of the time foliation $\T$
is also the hypersurface $\bT'(\zeta) = \bTh(\zeta)$ of the time
foliation $\T'$.

The wave functional $\bPsi[Z,\bT,\bg]$ does not provide a unique description
of the gravitational field on a given hypersurface for a second reason:
The Lagrangian coordinates $\zeta^k$ that label the dust worldlines are
not unique. A relabeling of the dust worldlines is brought about by a
diffeomorphism $\bXi:\S\to\S$ of the dust space, as discussed in Secs.~2e
and 5b. Under the action of $\bXi$, the dust time
$\bT$ and the metric $\bg$ change by the pullback mappings
\be
\bT' := \bT\circ\bXi
\ ,\qquad \bg' := \bXi^*\bg \ .\eqnum{6.22}
\ee
According to Eq.~(2.26), the action of a diffeomorphism $\bXi$ on the
dust frame $Z$ is given by $Z' := \bXi^{-1}\circ Z$.

The value of the state functional should not depend on the specification of
the hypersurface.  Thus, if $\bPsi[Z,\bT,\bg]$ describes the state of the
gravitational field relative to the dust time ${\T}$ and $\bPsi'[Z,\bT',\bg]$
describes the {\it same\/} state relative to the dust time ${\T}'$, then
these two wave functionals should be equal:
\be
\bPsi'[Z,\ \bT'\! =\!\bT + \bTh ,\ \bg] = \bPsi[Z,\bT,\bg] \ .\eqnum{6.23}
\ee
Likewise, the value of the state functional should not depend on the choice
of Lagrangian coordinate labels for the dust particles. Therefore,
\be
\bPsi'[Z'\! =\!\bXi^{-1}\circ Z ,\  \bT'\! =\!\bT\circ\bXi ,\
\bg'\! =\!\bXi^*\bg] = \bPsi[X,\bT,\bg]\eqnum{6.24}
\ee
where $\bPsi[Z,\bT,\bg]$ describes the state of the gravitational field with
respect to the Lagrangian coordinates $\zeta^k$ and $\bPsi'[Z',\bT',\bg']$
describes the {\it same\/} state with respect to the Lagrangian coordinates
$\zeta^{k'}$.

These changes in the description of the state are generated by the Noether
charges. Thus, the state $\bPsi'$ from Eq.~(6.23) can be obtained from the
state $\bPsi$ by the action of the operator
\be
{\hat U}[\bTh] := \exp\Bigl( -i {\hat Q}[\bTh]\Bigr) \eqnum{6.25}
\ee
generated by the Noether charge
\be
{\hat Q}[\bTh] = \int_\S d^3\zeta\; \bTh(\zeta) {\hat\bP}(\zeta)
\ .\eqnum{6.26}
\ee
Indeed, we have
\be
\bPsi'[Z,\bT',\bg] = \bPsi[Z,\ \bT'-\bTh,\ \bg] = {\hat U}[\bTh]
\;\bPsi[Z,\bT',\bg] \ .\eqnum{6.27}
\ee
Similarly, let $\bXi^k(\zeta,\sigma)$ denote a one--parameter group of
diffeomorphisms with $\bXi^k(\zeta,0) = \zeta^k$ corresponding to the
identify and $\bXi^k(\zeta,1) = \bXi^k(\zeta)$ corresponding to a
given final diffeomorphism.  As usual, define the infinitesimal generators
\be
\bxi^k_\sigma(\zeta) := \frac{d\bXi^k(\zeta,\sigma)}{d\sigma} \eqnum{6.28}
\ee
and the Noether charges ${\hat Q}[{\vec\bxi}_\sigma]$ corresponding to the
classical charges (5.11). Then the state $\bPsi'$ from Eq.~(6.24) can be
obtained from the state $\bPsi$,
\be
\bPsi'[Z',\bT',\bg'] = \bPsi[ \bXi\circ Z' ,\  \bT'\circ\bXi^{-1} , \
(\bXi^{-1})^*\bg' ] = {\hat U}[\bXi] \;\bPsi[Z',\bT',\bg'] \ ,\eqnum{6.29}
\ee
by the action of the operator
\be
{\hat U}[\bXi] := {\sf T}_\sigma \exp\biggl( -i \int_0^1 d\sigma\,
{\hat Q}[{\vec\bxi}_\sigma] \biggr) \ .\eqnum{6.30}
\ee
Here, ${\sf T}_\sigma$ demands that the operators ${\hat
Q}[{\vec\bxi}_\sigma]$ be ordered so that ${\hat
Q}[{\vec\bxi}_{\sigma_1}]$ precedes ${\hat Q}[{\vec\bxi}_{\sigma_2}]$
if $\sigma_{2} > \sigma_{1}$.

The consistency of the formalism requires that physical states be transformed
by the change of description again into physical states.  Thus, if $\bPsi$
satisfies the constraints (6.1) and (6.3), the state $\bPsi'$ defined by
Eq.~(6.23) should also satisfy the constraints.  This follows from the
commutator of the Noether charge (6.26) with the constraint operators:
\be
[ {\hat Q}[\bTh] , \hat\boldH_\uparrow(\zeta) ] = 0 = [ {\hat Q}[\bTh] ,
{\hat H}_{\uparrow k}(x) ] \ .\eqnum{6.31}
\ee
Equations (6.27) and (6.31) imply
\be
\hat\boldH_\uparrow(\zeta)\; \bPsi'[Z,\bT,\bg] = {\hat U}[\bTh]
\hat\boldH_\uparrow(\zeta)\; \bPsi[Z,\bT,\bg] \ ,\eqnum{6.32}
\ee
and similarly for ${\hat H}_{\uparrow k}(x)$. Hence, if $\bPsi$
satisfies the constraints, $\bPsi'$ also satisfies the constraints.

Similarly, if $\bPsi$ satisfies the constraints, then $\bPsi'$ as defined
in Eq.~(6.24) should also satisfy them. This follows from the commutation
relations
\bea
\bigl[\, {\hat H}_{\uparrow k}(x) ,{\hat Q}[\vec\bxi] \, \bigr] & = &
i {\bxi^\ell}_{,k}({\hat Z}(x)) \, {\hat H}_{\uparrow\ell}(x)
\ ,\eqnum{6.33a}\\
\bigl[\, {\hat\boldH}_\uparrow(\zeta) , {\hat{Q}}[\vec{\bxi}] \,
\bigr]  & =  & i \bigl( {\hat\boldH}_\uparrow(\zeta)  \,
\bxi^k(\zeta) \bigr)_{,k} \ ,\eqnum{6.33b}
\eea
which are the quantum counterparts of the classical Poisson brackets
(5.13b), (5.14) between $Q[\vec\bxi]$ and the constraints $H_{\uparrow
k}(x)$, $\boldH_\uparrow(\zeta)$.  Equations (6.33) show that if
$\bPsi[Z,\bT,\bg]$ satisfies the constraints, then ${\hat
Q}[\vec\bxi]\; \bPsi[Z,\bT,\bg]$ also satisfies the constraints. In
turn, if ${\hat Q}[\vec\bxi]\; \bPsi[Z,\bT,\bg]$ satisfies the
constraints, then $\bPsi'[Z,\bT,\bg] = {\hat U}[\bXi]\,
\bPsi[Z,\bT,\bg]$ also satisfies the constraints. We conclude that the
description of a physical state can be changed by the action of the
operators ${\hat U}[\bTh]$ and ${\hat U}[\bXi]$ generated by the
Noether charges.

\subsection{The Inner Product}

The transition from the Hamiltonian constraint $H_{\perp}(x)$ to the
new constraint $\boldH_\uparrow(\zeta)$ leads, upon quantization, to
the replacement of the Wheeler--DeWitt equation by the functional
Schr\"{o}dinger equation (6.5). Unlike the Wheeler--DeWitt equation,
the Schr\"{o}dinger equation has (at least formally) a conserved
positive--definite inner product \cite{K-Ox}. This, of course, is the
main motivation behind rewriting the Hamiltonian constraint in the
forms (4.6) and (5.8).

Let $\bPsi[\bg]$ be a functional of Riemannian metrics $\bg(\zeta)\in{\rm
Riem}\S$ in the dust space $\S$. The inner product between two such
functionals is defined by the functional integral
\be
\langle \bPsi_1 | \bPsi_2 \rangle := \int_{{\rm Riem}\S} {\rm D}\bg \,
{\bar\bPsi}_1[\bg] \bPsi_2[\bg] \ .\eqnum{6.34}
\ee
Those functionals $\bPsi[\bg]$ that are square integrable under the
product (6.34) form a Hilbert space ${\H}$.

The functionals $\bPsi[\bg]$ can be interpreted as states of the
gravitational field on a {\it fixed\/} initial hypersurface
$\bT_0(\zeta)$:
\be
\bPsi[\bT_0,\bg] := \bPsi[\bg] \ .\eqnum{6.35}
\ee
By virtue of their independence from $Z$, the state functionals
$\bPsi[\bT_0,\bg]$ automatically satisfy the operator
constraints (6.3). Let us evolve these states by the Schr\"{o}dinger
equation (6.5). The solutions $\bPsi[\bT,\bg]$, from Eq.~(6.19), form
a space which we call $\H_\uparrow$. By fixing $\bT(\zeta)$ to be the
given initial hypersurface $\bT_0(\zeta)$, a functional
$\bPsi[\bT,\bg]$ from $\H_\uparrow$ is mapped to a functional (6.35)
from ${\H}$.

We now ask whether the inner product
\be
\langle \bPsi_1 | \bPsi_2 \rangle := \int_{{\rm Riem}\S} {\rm D}\bg \,
{\bar\bPsi}_1[\bT,\bg] \bPsi_2[\bT,\bg]  \eqnum{6.36}
\ee
between two states from the solution space $\H_{\uparrow}$ depends on
the choice of hypersurface $\bT(\zeta)$. Formally, the answer is no,
provided that the Hamiltonian density $\boldh(\zeta;\,\hat\bg,\hat\bp]$
of Eq.~(6.5) is a self--adjoint operator under the inner product (6.34).
However, $\hat\boldh$ is not self--adjoint on the whole space ${\H}$,
but only on the linear subspace $\H^+\subset\H$ which is spanned by the
simultaneous eigenfunctions (6.13) of the operators
${\hat\boldG}(\zeta)$ with positive eigenvalues $\boldG(\zeta) > 0$.
Hence, if the inner product (6.36) is to be independent of the choice
of hypersurface, we must limit the solution space $\H_{\uparrow}$
to $\H_\uparrow^+$.

To summarize, the state of the gravitational field can be represented
either as an element of the space of solutions $\H_{\uparrow}^{+}$ or
as an element of the space of initial states $\H^+$. Conservation
of the inner product guarantees that the value obtained for the inner
product is the same irrespective of the representation.

\subsection{Observables}

Any functional $\boldF[\bg,\bp]$ of the metric $\bg(\zeta)$ and
extrinsic curvature $\bp(\zeta)$ represents a gravitational dynamical
variable that is invariant under $\diffsigma$. We will call the space
of such variables ${\cal F}$. If the dust particles are labeled by a
fixed set of Lagrangian coordinates $\zeta^k$, then in principle the
full metric $\bg_{k\ell}(\zeta)$ on a given hypersurface, not just the
intrinsic geometry of the hypersurface, can be measured. A
corresponding statement holds for the extrinsic curvature. If we want,
we can restrict our attention to variables in $\cal F$ that do not
depend on the choice of Lagrangian coordinates for the dust; such
variables have vanishing Poisson brackets with the generator
$Q[\vec\bxi]$ of dust space diffeomorphisms. In the following, it
makes no difference whether we take the position that all the
dynamical variables in ${\cal F}$ are measurable, or only those that
are invariant under $\diffs$. Thus, we leave open the issue of which
class of variables should be called observables.

In quantum theory, an observable from ${\cal F}$ is represented by
an operator
\be
\hat\boldF = \boldF[\hat\bg,\hat\bp] \eqnum{6.37}
\ee
that acts on ${\H}$ and is self--adjoint under the inner product (6.34).
The statistical interpretation of quantum geometrodynamics is based on
two fundamental assumptions. The first one is the mean value formula:
When ${\boldF}$ is measured on an ensemble of systems described by
the state functional $\bPsi[\bT_0,\bg] \in \H$ on a hypersurface
$\bT_0(\zeta)$, the mean value of the result is given by the expression
\be
\langle \boldF \rangle = \langle \bPsi | \hat\boldF\bPsi \rangle
\ ,\eqnum{6.38}
\ee
with the inner product (6.34). By the standard argument, it follows
that the permissible values of $\boldF$ are the eigenvalues of
$\hat\boldF$ and, if the system is in one of the eigenstates of
$\hat\boldF$, the measurement of $\hat\boldF$ on the hypersurface
$\bT_0(\zeta)$ yields the corresponding eigenvalue with certainty.
The second assumption is the {\it reduction postulate\/}: If $\hat\boldF$
is measured on a hypersurface $\bT_0(\zeta)$ and those systems for
which the measurement yields the eigenvalue $\boldF'$ are collected
into a new ensemble, then the state functional $\bPsi'[\bT_0,\bg]$
that describes the new ensemble is the orthogonal projection
(under the inner product (6.34)) of the original state $\bPsi[\bT_0,\bg]$
to the linear subspace of $\H$ spanned by the eigenstates of $\hat\boldF$
belonging to the eigenvalue $\boldF'$.

So far, observables have been represented by operators acting on a state
$\bPsi[\bT_0,\bg]\in\H$ at the time of measurement $\bT_0(\zeta)$. The
functional
\be
\bPhi[\bT_0,\bg] = \hat\boldF\, \bPsi[\bT_0,\bg] \eqnum{6.39}
\ee
lies in the same instantaneous Hilbert space $\H$ as the original
functional $\bPsi$. Let us consider the functionals $\bPsi$ and $\bPhi$
as initial data and evolve them by the Schr\"{o}dinger equation (6.5)
away from the initial hypersurface. In this way, we obtain states (6.19)
that belong to the solution space $\H_\uparrow$:
\bea
\bPsi[\bT,\bg] & = & \boldU[\bT-\bT_0;\;\hat\bg,\hat\bp]\; \bPsi[\bT_0,\bg]
\ ,\eqnum{6.40}\\
\bPhi[\bT,\bg] & = & \boldU[\bT-\bT_0;\;\hat\bg,\hat\bp]\; \bPhi[\bT_0,\bg]
\ .\eqnum{6.41}
\eea
The state (6.41) is obtained from the state (6.40) by the action of
the operator
\be
\boldF_{{\bf T}_0}[\bT;\;\hat\bg,\hat\bp] =
\boldU[\bT-\bT_0;\;\hat\bg,\hat\bp] \,
\boldF[\hat\bg,\hat\bp] \, \boldU^{-1}[\bT-\bT_0;\;\hat\bg,\hat\bp]
\ .\eqnum{6.42}
\ee
Equation (6.42) represents the observable $\boldF[\bg,\bp]$ which is
measured on the hypersurface $\bT_0(\zeta)$ as an operator
${\hat\boldF}_{{\bf T}_0}$ on the solution space $\H_{\uparrow}$.
The operator ${\hat\boldF}_{{\bf T}_0}$ is
constructed so that it commutes with the Hamiltonian constraint
${\hat\boldH}_\uparrow(\zeta)\,$:
\be
[ {\hat\boldF}_{{\bf T}_0} , {\hat\boldH}_\uparrow(\zeta) ] = 0
\ .\eqnum{6.43}
\ee
This property guarantees that the action of ${\hat\boldF}_{{\bf T}_0}$
on a state
in $\H_{\uparrow}$ again yields a state in $\H_{\uparrow}$.

We must now translate into the quantum theory the positivity condition
$\boldG(\zeta;\,\bg,\bp] > 0$ on the domain of classical observables
$\boldF[\bg,\bp]$. Quantum mechanically, the action of an operator
$\boldF[\hat\bg,\hat\bp]$ on a state $\bPsi[\bg]\in\H^+$ from the
subspace $\H^+$ spanned by the eigenfunctionals of the operators
$\boldG(\zeta;\,\hat\bg,\hat\bp]$ with positive eigenvalues
$\boldG(\zeta) > 0$ should leave the state in $\H^+$. If an observable
$\hat\boldF\in \hat{\F}$ does not have this property, the reduction
postulate leads us into difficulty. The reduced state $\bPsi'[\bT_0,\bg]$
obtained after the measurement of such an observable would not
lie in $\H^{+}$. Therefore, if we measure $\hat\boldG(\zeta)$ on the
reduced state $\bPsi'$, there would be a nonvanishing probability of finding
$\hat\boldG(\zeta)$ in the unphysical region $\boldG(\zeta) \leq 0$.
Furthermore, by allowing the reduced state $\bPsi'[\bT_0,\bg]$ to propagate
to a later time $\bT(\zeta) > \bT_0(\zeta)$ via the evolution operator
$\boldU[\bT - \bT_0;\;\hat\bg,\hat\bp]$, we would find that the norm
$\langle \bPsi'[\bT,\bg] | \bPsi'[\bT,\bg] \rangle$ depends on
$\bT(\zeta)$ and the total probability is not conserved.

These considerations force us to define observables as self--adjoint
operators $\hat\boldF \in {\hat\F}^+$ on $\H^{+}$ rather than on
$\H$.  That is, an operator $\hat\boldF$ is considered to be an observable
only if its action on the states $\bPsi$ from $\H^+$ again yield states
in $\H^+$. Correspondingly, the observables $\hat\boldF_{{\bf T}_0}$ are
self--adjoint operators on the space of solutions $\H^+_\uparrow$.
Thus, for $\bPsi$ an element of $\H^+_\uparrow$, the state
$\hat\boldF_{{\bf T}_0} \bPsi$ also is an element of $\H^+_\uparrow$. This
implies that the mean value of $\hat\boldF_{{\bf T}_0}$, defined by
\be
\langle \boldF_{{\bf T}_0} \rangle = \langle \bPsi |
\hat\boldF_{{\bf T}_0} \bPsi \rangle \ ,\eqnum{6.44}
\ee
can be evaluated on any hypersurface, not only at the instant
$\bT_0(\zeta)$ of the measurement.

The condition for an operator defined on $\H$ to be an
observable can be expressed in terms of the projection operator
\be
{\hat\P}^+ := \prod_{\zeta\in\S} \Theta(\hat\boldG(\zeta)) \eqnum{6.45}
\ee
from ${\H}$ to ${\H}^{+}$. The projector ${\hat\P}^+$ is a product of
step functions $\Theta$ of the commuting operators $\hat\boldG(\zeta)$.
The Hilbert space ${\H}^{+}$ is spanned by states of the form
${\hat\P}^+\bPsi$, where $\bPsi\in\H$. An operator $\hat\boldF$
on ${\H}$ is an observable if it satisfies the condition
\be
[\hat\boldF , {\hat\P}^+ ] {\hat\P}^+ = \bigl( 1 - {\hat\P}^+ \bigr)
\hat\boldF {\hat\P}^+ = 0 \ .\eqnum{6.46}
\ee
Thus, any variable $\boldF[\hat\bg,\hat\bp]$ that commutes with
${\hat\boldG}(\zeta)$, {\it i.e.\/}, that is a constant of motion,
is automatically an observable. However, $\hat\boldF$ does not need to
be a constant of motion to satisfy Eq.~(6.46); observables do not need
to be constants of motion.

The fundamental problem in interpreting canonical quantum gravity through
its coupling to dust is the construction of meaningful gravitational
observables. Ultimately, one would like to exhibit complete sets of
commuting observables.

On our way, we met some gravitational observables.  The operators
$G(z;\,{\bf \hat{g}},{\bf \hat{p}}]$, equivalent to the energy densities
$h(z;\,{\bf g},{\bf p}]$, form a commuting set of gravitational observables
on ${\cal H}^{+}$.  The Noether charges $\hat{Q}[{\vec\bxi}]$
and $\hat{Q}[{\bTh}]$ represent observables on $\H_{\uparrow}^{+}$.
By using the constraints, we can eliminate the
embedding momenta and express the charges in terms of the gravitational
variables.  There are thus some observables we can measure and subject to
statistical analysis.  All of these are constants of motion.

It is disappointing that the fundamental gravitational variables
$\hat\bg(z) = \bg(z)\times$ and $\hat\bp(z) = -i \delta/\delta\bg(z)$
are {\it not} gravitational observables. Although the condition (6.46)
is cumbersome to handle explicitly, it is nevertheless fairly obvious
that $\hat\bg(z)$ and $\hat\bp(z)$ do not necessarily leave the states
in ${\H}^{+}$. The underlying difficulty is that the gravitational
Hamiltonian $h(x;\,\hat\bg,\hat\bp]$ is not real {\em by its
construction} on the entire Hilbert space ${\cal H}$ which carries the
representation of the fundamental operators $\hat\bg(z)$ and $\hat\bp(z)$.

One can illustrate these problems on simple finite-dimensional models,
like a relativistic particle in the proper-time formalism (Appendix
G). In particular, one can explicitly verify that even on a flat
Minkowskian background the multiplication position operator is {\em
not} an observable. One may hope to clarify the status of fundamental
variables on such simple models.  For the time being, their
relation to quantum observables remains unclear.

\subsection{Quantum Theory in Terms of State
Functionals $\Psi[Z,T,g]$}

The quantum theory of gravity coupled to dust, as developed thus far,
can be expressed in terms of the original set of canonical variables
$Z^k(x)$, $P_k(x)$, $T(x)$, $P(x)$, $g_{ab}(x)$, and $p^{ab}(x)$. In
this case, the quantum states are
functionals $\Psi[Z,T,g]$ that satisfy the operator constraint equations
\bea
H_\uparrow(x;\,{\hat P},{\hat g}_{ab},{\hat p}^{ab}]\; \Psi[Z,T,g] & = & 0
\ ,\eqnum{6.47a} \\
H_{\uparrow k}(x;\,{\hat Z}^k,{\hat P}_k,{\hat T},{\hat g}_{ab},
{\hat p}^{ab}]\; \Psi[Z,T,g] & = & 0 \ .\eqnum{6.47b}
\eea
Let us assume that $\bPsi[\bT,\bg]$ is a solution to the functional
Schr\"{o}dinger equation (6.1). Recall that such a functional automatically
satisfies the operator constraint (6.3) by virtue of its
lack of dependence on $Z^k(x)$. Now choose a mapping $Z:\Sigma\to\S$
that associates the points in space with particles of dust, and consider
the functional defined by
\be
\Psi[Z,T,g] := \bPsi[(Z^{-1})^*T,\; (Z^{-1})^*g] \ .\eqnum{6.48}
\ee
Here, $\bT = (Z^{-1})^*T$ and $\bg = (Z^{-1})^*g$ denote the pullbacks
by $X=Z^{-1}$ of the space tensors $T$ and $g$ to the dust space $\S$.
As we show below, the functional (6.48) satisfies the constraints
(6.47) and represents the quantum state of the system in terms of the
original canonical coordinates.

Consider first the supermomentum operator constraint (6.47b).
Since ${\hat H}_{\uparrow k}(x) = {\hat Z}^a_k(x) {\hat H}_a(x)$,
this equation implies ${\hat H}_a(x)\,\Psi = 0$ which in turn implies
that $\Psi$ is invariant under spatial diffeomorphisms.
Indeed, let $\varphi_\sigma:\Sigma\to\Sigma$ denote a one--parameter family
of diffeomorphisms generated by the vector field ${\vec N}(x)$. Assuming
an appropriate factor ordering, we have
\be
\frac{d}{d\sigma} \Psi[\varphi_\sigma^*Z, \varphi_\sigma^*T,
\varphi_\sigma^*g]\biggr|_{\sigma=0} = i\int_\Sigma d^3x\, N^a(x)
{\hat H}_a(x)\, \Psi[Z,T,g] = 0  \eqnum{6.49}
\ee
for any wave functional that satisfies the supermomentum operator
constraint. By exponentiation in $\sigma$, this result yields
\be
\Psi[\varphi^*Z,\, \varphi^*T,\, \varphi^*g] = \Psi[Z,T,g] \ ,\eqnum{6.50}
\ee
and shows that $\Psi$ is invariant under finite diffeomorphisms of
$\Sigma$. Conversely, a diffeomorphism invariant functional (6.50)
will satisfy
the supermomentum constraint (6.47b). The functional $\Psi$ defined in
Eq.~(6.48) is just such an invariant functional. This follows from the
fact that $(Z^{-1})^*T$ and $(Z^{-1})^*g$ are invariant under the
action of $\diffsigma$: For $(Z^{-1})^*T$, we have
\bea
\Bigl( (\varphi^*Z)^{-1} \Bigr)^* (\varphi^*T) & = &
(T\circ\varphi)\circ(Z\circ\varphi)^{-1}  =  T\circ Z^{-1} \nonumber \\
& = &  (Z^{-1})^*T \ ,\eqnum{6.51}
\eea
and similarly for $(Z^{-1})^*g$. Therefore the state functional
$\Psi[Z,T,g]$ defined in Eq.~(6.48) satisfies the supermomentum
constraint (6.47b).

The functional (6.48), which we assume satisfies the functional
Schr\"{o}dinger equation (6.1), also satisfies the functional
Schr\"{o}dinger equation (6.47a). In order to understand this
result, it is only necessary to clarify
the relationship between the operators $\hat P$, $\hat g$, and $\hat p$
and their dust space counterparts $\hat\bP$, $\hat\bg$, and $\hat\bp$.
We will assume that these sets of operators are related by the familiar
mapping of tensor fields by $Z$. For example,
the action of $\hat P(x)$ and the action of $\hat\bP(\zeta)$
are related by
\be
{\hat P}(x)\; \bPsi[(Z^{-1})^*T,\; (Z^{-1})^*g] =
\biggl|\frac{\partial Z(x)}{\partial x}\biggr| \biggl( {\hat\bP}(\zeta)\,
\bPsi[\bT, \bg] \biggr) \biggr|_{\zeta=Z(x),{\ }
{\bf T} = (Z^{-1})^*{T},{\ } {\bf g} = (Z^{-1})^*{g}}
\ .\eqnum{6.52}
\ee
There are no factor ordering ambiguities in this relationship.  Using
an abbreviated notation, we can write ${\hat P}(x) = |\partial
Z(x)/\partial x|{\hat\bP}(Z(x))$. Observe that if the operator $\hat
P(x)$ is represented by $-i\delta/\delta T(x)$ then $\hat\bP(\zeta)$
is represented by $-i\delta/\delta \bT(\zeta)$.  A relationship
analogous to that expressed in Eq.~(6.52) can be used to define a
correspondence between functions of the operators $\hat\bP$,
$\hat\bg$, and $\hat\bp$, and functions of the operators $\hat P$,
$\hat g$, and $\hat p$. In particular, the new super--Hamiltonian
operator is (using the abbreviated notation)
\be
H_\uparrow(x;\,{\hat P},{\hat g},{\hat p}] =
\biggl|\frac{\partial Z(x)}{\partial x}\biggr|
\boldH_\uparrow(Z(x);\,{\hat\bP},{\hat\bg},{\hat\bp}] \ .\eqnum{6.53}
\ee
{}From this relationship we obtain the correspondence
\be
{\hat H}_\uparrow(x)\, \bPsi[(Z^{-1})^*T,\; (Z^{-1})^*g] = 0
\quad \Longleftrightarrow \quad
{\hat\boldH}_\uparrow(\zeta)\, \bPsi[\bT,\bg] = 0   \ .\eqnum{6.54}
\ee
Therefore, if $\bPsi[\bT,\bg]$ satisfies the Schr\"{o}dinger equation
(6.1), then $\Psi[Z,T,g]$ as defined in Eq.~(6.48) satisfies the
Schr\"{o}dinger equation (6.47a).

The preceding arguments show that the state functional $\bPsi[\bT,\bg]$
can be expressed as a functional $\Psi$ of $T$, $g$, and a mapping
$Z:\Sigma\to\S$ through definition (6.48). Different choices of mappings
$Z$ lead to different functionals $\Psi$; however, these functionals are
related by the unitary transformation ${\hat U}[\bXi]$ from Eq.~(6.30)
and represent the same physical quantum state.

The spectral analysis, the construction of the Hilbert space and inner
product, and the definition of observables can be translated into the
language of the original canonical variables in a straightforward manner.
In particular, the Hilbert space $\H^+$ is spanned by the eigenfunctionals
$\phi[(Z^{-1})^*G;\, (Z^{-1})^*g]$ of ${\hat G}(x)$ with positive
eigenvalues $G(x)>0$. The inner product between two states of the
form (6.48) is
\bea
\langle \Psi_1 | \Psi_2 \rangle & := & \int_{{\rm Riem}\S} {\rm D}\bg \;
{\bar\bPsi}_1[(Z^{-1})^*T,\;\bg]\; \bPsi_2[(Z^{-1})^*T,\;\bg]  \eqnum{6.55a} \\
& = & \int_{{\rm Riem}\Sigma} {\rm D}g \;
{\bar\Psi}_1[Z, T, g]\, \Psi_2[Z, T, g]  \ .\eqnum{6.55b}
\eea
Finally, the gravitational observables are operators
$F[{\hat Z},{\hat g},{\hat p}]$ that correspond, in the sense of
Eq.~(6.52), to operators $\boldF[\hat\bg,\hat\bp]$. Their action
on a state vector in the Hilbert space $\H^+$ again yields a state
in $\H^+$.
\section{Non--Rotating Dust}
\subsection{Rotation and Vorticity}

In general, no dust time ${\cal T}$ is better than any other. However, if the
dust does not rotate as it moves, so that the four--velocity field is
hypersurface orthogonal, then there is a privileged time foliation
${\cal T}_{\perp}$, namely, the foliation by hypersurfaces orthogonal to the
flow lines.  For such a foliation, $W_{k} = 0$.

The rotation of the dust is measured by the vorticity 2--form
\begin{equation}  \Omega_\ab := (\delta^\gamma_\alpha + U^\gamma U_\alpha)
     (\delta^\delta_\beta + U^\delta U_\beta)\, U_{[\gamma,\delta]} \ .
     \eqnum{7.1} \end{equation}
For $\Omega_{\ab} = 0$, the vector field $U^{\alpha}$ is
hypersurface orthogonal. The projectors in Eq.~(7.1) ensure that
$\Omega_{\ab} U^{\beta}\equiv 0$; hence, $\Omega_{\ab}$ is completely
characterized by its orthogonal projection $\bOm_{k\ell} :=
\Omega_{\ab} {\Upsilon^\alpha}_{,k} {\Upsilon^\beta}_{,\ell}$. The
equations of motion (2.17)--(2.18) imply that
\begin{equation}  \bOm_{k\ell} = {\boldW}_{[k,\ell]} \ . \eqnum{7.2}
     \end{equation}
Like ${\boldW}_{k}$ itself, $\bOm_{k\ell}$ is conserved along the flow lines
of $U^{\alpha}$:
\begin{equation} \pounds_{U}\bOm_{k\ell} = \bOm_{k\ell,\alpha} U^\alpha
    = 0 \ . \eqnum{7.3} \end{equation}
Hence, if the dust is irrotational at a spacetime point $y$,
then the equations of motion ensure that it is irrotational for all points
along the flow line that passes through $y$.

The projected vorticity (7.2) is an exact two--form in the dust space
${\cal S}$. Moreover, it is invariant under the symmetry transformation
(2.27).
If $\bOm_{k\ell}$ vanishes, the velocity field $\boldW_{k}$ is closed; thus
$\boldW_{k}$ can be (at least locally) transformed to zero by the symmetry
transformation (2.27). This transformation brings us from a foliation ${\cal
T}$
to the privileged time foliation ${\cal T}_{\perp}$.

The foliation ${\cal T}_{\perp}$ also can be constructed by choosing an
initial hypersurface orthogonal to the flow lines and then Lie propagating it
along $U^{\alpha}$. Because $U^{\alpha}$ is a geodesic vector field,
Eq.~(2.25),
the propagated hypersurface stays orthogonal to $U^{\alpha}$. This amounts to
the familiar construction of a Gaussian coordinate system \cite{Synge}. The
coordinate
system $(\tau, \zeta^{k})$ based on the time ${\cal T}_{\perp}$ and space
${\cal S}$ manifolds of an irrotational dust is a Gaussian coordinate system.
\subsection{Canonical Description of Non--Rotating Dust}

The pullback of the vorticity 2--form (7.2) from ${\S}$ to $\Sigma$ is
\begin{eqnarray}  \Omega_{ab} & := & \bOm_{k\ell} {Z^k}_{,a} {Z^\ell}_{,b}
          = ( W_k {Z^k}_{,{\sss [}a} )_{,b{\sss ]}}
          \nonumber \\   & = &
          U_{{\sss [}a,b{\sss ]}} \ . \eqnum{7.4} \end{eqnarray}
Note that in general the direct projection of the spacetime vorticity
tensor (7.1) onto $\Sigma$ differs from the spatial tensor $\Omega_{ab}$
by a term $A_{{\sss [}a} U_{b{\sss ]}}$, where $A_a$ is the spatial
projection of the dust acceleration $A_\alpha =
U^\beta\nabla_\beta U_\alpha$. When the Euler equations of motion hold,
the acceleration (2.25) vanishes and
the spatial projection of the spacetime vorticity coincides with the
vorticity $\Omega_{ab}$ of Eq.~(7.4).

Under spatial diffeomorphisms, $\Omega_{ab}(x)$ transforms as a
tensor; this determines its Poisson bracket with the supermomentum
$H^D_{a}(x)$. The Poisson bracket of $\Omega_{ab}(x)$ and
$\Omega_{cd}(x')$ vanishes:
\begin{equation} \{ \Omega_{ab}(x),\Omega_{cd}(x') \} = 0 \ .
     \eqnum{7.5} \end{equation}
Finally, the Poisson bracket of $\Omega_{ab}(x)$ and
the super--Hamiltonian is
\begin{equation} \{ \Omega_{ab}(x),H^D_\perp(x') \} = -2\frac{H^{Dc}(x')}
   {H^D_\perp(x')} \Omega_{c{\sss [}a}(x') \delta_{,b{\sss ]}}(x,x') \ .
      \eqnum{7.6} \end{equation}
This result is obtained as follows. Because neither $Z^{k}(x)$ nor
$W_{k}(x) = -P_k(x)/P(x)$ depend on $T(x)$, they have vanishing Poisson
brackets with $P(x')$. The only part of the super--Hamiltonian
$H^D_\perp(x')$
that affects these variables is the dust momentum $H_{a}^{D}(x')$ in the
square--root energy--density (3.19). The momentum $H_{a}^{D}(x')$ generates
the transformation of $Z^{k}(x)$ and $W_{k}(x)$ as scalars under
$\diffsigma$. Therefore,
\begin{eqnarray} \{ W_k(x), H^D_\perp(x') \} & = & W_{k,b}(x) S^b(x)
  \delta(x,x')  \ , \eqnum{7.7a}  \\
     \{ Z^k(x), H^D_\perp(x') \} & = & {Z^k}_{,b}(x) S^b(x) \delta(x,x')
     \ ,\eqnum{7.7b} \end{eqnarray}
with
\begin{equation} S^b(x) := \frac{H^{Db}(x)}{H^D_\perp(x)}
   \ . \eqnum{7.8} \end{equation}
Equations (7.7) allow us to calculate the Poisson bracket
\begin{eqnarray}  \lefteqn{ \{ W_k(x) {Z^k}_{,a}(x), H^D_\perp(x') \} =
   2\Omega_{ab}(x') S^b(x') \delta(x,x') }\qquad\qquad\qquad\qquad\qquad
   \nonumber\\ & & + \left( U_k(x) {Z^k}_{,b}(x) S^b(x)
      \delta(x,x') \right)_{,a} \ ,  \eqnum{7.9} \end{eqnarray}
where $\Omega_{ab}$ is given by Eq.~(7.4).  By taking the
exterior derivative of Eq.~(7.9), we obtain Eq.~(7.6) above.

The results (7.5)--(7.6) show that if a new constraint
\begin{equation} \Omega_{ab}(x) = 0 \eqnum{7.10} \end{equation}
is adjoined to the super--Hamiltonian and supermomentum constraints (3.20)
and (3.21), then the enlarged system of constraints is first class.
Moreover, from Eq. (3.14) we have
\begin{equation} U_a = -\frac{H^D_a}{P} \ . \eqnum{7.11} \end{equation}
It is easy to see that both $P$ and $H_{a}^{D}$ are invariant under the
symmetry transformations, {\it i.e.\/}, that their Poisson brackets with
the generators (3.24) and (3.25) vanish. Therefore $U_{a}$ and $\Omega_{ab}$
are also invariant under the symmetries:
\begin{equation} \{ \Omega_{ab}(x), Q[\bth] \} = 0 =
    \{ \Omega_{ab}(x), Q[\vec\bxi] \} \ . \eqnum{7.12} \end{equation}
Thus, the new constraint (7.10) is invariant under
the symmetry transformations (3.24), (3.25). The condition (7.10) constrains
the dust into irrotational flow. The first--class nature of the enlarged
system of constraints ensures that if the dust is irrotational on an
initial hypersurface, its dynamics prevents the evolution of vorticity.

Let us study the vorticity potential $W_{k}(x) := -P_{k}(x)/P(x)$ from the
canonical point of view. The Poisson bracket of $W_{k}(x)$ with the
super--Hamiltonian $H_{\perp}(x)$ is different from zero; on the other hand,
its Poisson bracket with the new constraint $H_{\uparrow}(x)$ vanishes:
\begin{equation} \{ W_k(x), H_\uparrow(x') \} = 0 \ .
     \eqnum{7.13} \end{equation}
This reflects the fact that $W_{k}(x)$ does not change along the worldlines
of the dust although it does change along the normal direction to an
embedding. We can now solve the Heisenberg equations of motion (4.18) for
the dynamical variables $W_{k}(x)\,$. Any such solution is a functional
$W_{k}(x;\,Z^{L}]$ of the embedding $Z^{L} = (T(x), Z^{\ell}(x))\,$, as
expressed in Eq.~(4.16). When we substitute Eq. (7.13) into the
$T$--component of the Heisenberg equations of motion, we learn that
$W_{k}(Z^{L}(x))$ cannot depend on $T(x)$, {\it i.e.\/}, that
$W_{k}=W_{k}(x;\,Z^{\ell}]\,$. Since $W_{k}(x) := -P_{k}(x)/P(x)$
is a spatial scalar, it follows that
\begin{equation}
\{ W_{k}(x), H_{a}(x')\} = W_{k,a}(x)\, \delta (x, x')\ .\eqnum{7.14}
\end{equation}
The definition (4.9) of $H_{\uparrow k}(x)$ then implies that
weakly, modulo the supermomentum constraint,
\begin{equation}
\{ W_{k}(x), H_{\uparrow \ell}(x')\} =  W_{k,a}(x)Z^{a}_{\ell}(x)
\, \delta (x, x')\ .\eqnum{7.15}
\end{equation}
The $Z^{\ell}$--component of the Heisenberg equations of motion
(4.18) thus reads
\begin{equation}
\frac{d W_{k}(x;\, Z^{m}]}{\delta Z^{\ell}(x')} =
 W_{k,a}(x)Z^{a}_{\ell}(x) \, \delta (x, x')\ .\eqnum{7.16}
\end{equation}
The delta function on the right-hand side of Eq.~(7.16) tells
us that $W_{k}(x;\,Z^{\ell}]$ is an ultralocal functional of
$Z^{\ell}(x)\,$, {\it i.e.\/}, that
\begin{equation} W_k(x;\,Z^L] = \boldW_k(Z^\ell(x)) \ .
    \eqnum{7.17} \end{equation}
It is easy to see that the expression (7.17) solves the Heisenberg
equations of motion (7.16). In this roundabout way, we recover the
canonical counterpart of the spacetime equation (2.18).

Now impose the condition that the dust does not rotate, so
that its vorticity (7.4) vanishes.  This implies that (locally)
$W_{k}(x){Z^k}_{,a}(x)$ must be a gradient of a spatial scalar. Therefore
the function $\boldW_{k}(\zeta^\ell)$ in Eq.~(7.17) must be the gradient
of a function $-{\bTh}(\zeta^{\ell})$:
\begin{equation} W_k(x;Z^L] = -\bTh_{,k}(Z^\ell(x)) \ .
       \eqnum{7.18} \end{equation}
The condition (7.18) enables us to transform to zero the particular
solution (7.17) of the Heisenberg equations of motion by successive
applications of the symmetry transformation (3.27) generated by the
Noether charge $Q[\bth]$.  This transforms the momentum
$P_{k}(x)= -P(x) W_k(x)$ to zero.

To summarize, any particular solution (7.18) of the constraint equations
and the Hamilton equations of motion of non--rotating dust can be cast
into the form in which the dust time foliation ${\T}_{\perp}$ is
orthogonal to the dust frame ${\S}$. When we demand from the outset not
only that the dust be non--rotating, but also that the time foliation
${\T}$ which is used for its description be orthogonal to the dust frame,
we can impose the additional constraint
\begin{equation} P_k(x) = 0 \ .
      \eqnum{7.19} \end{equation}
This constraint is stronger than the requirement (7.10) that the
vorticity $\Omega_{ab}(x)$ vanish; Eq.~(7.19) implies Eq.~(7.10),
while Eq.~(7.10) does not automatically enforce Eq.~(7.19).

Now consider the system of constraints $H_\uparrow(x)= 0$,
$H_{\uparrow k}(x)=0$, and $P_k(x) =0$. This constraint system is
first class: In Sec.~4 we showed that the new super--Hamiltonian
constraints $H_\uparrow (x)$ and the new supermomentum constraints
$H_{\uparrow k} (x)$ have vanishing Poisson brackets among themselves.
The momenta $P_{k}(x)$ also have vanishing Poisson brackets
among themselves and with the generators $H_{\uparrow}(x)$. From
Eq.~(4.9) we see that the Poisson brackets of $P_k$ and
$H_{\uparrow k}$ can be written as
\be
\{ P_k(x) , H_{\uparrow\ell}(x') \} =
Z^a_\ell(x') \{ P_k(x) , H_a(x') \} +
\{ P_k(x) , Z^a_\ell(x') \} H_a(x') \ .\eqnum{7.20}
\ee
Because $P_k(x)$ are scalars under $\diffsigma$, their Poisson brackets
with $H_{a}(x)$ are linear in $P_{k}(x)$. Thus, the Poisson
brackets $\{ P_k(x) , H_{\uparrow\ell}(x') \}$ also close.

In general, the functional $Q[\bth]$ from Eq.~(3.25) does not have vanishing
Poisson brackets with $P_k(x)$, and thus does not generate a symmetry of the
theory defined by the new system of constraints $H_\uparrow= 0$,
$H_{\uparrow k}=0$, and $P_k=0$. Indeed, for  $\bth_{,k}\neq 0$, $Q[\bth]$
generates a tilt of
the time foliation $\T$ that spoils the Gaussian character of the coordinate
system $(\tau,\zeta^k)$ based on the irrotational dust. For $\bth =
{\it const\/}$, $Q[\bth]$ generates a symmetry that consists of a relabeling
of the leaves of the privileged foliation $\T_\perp$. In this case, $Q[\bth]$
is a conserved charge whose value (for $\bth = 1$) is the total relativistic
mass of the dust. From  Eq.~(3.24), the functional $Q[\vec\bxi]$ is seen to
equal the new constraint (7.19) smeared with a prescribed dust space vector
field $-\bxi^k(Z(x))$. Thus, the Poisson brackets of $Q[\vec\bxi]$ with
$P_k(x)$ vanish weakly. The smeared constraint $Q[\vec\bxi]$ is the canonical
generator of dust space diffeomorphisms, in which the Lagrangian coordinate
labels for the dust particles are changed.

The additional constraint (7.19) enables us to replace the Hamiltonian
constraint $H_{\uparrow}$ by an equivalent constraint that does not
involve the square of the gravitational super--Hamiltonian $H^G_\perp$.
Modulo Eq.~(7.19), the new supermomentum constraint (4.9) implies
$H_a^G\approx -\sqrt{G} T_{,a}$. This can be combined with the definition
$G := (H_\perp^G)^2 - g^{ab} H_a^G H_b^G$ to yield $(H_\perp^G)^2 =
(1 + g^{ab} T_{,a} T_{,b} )G$. The Hamiltonian constraint
$H_\uparrow = P - \sqrt{G}$
then takes the form
\begin{equation} H_{\uparrow 0} := P(x) + h_0(x;\,T,g,p] = 0 \ ,
    \eqnum{7.21} \end{equation}
with the Hamiltonian
\begin{equation} h_0(x;\;T,g,p] := \left( 1 + g^{ab}(x) T_{,a}(x) T_{,b}(x)
    \right)^{-1/2} H^G_\perp(x) \ . \eqnum{7.22} \end{equation}
The new Hamiltonian (7.22) does not contain an inconvenient square
root and is quadratic in the gravitational
momentum $p^{ab}(x)$. Unlike the general Hamiltonian $h(x)$ of Eq.~(4.7),
however, it {\it does} depend on the Gaussian time $T(x)$ labeling
the leaves of ${\T}_{\perp}$. The constraint (7.21) was employed for the
quantization of gravity interacting with non--rotating dust in our previous
work [10].
\subsection{Quantizing a Non--Rotating Dust}

We have seen that any particular solution of the classical constraint
equations and the Hamilton equations of motion for non--rotating dust can
be described with respect to the time foliation ${\T}_{\perp}$ that is
orthogonal to the dust frame ${Z^k}_{,\alpha}(y)$. By imposing the
constraint $P_k(x)=0$, we demand both that the dust be non--rotating and
that its motion be described with respect to the frame--orthogonal
foliation. The canonical description of this system can be formulated
either in terms of the original canonical variables or in terms of the
new canonical dust--space variables. For the original variables, the new
constraint has the simple form $P_k(x)=0$, while the supermomentum
constraint has a relatively complicated form (4.9). With the
dust--space variables, the supermomentum constraint is simple---it
is the momentum conjugate to $Z^k(x)$---while the new constraint
(7.19) takes the form
\be
P_k(x) = H_{\uparrow k}(x) - \left|\frac{\partial Z(x)}
    {\partial x}\right| \biggl( \boldH_k^G(\zeta) +
    \bP(\zeta) \bT_{,k}(\zeta) \biggr)\biggr|_{\zeta = Z(x)}
\ .\eqnum{7.23}
\ee
In each case, the constraint that generates diffeomorphisms of the
manifold on which the tensor fields live ($\Sigma$ or $\S$) has a
relatively complicated form, while the constraint that generates
diffeomorphisms of the complementary manifold is trivial. Thus, we
see that for non--rotating dust with the frame--orthogonal foliation
there is no relative advantage in using one set of canonical variables
as opposed to the other. In what follows we will work with the original
canonical variables.

The system of constraints $H_{\uparrow k}=0$, $H_{\uparrow 0}=0$, and
$P_k=0$ is first class. Therefore these constraints can be imposed
consistently as restrictions on the state functional $\Psi[Z,T,g]$.
The constraint (7.19) in particular yields the restriction
\be
{\hat P}_k(x)\, \Psi[Z,T,g] = 0 \ ,\eqnum{7.24}
\ee
and implies that the state $\Psi$ does not depend on the frame variables
$Z^k(x)$. Our task then is to find solutions $\Psi[T,g]$ of the operator
equations that follow from the classical constraints $H_{\uparrow 0}(x)=0$
and $H_{\uparrow k}(x)=0$. These are the many--fingered time
Schr\"{o}dinger equation
\be
{\hat H}_{\uparrow 0}(x)\; \Psi[T,g] = 0 \quad\Longleftrightarrow\quad
i\frac{\delta\Psi[T,g]}{\delta T(x)} = h_0(x;\,T,\hat g,\hat p]\, \Psi[T,g] \ ,
\eqnum{7.25}
\ee
and the equation that requires $\Psi$ to be invariant under $\diffsigma$,
\be
\Bigl( T_{,a}(x) {\hat P}(x) + {\hat H}_a^G(x) \Bigr) \Psi[T,g] = 0
\ .\eqnum{7.26}
\ee
Functionals $\Psi[T,g]$ that satisfy Eqs.~(7.25) and (7.26) describe the
quantum states of non--rotating dust with respect to the frame--orthogonal
time foliation $\T_\perp$. We have explored this scheme in a previous
paper \cite{K+T}.

One might expect that any state $\Psi$ with vanishing vorticity,
\be
{\hat\Omega}_{ab}(x)\, \Psi[Z,T,g] = 0 \ ,\eqnum{7.27}
\ee
can be transformed into a state satisfying Eq.~(7.24) by the action of the
symmetry operator (6.25). However, this is not so. The underlying reason
is that a classical $P_{k}(x)$ can be brought to zero by a symmetry
transformation for any {\it particular\/} solution of the equations
describing a non--rotating dust, but the state functional $\Psi[Z,T,g]$
generally describes an ensemble of quantum systems corresponding to a {\it
superposition\/} of such individual solutions. Correspondingly, there is
in general no single function ${\bTh}(Z(x))$ whose charge $\hat{Q}[\bTh]$
would turn a state functional that satisfies Eq.~(7.27) into one that
satisfies the stronger equation (7.24).

We now ask whether every $Z$--independent solution $\Psi[T,g]$ of
Eqs.~(6.47) is also a solution of Eqs.~(7.25)  and (7.26), and inversely,
whether every solution of Eqs.~(7.25)  and (7.26) is also a solution of
Eqs.~(6.47). The proof of such a statement stumbles over a factor ordering
problem. Assume that $\Psi[T,g]$ solves Eq.~(6.47a) and hence also the
iterated equation
\be
{\hat P}^2(x)\, \Psi[T,g] = G(x;\,\hat g,\hat p]\, \Psi[T,g]
\ .\eqnum{7.28}
\ee
{\it If} the operator $\hat{G}(x)$ were factor ordered as
\be
G(x;\,\hat g,\hat p] = \Bigl({\hat H}^G_\perp(x)\Bigr)^2 -
{\hat g}^{ab}(x)
{\hat H}_a^G(x)  {\hat H}_b^G(x)  \ ,\eqnum{7.29}
\ee
then its application on a state $\Psi[T,g]$ which also satisfies
Eq.~(6.47b) would yield
\be
\Bigl(1 + {\hat g}^{ab}(x) T_{,a}(x) T_{,b}(x) \Bigr) {\hat P}^2(x)\;
\Psi = \Bigl({\hat H}^G_\perp(x)\Bigr)^2 \; \Psi \ .\eqnum{7.30}
\ee
The state $\Psi$ would then satisfy Eqs.~(7.25) and (7.26) with the
operator $\hat{h}_{0}(x)$ factor ordered as
\be
h_0(x;\;T,\hat g,\hat p] = \Bigl(1 + {\hat g}^{ab}(x) T_{,a}(x) T_{,b}(x)
\Bigr)^{-1} {\hat H}^G_\perp(x) \ .\eqnum{7.31}
\ee
However, the factor ordering of the operator $G(x;\;\hat{g},\hat{p}]$ is
dictated by the requirement that the operators $\hat{G}(x)$ and
$\hat{G}(x')$ commute, thereby replicating the vanishing of the
corresponding classical Poisson brackets, Eq.~(4.22). One can hardly
expect that any ordering of the factors $\hat{g}$ and $\hat{p}$ in the
individual operators $H^{G}_\perp(x;\,\hat{g},\hat{p}]$ and
$H_{a}^{G}(x;\,\hat{g},\hat{p}]$, followed by the ordering (7.29) of
$\hat{G}(x)$, can achieve this aim. Similarly, one cannot expect that
the ordering (7.31) would yield  Hamiltonian densities
$\hat{h}_{0}(x)$ that are self--adjoint.

Formal considerations like these indicate that, {\it even if\/} we
were able to construct consistent quantum field theories based
respectively on the Schr\"{o}dinger equations (6.47a) and (7.25), these
theories would likely be non--equivalent.

\section{Acknowledgments}

The work on this paper was partially supported by the NSF grant PHY--9207225
to the University of Utah.

\appendix
\section{Notation}

{\bf Manifolds.} Manifolds play an important role in our study:
spacetime $\M$, space $\Sigma$, dust space $\S$, and dust time ${\cal
T}\,$.  Points $y\in\M$ are labeled by local coordinates $y^{\alpha}$
which carry Greek indices. Points $x\in\Sigma$ are labeled by local
coordinates $x^{a}$ with indices from the beginning of the Latin
alphabet. Points $z\in\S$ are labeled by local coordinates $z^{k}$
with indices from the middle of the Latin alphabet. Instants
$\tau\in{\cal T}$ are always labeled by the Cartesian coordinate
which (a minor ambiguity) is also called $\tau$.

{\bf Mappings.} Geometric properties of spacetime and physical
properties of the dust are studied in the canonical formalism on
(spacelike) embeddings $Y:\Sigma \rightarrow \M$ by $x\in\Sigma
\mapsto y=Y(x) \in \M\,$. The spacetime metric $\gamma$ on $\M$
is pulled back by the embedding mapping $Y$ into the spatial metric
$g$ on $\Sigma \,$. The mapping $\Upsilon : {\cal T} \times \S
\rightarrow \M$ by $ \Big( \tau \in {\cal T},\,z\in\S \Big) \mapsto
y=\Upsilon (\tau, z) \in \M$ foliates $\M$ into the dust time $\cal T$
and the dust space $\S$. The inverse mappings are $T:\M \rightarrow
{\cal T}$ by $y \in \M \mapsto \tau = T(y) \in {\cal T} $ and $Z: \M
\rightarrow \S$ by $y\in\M \mapsto z=Z(y) \in \S\,$. A point $x\in\Sigma$
of the embedded hypersurface $Y(\Sigma)$ lies at the instant $\tau = T
\circ Y(x)$ of the dust time and at the dust space point $z=Z \circ
Y(x)\,$. The proliferation of symbols is avoided by a slight misuse
of notation: the mappings $T \circ Y : \Sigma \rightarrow {\cal T}$
and $Z \circ Y : \Sigma \rightarrow \S$ are simply called $T$ and $Z$.
We thus write $T: \Sigma \rightarrow {\cal T} $ by $ x \in \Sigma
\mapsto \tau = T(x) \in {\cal T}$ and $Z: \Sigma \rightarrow \S$ by
 $ x \in \Sigma \mapsto z= Z(x) \in \S$. It is fairly clear from
the context what mappings the symbols $T$ and $Z$ denote.

The mapping $Z:\Sigma \rightarrow \S$ has the inverse $X=Z^{-1} :
\S \rightarrow \Sigma$ by $z\in\S \mapsto x=X(z) \in \Sigma \,$.
Tensor fields on $\Sigma$ are mapped to $\S$ by $Z^{\ast}$, and tensor
fields on $\S$ are mapped to $\Sigma$ by $X^{\ast}$. The corresponding
tensor fields are given lightface kernel symbols on $\Sigma$ and
boldface kernel symbols on $\S$.  For example, the spatial metric
$g_{ab}(x)$ on $\Sigma$ is mapped to $\bg_{k\ell}(\zeta)$ on $\S$ and
the dust time $T(x)$ on $\Sigma$  is mapped to $\bT(\zeta)$ on $\S\,$.

Notice that in all these examples, the {\em mappings} are denoted by
capital letters, and their {\em values} by the corresponding lower
case letters.

{\bf Derivatives.} Partial derivatives are denoted by comma followed
by the appropriate coordinate index: ${}_{,\alpha}\,,\;{}_{ ,a}$ and
${}_{,k}\,$. The covariant derivative on $\big( \M, \gamma \big)$ is
written as $\nabla_{\alpha}\,$, the covariant derivative on $\big(
\Sigma, g
\big)$ as $D_{a}\,$. The pound symbol $\pounds$ is used for
the Lie derivative.

{\bf Brackets.} The arguments of functions are enclosed in round
brackets $(\ \ )$, the arguments of functionals in square brackets $[\
\ ]\,$. Mixed brackets $(\ \  ]$ indicate that an object is a function of
one argument and a functional of another: e.g., $R(x;\,g]$ means that
the scalar curvature $R$ on $\Sigma$ is a function of $x\in\Sigma$ and
a functional of the induced metric $g_{ab}(x')\,$.

     The Lie brackets of vector fields and the commutators of
operators are denoted by square brackets $[\  ,\  ]\,$. Braces $\{\  ,\  \}$
are reserved for the Poisson brackets.

{\bf Densities and operators.} When the same quantity appears both as
a tensor and as a tensor density, the density is marked by a tilde.
Thus, $J^{\alpha}$ is a vector while $\Jtilde$ is a vector density of
weight 1. All operators wear hats, as $\hat{g}(x)$ or $\hat{p}(x)$.

{\bf Units.} We are using natural units in which the speed of
light $c$ and the rationalized Newton's constant $16 \pi G$ are set
equal to $1\,$: $c=1=16 \pi G\,$. In these units, the gravitational
Lagrangian is simply the densitized scalar curvature $|\gamma|^{1/2}R
(y;\,\gamma]$, and the Einstein law of gravitation takes the form
$R^{\alpha \beta} - \frac{1}{2} R \gamma ^{\alpha \beta} = \frac{1}{2}
T^{\alpha \beta}$.

\section{Alternative Forms of the Spacetime Action}

The action (2.4) is a functional of eight state variables, $T,Z^{k}$ and
$M,W_{k}$.  It can be cast into an equivalent form that depends on
fewer state variables by solving the Euler equation (2.7), obtained by
varying $W_{k}$, with respect to $W_{k}$. This yields
\begin{equation} W_k = h_{k\ell} {Z^\ell}_{,\alpha} \gamma^{\ab} T_{,\beta}
     \ , \eqnum{B.1} \end{equation}
where $h_{k\ell}$ is the inverse of the matrix
\begin{equation} h^{k\ell} := {Z^k}_{,\alpha} \gamma^\ab {Z^\ell}_{,\beta}
     \ . \eqnum{B.2} \end{equation}
The metric $h_{k\ell}$ measures the distance $ds$ orthogonal to the flow lines
between neighboring flow lines with Lagrangian coordinates $\zeta^{k}$
and $\zeta^{k}+d\zeta^{k}$:
\begin{equation} ds^2 = h_{k\ell} d\zeta^k d\zeta^\ell \ .
    \eqnum{B.3}\end{equation}
By using Eq.~(B.1) to eliminate $W_{k}$ from the action, we obtain the
equivalent action
\begin{equation}  S^D[T,Z^k;M;\gamma_\ab] :=
     S^D \left[ T,Z^k;M,W_k=W_k[T,Z^\ell ,\gamma_\ab ]; \gamma_\ab \right]
     \eqnum{B.4} \end{equation}
which depends on the four dynamical variables $T$, $Z^{k}$, and a single
multiplier $M$.

The mass multiplier cannot be eliminated at the spacetime level by an
algebraic
process, because the equation (2.6) obtained by varying $M$ does not
depend on $M$.  In Sec.~3a we showed how $M$ can be eliminated from the
canonical action (or, rather, how $M$ can be replaced by the canonical
momentum conjugate to the time function $T$).
Instead of decreasing the number of variables by solving the field equations,
one can increase the number of variables by adjoining valid equations to the
action by additional Lagrange multipliers.  As an example, the definition
(2.2) of $U_{\alpha}$ can be adjoined to the action (2.4) by a set of four
multipliers $\Jtilde$ that are densities of weight 1. One obtains thereby
the action
\begin{eqnarray} \lefteqn{S^D[T,Z^k;M,W_k;U_\alpha,\Jtilde;\gamma_\ab]}
                \nonumber \qquad \\
    & = & \int_\M d^4y \left( -\frac{1}{2} |\gamma|^{1/2} M (\gamma^{\ab}
           U_\alpha U_\beta + 1) + \Jtilde (U_\alpha + T_{,\alpha}
           - W_k {Z^k}_{,\alpha} ) \right)
    \eqnum{B.5} \end{eqnarray}
which depends on sixteen state variables $T$, $Z^{k}$, $M$, $W_{k}$ and
$U_{\alpha}$, $\Jtilde$.  By varying the action (B.5) with respect to
$\Jtilde$, we recover the definition (2.2) of $U_{\alpha}$.
By varying $U_{\alpha}$, we obtain
\begin{equation} \Jtilde - |\gamma|^{1/2} M U^\alpha = 0
     \ .\eqnum{B.6} \end{equation}
This shows that the multipliers $\Jtilde$ are the components of the
(densitized) mass current (2.15).  By calculating $U^{\alpha}$
from Eq.~(B.6),
\begin{equation} U^{\alpha} = M^{-1} |\gamma|^{-1/2} \Jtilde \ ,
     \eqnum{B.7} \end{equation}
and substituting it back into the action (B.5), we obtain an
equivalent action
\begin{eqnarray} \lefteqn{S^D[T,Z^k;M,W_k;\Jtilde;\gamma_\ab]}
       \nonumber \quad \\ & = & \int_\M d^4y \left( \frac{1}{2}
     |\gamma|^{-1/2} M^{-1} \gamma_{\ab} \Jtilde\Jtildeb - \frac{1}{2}
     |\gamma|^{1/2} M + \Jtilde ( T_{,\alpha} - W_k {Z^k}_{,\alpha} )
     \right) \quad \eqnum{B.8} \end{eqnarray}
 that depends on twelve state variables $T$, $Z^{k}$, $M$, $W_{k}$ and
 $\Jtilde$. The variation of the action (B.8) with respect to M
 yields an equation that determines M, namely,
\begin{equation}  M = |\gamma|^{-1/2} \sqrt{-\gamma_\ab\Jtilde\Jtildeb}
    \ . \eqnum{B.9} \end{equation}
By using Eq.~(B.9), M can be eliminated from the action (B.8).  The
result is the action
\begin{eqnarray} \lefteqn{S^D[T,Z^k;W_k;\Jtilde;\gamma_\ab]}
       \nonumber \qquad\qquad \\
     & = & \int_\M d^4y \left(  - \sqrt{-\gamma_\ab\Jtilde\Jtildeb}
          + \Jtilde ( T_{,\alpha} - W_k {Z^k}_{,\alpha} ) \right)
     \eqnum{B.10} \end{eqnarray}
that depends on eleven state variables $T,Z^{k}, W_{k}$ and
$\Jtilde$.  This action for dust was introduced in Ref~\cite{Brown}
in the study of perfect fluid actions.

All of the actions (2.4), (B.4), (B.5), (B.8), and (B.10) lead to the same
canonical action.  We have chosen the action (2.4) because the number of
state variables exactly matches the number of canonical coordinates plus
conjugate momenta. The physical meaning of the momenta then clearly emerges
from the Legendre transformation.

\section{Symmetry Transformations}

Transformations of the dust variables $(T, Z^k, W_k) \to (T', {Z}^{k'},
W_{k'})$ that satisfy
\begin{equation} S^D\bigl[ T'(T,Z^\ell,W_\ell),\; {Z}^{k'}(T,Z^\ell,W_\ell), \;
    W_{k'}(T,Z^\ell,W_\ell) \bigr] =  S^D[T, Z^k, W_k]
   \eqnum{C.1} \end{equation}
constitute symmetries of the dust action. Inspection of the action
(2.4)--(2.5) shows that transformations that
preserve the Pfaff form (2.3),
\begin{equation} -dT + W_k dZ^k = -dT' + W_{k'} d{Z}^{k'}
     \ ,\eqnum{C.2}\end{equation}
are symmetries.
A Pfaff form $-dT + W_k dZ^k$, with $k=1$, $2$, $3$, always
can be reduced to $dA + B\,dC$ for some functions $A$, $B$, and
$C$ \cite{Caratheodory}; however, here we are interested in invertible
transformations that leave the action functional invariant, as
expressed though Eq.~(C.1). (See Refs.~\cite{Brown,Schutz} for
discussions of the use of the reduced Pfaff form in the construction
of perfect fluid action functionals.) We shall confine our attention to
transformations that are ultralocal in spacetime, so the primed variables
at a given spacetime point $y$ depend only on the unprimed variables at
the point $y$. In particular, the primed variables are not allowed to
depend on derivatives of the unprimed variables.

For the case of ultralocal transformations, the relationship (C.2)
can be written as
\begin{equation}  W_{k'}(T,Z^\ell,W_\ell) \, d{Z}^{k'}(T,Z^\ell,W_\ell)
     =   W_k dZ^k
     +  d{\cal F}(T,Z^\ell,W_\ell)   \eqnum{C.3}
     \end{equation}
and viewed as a time $T$--dependent canonical transformation for a
fictitious phase space with coordinates $Z^k$ and momenta $W_k$
\cite{Brown,Schutz}. Here, ${\cal F}$ is the function of $T$ and of the
`canonical variables' $Z^k$, $W_k\,$, defined by ${\cal F} := T' - T$.
Now, for those transformations that are connected to the identity, the
dependence $Z^{k'} = Z^{k'}(T,Z^\ell,W_\ell)$ of the primed
`coordinates' on the unprimed variables can be inverted for the
unprimed `coordinates', $Z^{k} = Z^{k}(T,Z^{\ell'},W_\ell)$. Then $T$,
$Z^{k'}$, and $W_{k}$ can be treated as independent variables by
substituting the functional relationship $Z^{k} =
Z^{k}(T,Z^{\ell'},W_{\ell})$ into Eq.~(C.3). This leads to the
expression
\begin{equation} W_{k'}(T,Z^{\ell'},W_\ell)\, d{Z}^{k'} =
       W_k\, dZ^k(T,Z^{\ell'},W_\ell) + d{\cal F}(T,Z^{\ell'},W_\ell)
     \ ,\eqnum{C.4} \end{equation}
with a slight abuse of notation in which
$W_{k'}(T,\;Z^\ell(T,Z^{m'},W_m),\;W_\ell)$ is set equal to
$W_{k'}(T,Z^{\ell'},W_\ell)$, and likewise for ${\cal F}$. In terms
of the generating function $F(T,Z^{\ell'},W_\ell) :=
{\cal F}(T,Z^{\ell'},W_\ell) + W_{k} {Z}^{k}(T,Z^{\ell'},W_\ell)$,
Eq.~(C.4) can be rewritten as
\begin{equation} W_{k'}(T,Z^{\ell'},W_\ell) \,dZ^{k'} +
       {Z}^{k}(T,Z^{\ell'},W_\ell) \,dW_{k} =
        dF(T,Z^{\ell'},W_\ell)
      \ .\eqnum{C.5}\end{equation}
This expression along with the definitions of ${\cal F}$ and $F$
yield the results
\begin{eqnarray} T' - T & = & F - W_{k}
          \frac{\partial F}{\partial W_{k}}  \ ,\eqnum{C.6a}\\
    {Z}^{k} & = & \frac{\partial F}{\partial W_{k}}
        \ ,\eqnum{C.6b}\\
     W_{k'} & = & \frac{\partial F}{\partial Z^{k'}} \ ,\eqnum{C.6c}
    \end{eqnarray}
where $F = F(Z^{\ell'},W_{\ell})$ is independent of $T$.

The Noether currents (2.29) associated with one--parameter subgroups
of symmetries $F(Z^{\ell'},W_{\ell},\sigma)$ connected to the identity
$F(Z^{\ell'},W_{\ell},0) = Z^{k'} W_{k} $ are given by
\begin{equation} \Jtilde[f] = |\gamma|^{1/2} J^\alpha f(Z^\ell,W_\ell)
      \ ,\eqnum{C.7}\end{equation}
where $J^\alpha$ is the mass current (2.15) and
\begin{equation} f(Z^{\ell},W_{\ell}) := \left.
     \frac{\partial F(Z^{\ell'},W_{\ell},\sigma)}{\partial\sigma}
     \right|_{\sigma=0}  \ .\eqnum{C.8} \end{equation}
The corresponding Noether charges (2.28) can be written as
\begin{eqnarray} Q[f] & = & - \int_\Sigma d^3x\, |g|^{1/2} \,n_\alpha
    J^\alpha \,f(Z^\ell,W_\ell)  \eqnum{C.9a} \\
    & = & \int_\Sigma d^3x\, P\,f(Z^\ell,-P_\ell/P)  \ .\eqnum{C.9b}
    \end{eqnarray}
In expression (C.9b), we have made use of the canonical
variables $P$ and $P_k$ as defined by Eqs. (3.10) and (3.12).

In Sec.~2e of the main text we restrict our attention to those
symmetry transformations that mix the dust spacetime variables $T$
and $Z^k$ with one another. That is, we consider transformations
for which $T'$ and ${Z}^{k'}$ are functions of $T$ and $Z^k$ only.
If this restriction is imposed on the transformations (C.6), it
follows that the most general generating function $F$ has the form
\begin{equation} F(Z^{\ell'},W_{\ell}) = W_{k}\, \bXi^{k}(Z^{\ell'}) +
     \bTh(\bXi^\ell(Z^{m'})) \ .\eqnum{C.10}\end{equation}
The corresponding symmetry transformations (C.6) are given by
\begin{eqnarray} T'  & = & T + \bTh(Z^{m})  \ ,\eqnum{C.11a}\\
           {Z}^{k} & = &  \bXi^{k}(Z^{m'}) \ ,\eqnum{C.11b}\\
           W_{k'}  & =  & {\bXi^{\ell}}_{,k'}(Z^{m'}) \Bigl( W_\ell
           + \bTh_{,\ell}(Z^{m}) \Bigr) \ ,\eqnum{C.11c}
    \end{eqnarray}
which coincide with the transformations (2.26)--(2.27) discussed
in Sec.~2e.

The Lie algebra of one--parameter groups of symmetries generated
by $F$ of the form (C.10) is labeled by the functions
\begin{equation} f(Z^{\ell}, W_{\ell}) :=  \left.
     \frac{\partial F(Z^{\ell'},W_{\ell},\sigma)}{\partial\sigma}
     \right|_{\sigma=0} = W_{k}\, \bxi^{k}(Z^{\ell}) + \bth(Z^{\ell})
     \ ,\eqnum{C.12} \end{equation}
where $\bxi^{k}(Z^{\ell})$ and $\bth(Z^{\ell})$ are defined in
Eq.~(2.32). Insertion of this expression for $f$ into the Noether charge
(C.9) yields
\begin{eqnarray} Q[\bxi, \bth] & = & -\int_\Sigma d^3x\;|g|^{1/2}
     n_\alpha \Bigl(J^\alpha_k\, \bxi^k (Z^{\ell}) +
     J^\alpha \bth(Z^{\ell}) \Bigr) \eqnum{C.13a}\\
     & = & \int_\Sigma d^3x\; \Bigl(- P_k\, \bxi^k(Z^{\ell}) +
     P\,\bth(Z^{\ell}) \Bigr) \ ,\eqnum{C.13b} \end{eqnarray}
where $J^\alpha$ and $J^\alpha_k$ are the mass current (2.15) and
the momentum current (2.16), respectively. The charges
$Q[\bxi, \bth]$ coincide with the Noether charges (2.36)--(2.37)
and (3.24)--(3.25) presented in the main text.

\section{The Action for Relativistic Particles}

The geodesic motion of a single relativistic particle of rest mass $m$ is
generated by the super--Hamiltonian
\begin{equation} H := \frac{1}{2m} \left( \gamma^\ab(y) P_\alpha P_\beta +
    m^2 \right) \ .\eqnum{D.1} \end{equation}
The spacetime coordinates $y^\alpha$ of the particle and the four--momentum
$P_{\alpha}$ are canonically conjugate variables. These variables can be
split into their space and time components with respect to the foliation
(3.1)--(3.2). The momenta conjugate to $t$ and $x^{a}$ are
\begin{equation} p_t := \dot X^\alpha P_\alpha \quad{\rm and} \quad
    p_a := {X^\alpha}_{,a} P_\alpha \ . \eqnum{D.2} \end{equation}
By using Eqs.~(3.3) and (3.4), the super--Hamiltonian constraint $H =
0$ can be resolved with respect to the energy variable $- p_{t}$:
\begin{eqnarray} -p_t & = & h(t,x^a,p_a)  \nonumber  \\
       & = & -N^a(t,x) p_a + N^\perp(t,x) \sqrt{m^2 + g^{ab}(t,x) p_ap_b}
          \ .  \eqnum{D.3} \end{eqnarray}
The expression $h(t,x^{a},p_{a})$ is the Hamiltonian of the particle.

The action associated with the Hamiltonian (D.3) is
\begin{equation} S[x^a,p_a] = \int dt\;L(x^{a},p_{a}, \dot{x}^{a})=
 \int dt \left( p_a \frac{dx^a}{dt} - h(t,x^a,p_a) \right)
 \ .\eqnum{D.4}\end{equation}
Observe that the derivative of the canonical Lagrangian $L$ with
respect to mass $m$ is
\begin{equation} \frac{\partial L}{\partial m} = -\frac{N^\perp}{ \sqrt{ 1 +
      g^{ab} p_ap_b/m^2}}  \ .\eqnum{D.5}\end{equation}
Since $N^\perp dt$ is the orthogonal proper time separation between
$t={\it const\/}$ surfaces, and $\sqrt{ 1 + g^{ab} p_ap_b/m^2}$ is the
relativistic `gamma factor', $\partial L/\partial m$ equals
the ratio $d\tau/dt$, where $\tau$ is the proper time along the particle
worldline. This shows that the action (D.4) can be extended to a functional
of $m(t)$ and $\tau(t)$,
\begin{eqnarray} \lefteqn{ S[\tau,m;\;x^a,p_a] = \int dt \biggl(
    m\frac{d\tau}{dt} + p_a \frac{dx^a}{dt} -N^a(t,x) (-p_a)  }
    \nonumber \qquad\qquad\qquad\qquad\qquad \\
    & &   - N^\perp(t,x) \sqrt{m^2 + g^{ab}(t,x) p_ap_b}\; \biggr)
    \ .\eqnum{D.6}\end{eqnarray}
Extremization of $S$ with respect to $m$ relates $\partial L /
\partial m$ to $d \tau / d t \,$. Extremization of $S$
with respect to $\tau$ yields the condition $dm/dt = 0$, that the
particle mass is a constant of motion.

We now compare the relativistic particle action (D.6) with the canonical
dust action (3.16). For the purpose of comparison, the metric tensor
components $N^\perp(t,x)$, $N^a(t,x)$, $g_{ab}(t,x)$ are treated as
prescribed external fields. From the analysis of Sec.~5 it follows that
the dust action (3.16) can be written as a functional of canonical
variables $\bT(\zeta)$, $\bP(\zeta)$, $Z^k(x)$, and $H_{\uparrow k}(x)
:= Z^a_k(x) H^D_a(x)$. We can make a further canonical transformation
[15] by replacing the dust frame $Z^k(x)$ with its inverse, $X^a(\zeta)$.
The new variable $X^a(\zeta)$ specifies the spatial location of the
dust particle with Lagrangian coordinate label $\zeta^k$.
The momentum conjugate to $X^a(\zeta)$ is found from the relationship
\be
\int_\Sigma d^3x \, H_{\uparrow k}(x) {\dot Z}^k(x)  =
- \int_\S d^3\zeta \, \left| \frac{\partial X(\zeta)}{\partial \zeta}
\right| H_{\uparrow k}(X(\zeta)) {Z^k}_{,a}\Big(X(\zeta)\Big)
{\dot X}^a(\zeta)  \ .\eqnum{D.7}
\ee
With the momentum conjugate to $X^a(\zeta)$ denoted by $P_a(\zeta)$,
the action (3.16) becomes
\bea
\lefteqn{ S^D[X^a,P_a;\;\bT,\bP] = \int_{\rr} dt \int_S d^3\zeta \biggl(
\bP(\zeta) {\dot\bT}(\zeta) + P_a(\zeta) {\dot X}^a(\zeta)
- N^a\Big(t,X(\zeta)\Big)\, \Bigl( -P_a(\zeta)\Bigr) }
\qquad\qquad\qquad\qquad\qquad\qquad\qquad\nonumber\\
& & - N^\perp\Big(t,X(\zeta)\Big)
\sqrt{ \bP^2(\zeta) + g^{ab}(t,X(\zeta)) P_a(\zeta) P_b(\zeta) }\; \biggr)
\ .\eqnum{D.8}
\eea
This action coincides  with the relativistic particle action in the following
sense. The action for many relativistic particles is obtained from the single
particle action (D.6) by summing over the individual particles. With the
particles labeled by $\zeta = 1,2,\ldots$, the canonical variables are given
by the set ($\tau_\zeta$, $m_\zeta$, $x^a_\zeta$, $p_{\zeta a}$). In the
continuum limit of an infinite number of particles scattered throughout
space, the sum $\sum_\zeta$ over individual particles is replaced by an
integral $\int d^3\zeta$ over the space of particle labels, the dust space
$\S$. At the same time, the canonical momenta $m_\zeta$ and $p_{\zeta a}$
become densities in the space of labels. The resulting action functional
is precisely the dust action (D.8), with the correspondences
$\tau_\zeta\to \bT(\zeta)$, $m_\zeta\to \bP(\zeta)\, d^3\zeta$,
$x^a_\zeta\to X^a(\zeta)$, and $p_{\zeta a}\to P_a(\zeta)\, d^3\zeta$.

\section{The Poisson Algebra of $G(\lowercase{x})$}

We check by direct calculation that the quadratic combinations $G(x)$,
Eq.~(4.7b), of the gravitational super--Hamiltonian $H^{G}_\perp(x)$ and
supermomentum $H_{a}^{G}(x)$ have strongly vanishing Poisson brackets,
Eq. (4.22).  The result follows from the fact that the gravitational
generators $H^{G}_\perp(x)$ and $H_{a}^{G}(x)$ satisfy the Dirac `algebra'
Eqs.~(4.1)--(4.3), and from the action of these generators on the
metric $g^{ab}(x)$.

Let us first show that the expressions
\begin{equation} F(x) := g^{ab}(x) H^G_a(x) H^G_b(x)
        \eqnum{E.1} \end{equation}
commute among themselves,
\begin{equation} \{ F(x), F(x') \} = 0 \ .
      \eqnum{E.2} \end{equation}
{}From the definition (E.1), it holds that
\begin{eqnarray}  \lefteqn{ \{ F(x), F(x') \} =
\left( 2\{ F(x), H^G_a(x') \}
      H^{Ga}(x') - (x\leftrightarrow x') \right) }
       \qquad\qquad\qquad\qquad\nonumber\\
       & & - 4 \{ H^G_a(x), H^G_b(x') \} H^{Ga}(x) H^{Gb}(x') \ .
       \eqnum{E.3} \end{eqnarray}
Now smear Eq.~(E.3) by two externally prescribed scalar densities of
weight $-1$, $M(x)$ and $N(x')$. With the abbreviations
\begin{equation} M^a(x) := M(x) H^{Ga}(x) \ ,\qquad N^a(x) := N(x) H^{Ga}(x)
      \ , \eqnum{E.4} \end{equation}
Eq.~(E.3) becomes
\begin{equation} \{ F[M], F[N] \} = \left( 2\{ F[M], H^G[\vec N] \} -
        (N\leftrightarrow M) \right) - 4 \{ H^G[\vec M], H^G[\vec N] \}
        \ .  \eqnum{E.5} \end{equation}
Note that in Eq.~(E.5), $\vec {M}$ and $\vec {N}$ must be treated as
externally prescribed smearing vectors. This is because the supermomenta
$H^{Ga}(x)$, which combined with $M(x)$ and $N(x)$ into the smearing
vectors (E.4), stand {\it outside} the Poisson brackets in Eq.~(E.3).

Since $F(x)$ is a scalar density of weight $2$ under $\diffsigma$, its
Poisson bracket with $H^G[\vec N]$ is given by
\begin{eqnarray}  \{ F[M], H^G[\vec N] \} & = & \int_\Sigma d^3x\, M(x)
    \pounds_{\vec N} F(x) \nonumber  \\
    & = & \int_\Sigma d^3x\, M \left( F_{,a} N^a + 2 F {N^a}_{,a} \right)
     \ . \eqnum{E.6} \end{eqnarray}
Therefore the first term on the right--hand side of Eq.~(E.5) yields
\begin{equation}  2 \{ F[M], H^G[\vec N] \} - (N\leftrightarrow M) =
      4\int_\Sigma d^3x\, F\, (M N_{,a} - N M_{,a} )\, H^{Ga} \ .
      \eqnum{E.7} \end{equation}
The second term on the right--hand side of Eq.~(E.5) can be evaluated from
the smeared form of Eq.~(4.3), which asserts that $H^{G}[\vec M]$ represent
the Lie algebra $\diffsigma$:
\begin{equation} \{ H^G[\vec M], H^G[\vec N] \} = H^G\left[ [\vec M, \vec N]
     \right] \ .  \eqnum{E.8} \end{equation}
Here,  $[\vec M, \vec N]$ is the commutator of the vector fields $\vec {M}$
and $\vec {N}$,
\begin{eqnarray}  [\vec M, \vec N]^b & = & {N^b}_{,a} M^a - {M^b}_{,a} N^a
      \nonumber  \\
      & = & (M N_{,a} - N M_{,a}) H^{Ga} H^{Gb} \ .
      \eqnum{E.9} \end{eqnarray}
Equations (E.8) and (E.9) imply that the second term on the right--hand side
of Eq.~(E.5) exactly compensates the first term given by the integral (E.7).
This proves Eq.~(E.2).

The Poisson bracket that interests us is
\begin{eqnarray}  \lefteqn{ \{ G(x), G(x') \} = \{ F(x), F(x') \} }
      \qquad\qquad\qquad \nonumber \\
      & & \quad - \left( 2 H^G_\perp(x) \{ H^G_\perp(x), F(x') \} -
        (x\leftrightarrow x') \right)  \nonumber\\
      & & \quad + 4 H^G_\perp(x) H^G_\perp(x') \{ H^G_\perp(x),
         H^G_\perp(x') \} \ .
      \eqnum{E.10} \end{eqnarray}
We have just proved that the first term on the right--hand side of Eq.~(E.10)
vanishes. Because $H^{G}_\perp(x)$ induces an ultralocal change of the metric,
\begin{equation} \{ g^{ab}(x), H^G_\perp(x') \} = 2 K^{ab}(x) \delta(x,x')
       \eqnum{E.11} \end{equation}
(the coefficient $K^{ab}$ is the extrinsic curvature), the action
of $H^{G}_\perp$ on $g^{ab}$ in $F$ cancels out under the interchange
$x\leftrightarrow x'$.  Therefore, by Eq.~(4.2), the second term on
the right--hand side of Eq.~(E.10) yields
\begin{eqnarray} \lefteqn{ -4 H^G_\perp(x) \{ H^G_\perp(x), H^G_a(x') \}
      H^{Ga}(x') -  (x\leftrightarrow x')  }\qquad\qquad \nonumber  \\
     & = & 4 H^G_\perp(x) H^G_\perp(x') H^{Ga}(x') \delta_{,a'}(x,x') -
          (x\leftrightarrow x') \ .
           \eqnum{E.12} \end{eqnarray}
Up to a sign, this is what the last line of Eq.~(E.10) yields when we
substitute into it the Poisson bracket (4.1).  As a result, the
right--hand side of Eq.~(E.10) vanishes.  We have thereby checked the
pivotal equation (4.22) of the main text.

\section{ D\lowercase{iff}$\S$ and
the Eigenfunctionals of  ${\hat{\bf G}}(\zeta)$ }

We want to prove that if $\bphitoo[\bg]$ is an eigenfunctional of
${\hat\boldG}(\zeta)$ with eigenvalue $\boldG(\zeta)$,
\be
{\hat\boldG}(\zeta)\, \bphitoo[\bg] = \boldG(\zeta)\, \bphitoo[\bg]
\ ,\eqnum{F.1}
\ee
then $\bphitoo[\bXi^*\bg]$ is an eigenfunctional of ${\hat\boldG}(\zeta)$
with eigenvalue $(\bXi^{-1})^*\boldG(\zeta)\,$:
\be
{\hat\boldG}(\zeta)\, \bphitoo[\bXi^*\bg] = (\bXi^{-1})^*\boldG(\zeta)\;
\bphitoo[\bXi^*\bg] \ .\eqnum{F.2}
\ee
It is sufficient to prove the infinitesimal version of Eq.~(F.2), namely
\be
\frac{d}{d\sigma} \biggl( {\hat\boldG}(\zeta) \bphitoo[\bXi_\sigma^*\bg]
\biggr)\biggr|_{\sigma=0}  = \frac{d}{d\sigma} \biggl(
(\bXi_\sigma^{-1})^*\boldG(\zeta)
\bphitoo[\bXi_\sigma^*\bg] \biggr)\biggr|_{\sigma=0} \ ,\eqnum{F.3}
\ee
and then to pass to an arbitrary finite diffeomorphism $\bXi$ by
exponentiation in $\sigma$.

To begin, recall that the classical charge $Q[\vec\bxi]$ discussed in
Sec.~5b is the canonical generator of $\diffs$. Thus, the action of
$\diffs$ on the operator ${\hat\boldG}(\zeta)$ is given by the commutator
\be
\frac{1}{i} \bigl[\, {\hat\boldG}(\zeta) , {\hat Q}[\vec\bxi]\, \bigr] =
\pounds_{\vec\bxi} \,{\hat\boldG}(\zeta) \ .\eqnum{F.4}
\ee
Also observe that the change in $\bphitoo[\bg]$ under an infinitesimal
diffeomorphism of its argument $\bg$ is determined by the operator
${\hat Q}[\vec\bxi]$ acting on $\bphitoo[\bg]$:
\bea
\frac{d}{d\sigma} \bphitoo[\bXi_\sigma^*\bg] \biggr|_{\sigma=0} & = &
\int_\S d^3\zeta
\Bigl( \pounds_{\vec\bxi}\, \bg_{k\ell}(\zeta) \Bigr) \frac{\delta
\bphitoo[\bg]} {\delta \bg_{k\ell}(\zeta)}  \nonumber\\
& = & i \int_\S d^3\zeta \Bigl( \pounds_{\vec\bxi}\,
{\hat\bg}_{k\ell}(\zeta)
\Bigr) {\hat\bp}^{k\ell}(\zeta) \bphitoo[\bg]  \nonumber\\
& = & i {\hat Q}[\vec\bxi] \bphitoo[\bg] \ .\eqnum{F.5}
\eea
By letting ${\hat\boldG}(\zeta)$ act on equation (F.5) and using the
commutator
(F.4), we obtain
\bea
\frac{d}{d\sigma} \biggl( {\hat\boldG}(\zeta) \bphitoo[\bXi_\sigma^*\bg]
\biggr)\biggr|_{\sigma=0} & = & i \bigl[\, {\hat\boldG}(\zeta) ,
{\hat Q}[\vec\bxi] \, \bigr] \bphitoo[\bg]  + i {\hat Q}[\vec\bxi]
{\hat\boldG}(\zeta) \bphitoo[\bg]  \nonumber\\
& = & - \Bigl(\pounds_{\vec\bxi}\, {\hat\boldG}(\zeta)\Bigr) \bphitoo[\bg]
+ i {\hat Q}[\vec\bxi]  {\hat\boldG}(\zeta) \bphitoo[\bg] \ .\eqnum{F.6}
\eea
Now use Eq.~(F.5) along with the fact that $\bphitoo[\bg]$ is an
eigenfunctional
of ${\hat\boldG}(\zeta)$. Equation (F.6) then becomes
\be
\frac{d}{d\sigma} \biggl( {\hat\boldG}(\zeta) \bphitoo[\bXi_\sigma^*\bg]
\biggr)\biggr|_{\sigma=0} = \frac{d}{d\sigma} \biggl(
(\bXi_\sigma^{-1})^*\boldG(\zeta)
\biggr)\biggr|_{\sigma=0} \bphitoo[\bg] + \boldG(\zeta) \frac{d}{d\sigma}
\biggl( \bphitoo[\bXi_\sigma^*\bg] \biggr)\biggr|_{\sigma=0} \ ,\eqnum{F.7}
\ee
which is equivalent to Eq.~(F.3). This proves the theorem expressed in
Eq.~(F.2).

\section{Relativistic Particle in the Proper Time Formalism}

As in Appendix D, let us regard the rest mass $m$ for the relativistic
particle
as a dynamical variable canonically conjugate to the proper time $\tau$ along
the particle's worldline:
\be
\{ \tau, m \} = 1 \ .\eqnum{G.1}
\ee
The proper time formalism for the relativistic particle is obtained by
resolving
the super-Hamiltonian constraint in the extended phase space $\Big(
\tau , y^{\alpha} ;\; m, P_{\alpha} \Big)\,$,
\be
H_{\rm ext} := m^2 + \gamma^{\ab}(y) P_\alpha P_\beta = 0 \ ,\eqnum{G.2}
\ee
with respect to $m$. This yields
\be
h_\uparrow := m + h = 0 \ ,\qquad h := -\sqrt{ -\gamma^\ab(y) P_\alpha
P_\beta}
\ ,\eqnum{G.3}
\ee
where the choice of the negative square root in the Hamiltonian $h$ ensures
the
positivity of $m$. The reality of $m$ requires that
\be
G := -\gamma^\ab(y) P_\alpha P_\beta > 0 \ ,\eqnum{G.4}
\ee
{\it i.e.\/}, that the particle moves along a timelike worldline.

The imposition of the constraint (G.3) as an operator restriction on the state
function $\Psi(\tau,y^{\alpha})$ yields the Schr\"{o}dinger equation in
proper time $\tau$:
\be
{\hat H}_\uparrow\, \Psi = 0 \quad\Longleftrightarrow \quad i\partial_\tau
\Psi(\tau,y) = {\hat h}\, \Psi(\tau,y) \ .\eqnum{G.5}
\ee
The Hamiltonian
\be
{\hat h} = -\sqrt{-{\hat G}} \ ,\qquad {\hat G} =
{\framebox[11pt]{\rule[1pt]{0pt}{4pt}}}_\gamma \eqnum{G.6}
\ee
must be defined by spectral analysis. It is self--adjoint on the
Hilbert space ${\cal H}^{+}$ spanned by the eigenfunctions $\phi_{G}$ of
$\hat{G}$ belonging to positive eigenvalues $G\,$:
\be
{\hat G}\, \phi_G(y) = G \, \phi_G(y) \ ,\qquad G>0 \ .\eqnum{G.7}
\ee
On ${\cal H}^{+}$, the Schr\"{o}dinger inner product
\be
\langle \Psi | \Psi \rangle := \int d^4y \, {\bar\Psi}(\tau,y) \Psi(\tau,y)
\eqnum{G.8}
\ee
is conserved in $\tau$. Observables must leave the states in ${\cal H}^{+}$,
{i.e.\/}, they must satisfy the condition (6.46), where
\be
{\hat\P}^+ = \Theta({\hat G}) \eqnum{G.9}
\ee
is the projector to ${\cal H}^{+}$.

In general, neither ${\hat y}^{\alpha} =
y^{\alpha}\times$ nor ${\hat P}_{\alpha} = - i\partial /\partial y^{\alpha}$
are observables. This prevents us from interpreting the multiplication
operator ${\hat y}^{\alpha}$ as an observable whose measurement at the proper
time $\tau$ would localize the particle at a spacetime event $y^{\alpha}$,
and $\bar{\Psi}(\tau,y) \Psi(\tau,y)$ as the probability density that the
localization will occur at $y^{\alpha}$.

    This difficulty persists even for a particle moving in flat
spacetime $\gamma^{\alpha \beta}(y) = \eta^{\alpha \beta} \,$.  In
this case, $\hat{P}_{\alpha}$ commutes with $\hat{G}$ and hence it is
an observable. On the other hand, one can explicitly show that
$\hat{y} ^{\alpha} = y^{\alpha} \times$ is not an observable since
one can always choose a constant covector $\pi _{\alpha}$ such that the
action of the unitary operator
\be
\hat{U}_{\pi} := \exp\big( i \pi_{\alpha} \hat{y}^{\alpha}\big) \eqnum{G.10}
\ee
leads the state $\Psi$ out of ${\cal H}^{+}$.

The demonstration follows the general outline of an old argument by
Pauli \cite{Pauli} that if time were represented by the
multiplication operator, the energy spectrum could not be bounded from
below. In our example, the Hilbert space $\cal H$ is spanned by the
joint eigenfunctions
\be
\Phi _{P} (y) = (2 \pi)^{-2}\, \exp\big( i P_{\alpha} y^{\alpha}\big)
\eqnum{G.11}
\ee
of the momentum operators $\hat{P}_{\alpha}\,$. Equation
\be
\Psi(y)= \int d^{4}P\; \tilde{ \Psi}(P)\, \Phi _{P} (y)
\eqnum{G.12}
\ee
relates the $y$--representation $
\Psi(y^{\alpha})$ of a state $\Psi\in{\cal H}$ to its $P$--representation
$ \tilde {\Psi}(P_{\alpha})\, $. The momentum eigenstates (G.11) are
also eigenstates of $\hat{G}$ with eigenvalues $ G = -
\eta ^{\alpha \beta} P_{\alpha} P_{\beta} \,$. The state $\Psi$
lies in ${\cal H}^{+}$ if the state function $\tilde{\Psi} (P_{\alpha})$
has its support only on timelike vectors $P_{\alpha}\,$:
\be
\tilde{\Psi} (P_{\alpha}) =0 \ \  {\rm for} \ \ \eta^{\alpha \beta}
P_{\alpha} P_{\beta} \geq 0 \,.  \eqnum{G.13}
\ee
The action of $\hat{U}_{\pi}$ on the state function $
\tilde{\Psi} (P_{\alpha})$ translates its argument by $- \pi_{\alpha}\,$:
\be
\big(\hat{U}_{\pi}\tilde{\Psi}\big)\, (P_{\alpha}) =
\tilde{\Psi} (P_{\alpha} - \pi_{\alpha}) \,. \eqnum{G.14}
\ee
The support of $\big(\hat{U}_{\pi}\tilde{\Psi}\big)\,(P_{\alpha})$
is thereby shifted by the amount $\pi_{\alpha}$ relative to the
support of $\tilde{\Psi} (P_{\alpha})$.
Because $\pi _{\alpha}$ is arbitrary, the support of
$\big(\hat{U}_{\pi}\tilde{\Psi}\big)\, (P_{\alpha})$ cannot always
remain timelike. This means that $\hat{y}^{\alpha}$ is not an
observable.

This simple example illustrates why the fundamental gravitational
variables cannot be expected to be observables.


\vfill\eject
\vspace*{1in}
\begin{center}
\leavevmode
\epsffile{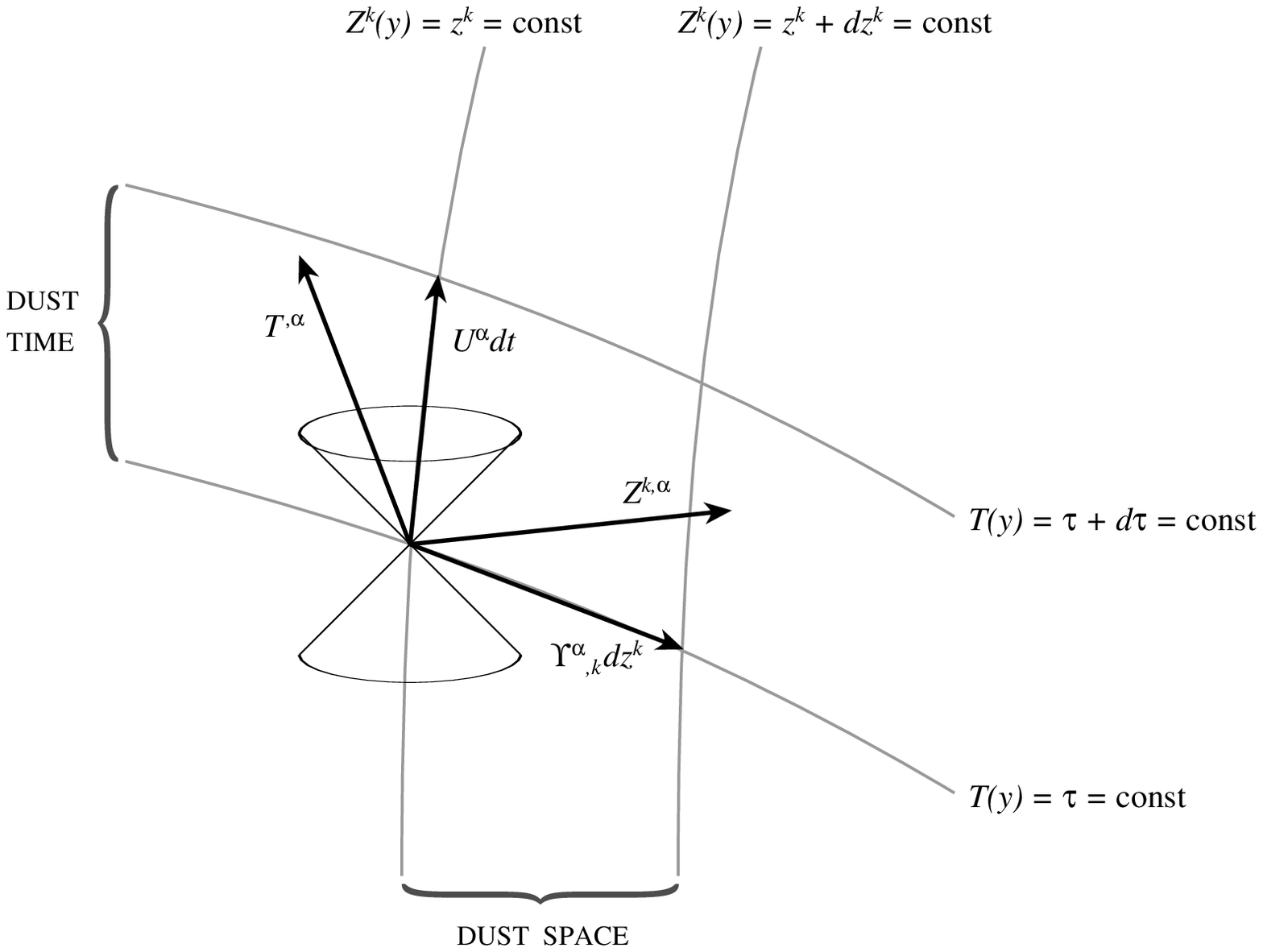}
\end{center}
\noindent {\bf Figure 1:} $U^\alpha \partial/\partial y^\alpha =
\partial/\partial\tau$ and ${\Upsilon^\alpha}_{,k} \partial/\partial
y^\alpha = \partial/\partial\zeta^k$ are the coordinate basis vectors
for the dust coordinates ($\tau$, $\zeta^k$). The covectors $T_{,\alpha}
dy^\alpha = d\tau$ and ${Z^k}_{,\alpha} dy^\alpha = d\zeta^k$ correspond
to the stacks of surfaces $\tau = {\rm const}$ and $\zeta^k = {\rm const}$,
respectively; the  associated vectors are
$T^{,\alpha} := g^{\alpha\beta} T_{,\beta}$ and $Z^{k,\alpha} :=
g^{\alpha\beta} {Z^k}_{,\beta}$. The vectors $U^{\alpha}$ and
$Z^{k,\alpha}$ are orthogonal, $U^{\alpha} Z^{k}{}_{,\alpha} = 0$, and
are therefore drawn symmetrically about the light cone. Likewise, the
vectors $T^{,\alpha}$ and ${\Upsilon^\alpha}_{,k}$ are orthogonal,
$T_{,\alpha} {\Upsilon^\alpha}_{,k} = 0$, and are drawn symmetrically
about the light cone.
\vfill\eject
\vspace*{1in}
\begin{center}
\leavevmode
\epsffile{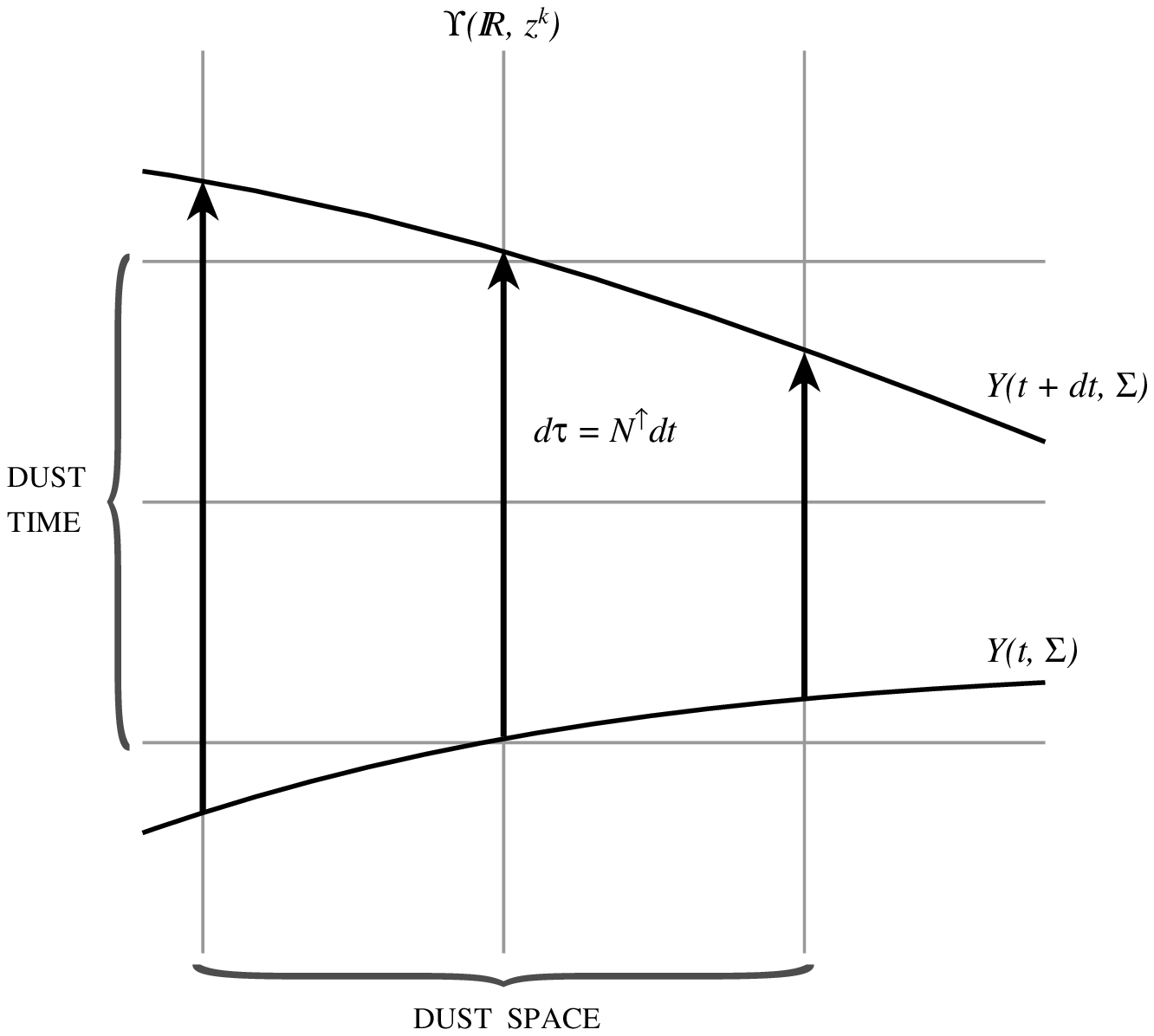}
\end{center}
\noindent {\bf Figure 2:} The hypersurface $Y(t,\Sigma)$ is displaced
to the hypersurface $Y(t+d t, \Sigma)$
along the dust flow lines $z^{k} = {\rm const}$
by the proper time $d \tau = N^{\uparrow}\Big(x=Z^{-1}(z)\Big)d t \,$.
The displacement is generated by the Poisson brackets with
$H[N^\uparrow]\,dt$.
\vfill\eject
\vspace*{1in}
\begin{center}
\leavevmode
\epsffile{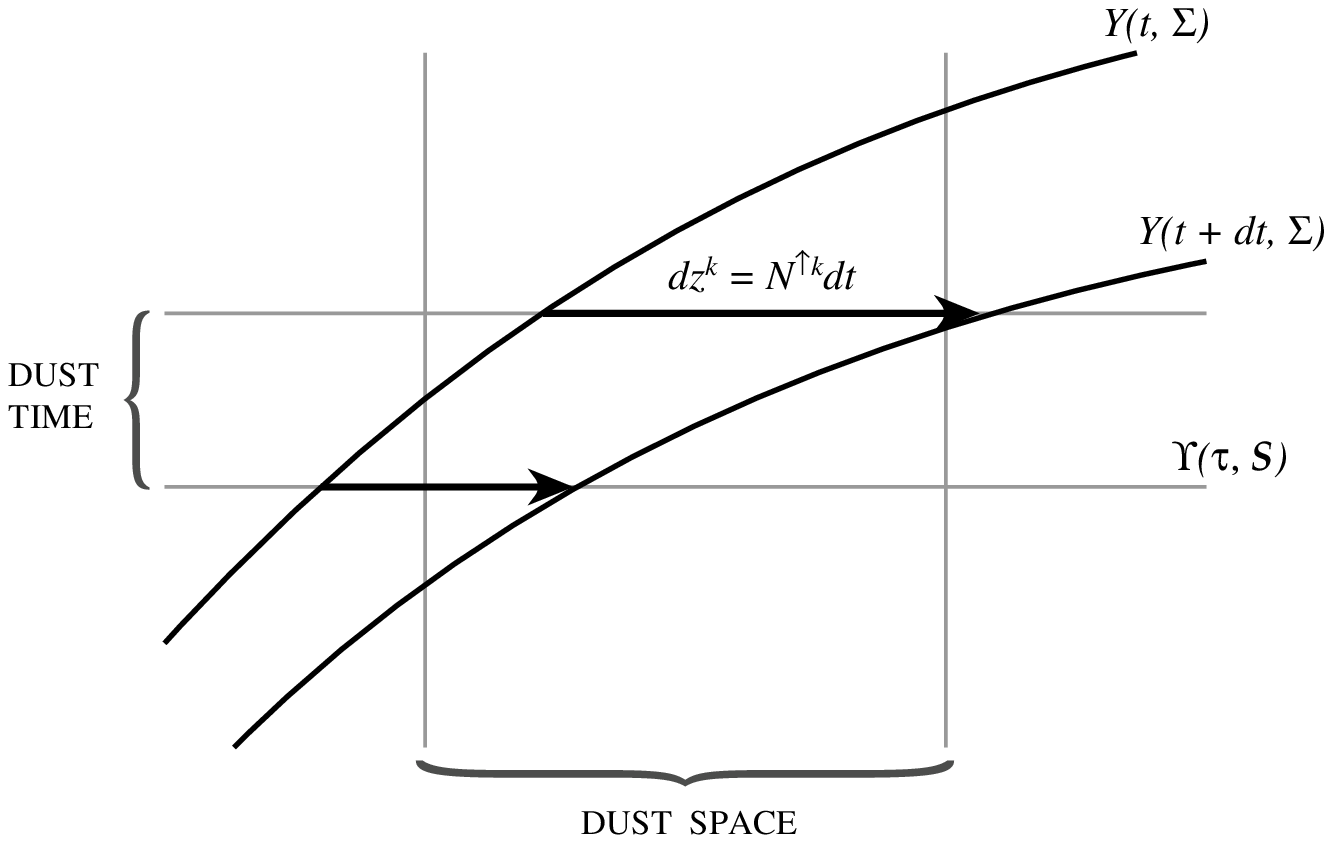}
\end{center}
\noindent {\bf Figure 3:} The hypersurface $Y(t,\Sigma)$ is displaced
to the hypersurface $Y(t+d t, \Sigma)$ along the leaves $\Upsilon
(\tau, {\cal S})$ of the dust time foliation $\cal T$ by the amounts
$d z^{k} = N^{\uparrow k} \Big ( x=Z^{-1}(z) \Big)\, d t$ in the
dust space $\cal S$. The displacement is generated by the Poisson
brackets with $H[{\vec N}^\uparrow]\,dt \,$.
\vfill\eject

\begin{references}
\bibitem{Wheeler} J. A. Wheeler, in {\em Batelle Rencontres: 1967 Lectures
in Mathematics and Physics\/}, edited by C. DeWitt and J. A. Wheeler
(Benjamin, New York, 1968).
\bibitem{DW} B. S. DeWitt, Phys. Rev. {\bf 160}, 1113 (1967).
\bibitem{K-Ox} K. V. Kucha\v{r}, in
  {\em Quantum Gravity 2: A Second Oxford Symposium\/}, edited by
  C. J. Isham, R. Penrose, and D. W. Sciama (Clarendon, Oxford, 1981).
\bibitem{k-nosymmetry} K. V. Kucha\v{r}, J. Math. Phys. {\bf 22},
  2640 (1981).
\bibitem{KonT} K. V. Kucha\v{r}, in {\em Proceedings of the 4th
  Canadian Conference on General Relativity and Relativistic
  Astrophysics\/}, edited by G. Kunstatter, D. Vincent, and J. Williams
  (World Scientific, Singapore, 1992).
\bibitem{Einstein} A. Einstein, {\em \"{U}ber die Spezielle und die
  Allgemeine Relativit\"{a}tstheorie} (Vieweg, Braunschweig, 1920); English
  translation by R. W. Lawson, {\em Relativity: The Special and the General
  Theory} (Crown, New York, 1961).
\bibitem{Hilbert} D. Hilbert, 2 Mitt., Nachr. Ges. Wiss. G\"{o}ttingen
  {\bf 53} (1917).
\bibitem{Moller} C. M{\o}ller, {\em The Theory of Relativity}
  (Oxford University Press, Oxford, 1952).
\bibitem{Landau} L. D. Landau  and E. M. Lifshitz, {\em The Classical
  Theory of Fields}, 4th edition (Pergamon Press, Oxford, 1989).
\bibitem{K+T} K. V. Kucha\v{r} and C. G. Torre, Phys. Rev. D {\bf
  43}, 419 (1991).
\bibitem{K+S+T} K. V. Kucha\v{r} and C. G. Torre, Phys. Rev. D {\bf
  44}, 3116 (1991).
\bibitem{Stone+K} C. L. Stone and K. V. Kucha\v{r}, Class. Quantum
  Grav. {\bf 9}, 757 (1991).
\bibitem{K=con} K. V. Kucha\v{r}, Phys. Rev. D {\bf 44}, 43 (1992).
\bibitem{DW-medium} B. S. DeWitt, in {\em Gravitation: An
  Introduction to Current Research}, edited by L. Witten (Wiley, New
  York,1962).
\bibitem{Brown2} J. D. Brown, ``On variational principles for
  gravitating perfect fluids", submitted to Class. Quantum Grav.
\bibitem{FluidS} Our dust action is based on the canonical
relativistic perfect fluid action discussed in J. D. Brown, Class.
Quantum Grav. {\bf 10}, 1579 (1993). This action makes use of a
velocity--potential or ``Clebsch" representation of the fluid
velocity (A. Clebsch, J. Reine Angew. Math. {\bf 56}, 1 (1859)).
Clebsch potentials were used in the development of action
principles for nonrelativistic fluids by Lin (C. C. Lin, in
{\em Liquid Helium}, edited by G. Careri (Academic Press, New
York, 1963); see also J. Serrin, in {\em Handbuch der Physik},
volume 8, edited by S. Fl{\"u}gge and C. Truesdell (Springer,
Berlin, 1959)) and by Seliger and Whitham (R. L. Seliger and G. B.
Whitham, Proc. Roy. Soc. A {\bf 305}, 1 (1968)). Tam used Clebsch
potentials to formulate an action principle for an ideal, charged
fluid in the context of special relativity (K. Tam, Can. J. Phys.
{\bf 44}, 2403 (1966)). Action principles for perfect fluids in
general relativity were developed independently by Tam and
O'Hanlon (K. Tam and J. O'Hanlon, Nuovo Cimento {\bf 62}B, 351
(1969)) and by Schutz (B. F. Schutz, Phys. Rev. D {\bf 2},
2762 (1970). The actions discussed by Tam and O'Hanlon and by
Schutz use a minimal set of velocity potentials, which precludes
the interpretation of the appropriate potential fields as
Lagrangian coordinates for the fluid. The Hamiltonian form of
Schutz's action was developed by Schutz (B. F. Schutz, Phys.
Rev. D {\bf 4}, 3559 (1971)) and by Demaret and Moncrief
(J. Demaret and V. Moncrief, Phys. Rev. D {\bf 21}, 2785 (1980)).
\bibitem{Misnerfest} K. V. Kucha\v{r}, in {\em Directions in General
  Relativity}, volume 1, edited by B. L. Hu, M. P. Ryan, and C. V.
  Vishveshwara (Cambridge University Press, Cambridge, 1993);
  J. D. Brown, in {\em Proceedings of the Cornelius Lanczos Centenary
  Conference}, edited by J. D. Brown, M. T. Chu, D. C. Ellison, and
  R. J. Plemmons (SIAM, Philadelphia, in press).
\bibitem{Caratheodory} C. Carath\`{e}odory, {\em Variationsrechnung
  und Partielle Differentialgleichungen Erster Ordnung I} (Teubner,
  Leipzig, 1956).
\bibitem{Born+Infeld} M. Born, Proc. Roy. Soc. A. {\bf 143}, 410 (1934);
  M. Born and L. Infeld, Proc. Roy. Soc. A. {\bf 144}, 425 (1934).
\bibitem{I+K} C. J. Isham and K. V. Kucha\v{r}, Ann. Phys. (N.Y.) {\bf 164},
  288 and 316 (1985).
\bibitem{H+K+T} S. Hojman, K. V. Kucha\v{r} and C. Teitelboim,
  Ann. Phys. (N.Y.) {\bf 96}, 88 (1974).
\bibitem{Synge} J. L. Synge, {\em Relativity: The General Theory}
  (North Holland, Amsterdam, 1964).
\bibitem{KucharScalar} K. V. Kucha\v{r}, Phys. Rev. D {\bf 39},
  1579 (1989).
\bibitem{HRTetc} A. Hanson, T. Regge, and C. Teitelboim,
  {\em Constrained Hamiltonian Systems} (Accad. Naz. dei Lincei,
  Rome, 1976); L. Brink and M. Henneaux, {\em Principles of String
  Theory} (Plenum, New York, 1988); K. V. Kucha\v{r} and C. G. Torre,
  J. Math. Phys. {\bf 30}, 1769 (1989).
\bibitem{TandT} C. Teitelboim, in {\em Quantum Theory of Gravity},
  edited by S. M. Christensen (Adam Hilger, Bristol, 1984);
  C. G. Torre, Phys. Rev. D {\bf 40}, 2588 (1989).
\bibitem{Brown} J. D. Brown, Ref.~[15].
\bibitem{Schutz} B. F. Schutz, first citation in Ref.~[15].
\bibitem{Pauli} W. Pauli, in {\em Die Allgemeinen Prinzipien der
Wellenmechanik, Handbuch der Physik V.1\/}, edited by S. Fl\"{u}gge
(Springer, Berlin, 1958).
\end{references}
\end{document}